\documentclass[reprint,amsmath,amssymb,aps]{revtex4-2}
\usepackage{graphicx}
\usepackage{dcolumn}
\usepackage{braket}
\usepackage{bm}
\usepackage{subfigure}

\begin{document}
\preprint{APS/123-QED}

\title{The Breaking of Geometric Constraint of Classical Dimers on the Square Lattice}
\author{Hongxu Yao}\altaffiliation{Corresponding author;18307110355@fudan.edu.cn}\affiliation{Department of Physics and State Key Laboratory of Surface Physics, Fudan University, Shanghai 200438, China}
\author{Jiaze Li}\affiliation{Department of Physics and State Key Laboratory of Surface Physics, Fudan University, Shanghai 200438, China}
\author{Jintao Hou}\affiliation{Department of Physics and State Key Laboratory of Surface Physics, Fudan University, Shanghai 200438, China}
\date{\today}
	
\begin{abstract}
	We study a model of two-dimensional classical dimers on the square lattice with strong geometric constraints (there is exactly one bond with the nearest point for every point in the lattice). This model corresponds to the quantum dimer model suggested by D.S. Rokhsar and S.A. Kivelson (1988). We use the directed-loop algorithm to show the system undergoes a Berezinskii-Kostelitz Thousless transition (BKT transition) in finite temperatures. After that, if we destroy the geometric constraint of dimers, the topological transition will transfer to a quasi one-order transition. For the dimer updates, we also introduce a new cluster updating algorithm called the edged cluster algorithm. By this method, we succeed in rapidly traversing the winding (topological) sections uniformly and widening the effective matrical ensemble to include more topological sections.
\end{abstract}

\maketitle
\section{Introduction}
	The model of lattice covering hard-core dimers has been proven to be of great value by statistical mechanics in many-body physics. Initially, the quantum dimer model (QDM) was introduced by D.S. Rokhsar and S.A. Kivelson (RK) \cite{1988Superconductivity} as a description of non-Neel phase of the spin-1/2 Heisenberg quantum antiferromagnet phenomenologically. Interest of the suggestion on the lattice with dimers was motivated by P.W. Anderson who proposed the resonating valence-bond (RVB) theory to explain the properties of high temperature superconductivity in $La_{2-x}Sr_xCuO_4$ and $YBa_2Cu_3O_{7-x}$ \cite{1973Resonating,ANDERSON1987The}   with exponentially decaying two-spin correlation functions. For lowering the energy, two electrons couple nearest as a bond called dimer. Furthermore, these singlet bonds tend to form a quasi long-range ordered superposition of RVB so that the energy could be lowered ulteriorly. We could consider there exits one electron in every site. If some of them are removed by doping, superconducting states will appear as RVBs condense. Thus, RK proposed a reasonable approximation to rigidly restrict bonding conditions and completely cover the whole lattice. Eventually, they derive a transition from a dimer crystal state to an insulation quantum liquid state \cite{1988Superconductivity}.

	Based on RK’s early research, R. Moessner, S.L. Sondhi et al. demonstrate the relation between the quantum dimer model and the frustrated quantum antiferromagnet \cite{2001Ising}. Furthermore, they have shown limiting case of 3-dimensional Ising gauge theories are dual to frustrated Ising model \cite{PhysRevB.65.024504}. More following studies by means of series expansion \cite{1989Zero}, large-N expansion \cite{1989Valence} and trick of statistical field theory \cite{2010Statistical} give a comprehensive look of the dual correspondence together. Early authors also have argued that quantum dimer models have been derived from a spin-orbital model to describe $LiNiO_2$ \cite{2007Identification}, the kagome antiferromagnet \cite{2005Effective}  and so on.
	
	In RK’s article \cite{1988Superconductivity}, they argued a phenomenological Hamiltonian in the following form:
	\begin{equation}
		H =  \sum_{plaquettes}[-J(\ket{\mathop{\rule[0.1pt]{0.23cm}{0.05cm}}^{\rule[-0.1pt]{0.23cm}{0.05cm}}}\bra{\rule[-0.6pt]{0.05cm}{0.23cm} \kern0.08cm\rule[-0.6pt]{0.05cm}{0.23cm} }+H.c.)+V(\ket{\mathop{\rule[0.1pt]{0.23cm}{0.05cm}}^{\rule[-0.1pt]{0.23cm}{0.05cm}}}\bra{\mathop{\rule[0.1pt]{0.23cm}{0.05cm}}^{\rule[-0.1pt]{0.23cm}{0.05cm}}}+\ket{\rule[-0.6pt]{0.05cm}{0.23cm}            \kern0.08cm\rule[-0.6pt]{0.05cm}{0.23cm}}\bra{\rule[-0.6pt]{0.05cm}{0.23cm}\kern0.08cm\rule[-0.6pt]{0.05cm}{0.23cm}}] 
	\end{equation}
	Where $J$ and $V$ are regarded as coupling constants, more detailed, we consider $J$ as interacting intensity and $V$ as potential intensity. $N(\rule[-0.6pt]{0.05cm}{0.23cm}\kern0.08cm\rule[-0.6pt]{0.05cm}{0.23cm} )$ and $N(\pmb{=})$ represent parallel dimers in a plaquette. The first term of $H$ is the dimer kinetic energy with complex conjugate terms and the second term is the potential energy. Every states in Hilbert space could be linearly represented by all close-packed dimer configurations. Meanwhile, each state could be divided into different classes, and if they are conserved by winding numbers ($W_x$,$W_y$), we can define that they are in a same topological section \cite{2020Improved} (refer to Appendix \ref{app1}).
	
	Stemming from the quantum dimer model, classical dimer model also plays a significant theoretical role due to the following two reasons: one is the fact that the quantum wave function is a superposition of all classical dimer configurations; the other is that the limit of the quantum model at high temperature and low interaction limit is directly correlated to the classical dimers. Specially, classical dimer model is the case of $V>>J$ in QDM, and the Hamiltonian will be denoted as the following:
	\begin{equation}
		H=\sum_{plaquettes}V(\ket{\mathop{\rule[0.1pt]{0.23cm}{0.05cm}}^{\rule[-0.1pt]{0.23cm}{0.05cm}}}\bra{\mathop{\rule[0.1pt]{0.23cm}{0.05cm}}^{\rule[-0.1pt]{0.23cm}{0.05cm}}}+\ket{\rule[-0.6pt]{0.05cm}{0.23cm}\kern0.08cm\rule[-0.6pt]{0.05cm}{0.23cm}}\bra{\rule[-0.6pt]{0.05cm}{0.23cm}\kern0.08cm\rule[-0.6pt]{0.05cm}{0.23cm}})
	\end{equation}

	Recently, researchers have used the method of transfer matrix to give a rigid mathematics solution and they have derived some good conclusions of the quantum dimer transition in finite temperatures on square \cite{2005Interacting}, triangular \cite{2007Criticality}  and hexagonal lattices in approximate ways. However, for classical dimers, transfer matrix needs enormous computing power on a huge lattice and it can not change geometric conditions for the elementary dimers. In principle, we think the question could be solved if we cosider the theories such as conformal field theory. But before the miscellaneous theories, numerical simulation is urgently needed.
	
	In our work, one of the main targets is to present the BKT transition process of dimer model using the method of Monte Carlo \cite{2011A} and directed loop algorithm \cite{2003The} on the square lattice. Different from early articles, we measure the thermodynamic quantities in every microstates and the whole process of simulations is high-speed. After that, simplifying the actual doped high superconducting system, we introduced a geometrical breaking for classical dimer model on the square lattice. The conclusion shows that topological properties disappear and some divergent properties emerge. This evidence proves that the new transition transfer to a quasi first-order phase transition. We also find that the directed loop algorithm could bring relatively large error cumulants in high temperatures. We confirm the reason is the lock effect of topology(it means the directed loop is hard to cross topological sections). For more actual measurements, we introduce edged cluster algorithm in our classical dimer model. This algorithm is a melioration of the pocket algorithm \cite{Werner2003Pocket}  analogizing to the Wolff algorithm \cite{Wolff1989Collective}  in the traditional Ising model. By this algorithm, we widen the effective matrical ensemble that could includes more topological sections. 
\section{Foundamentions}
\subsection{Model and the geometric constraint}	
\begin{figure}[htbp]
		\centering
		\subfigure[]{ \label{a} \includegraphics[scale=0.15]{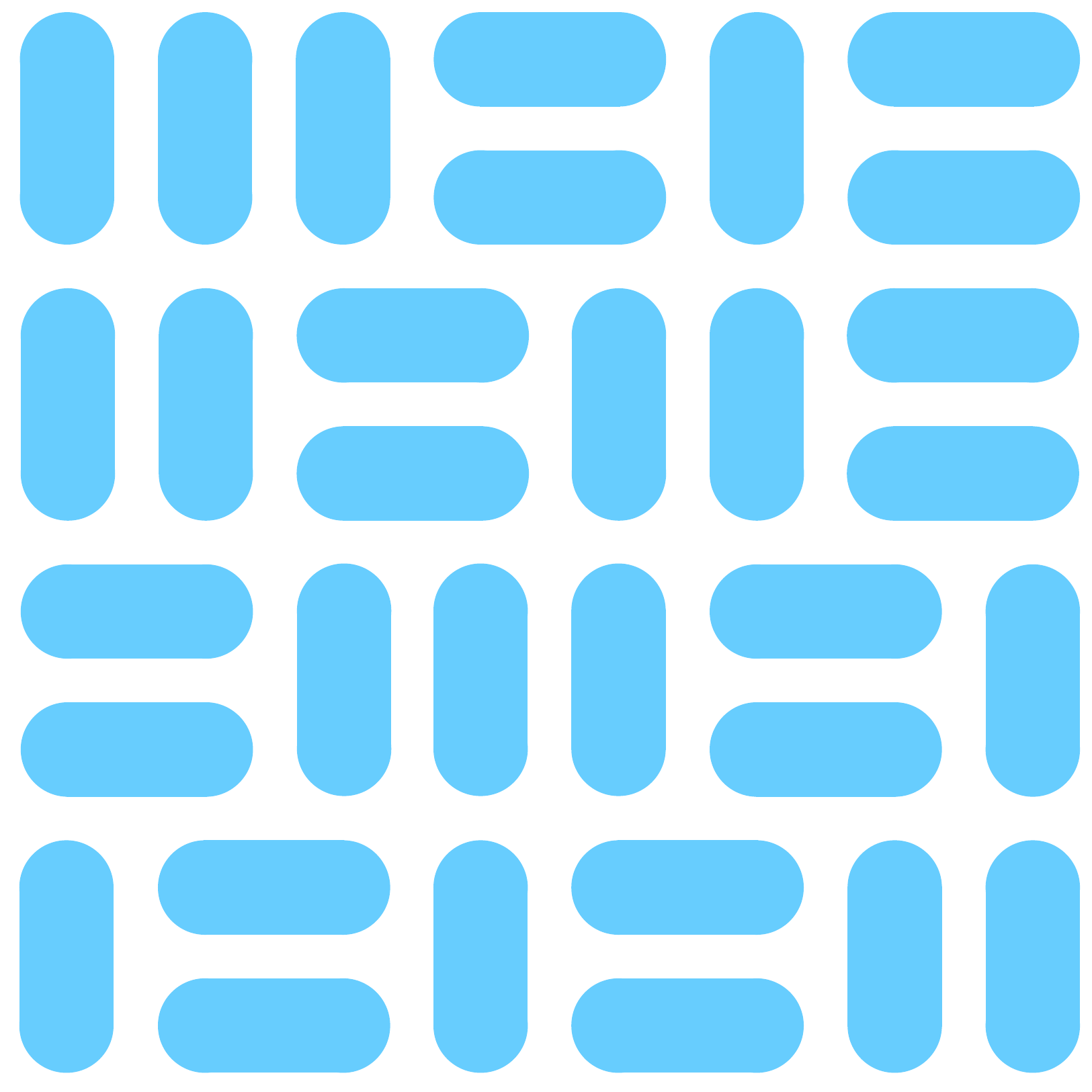}}
    	\quad
    	\subfigure[]{ \label{b} \includegraphics[scale=0.19]{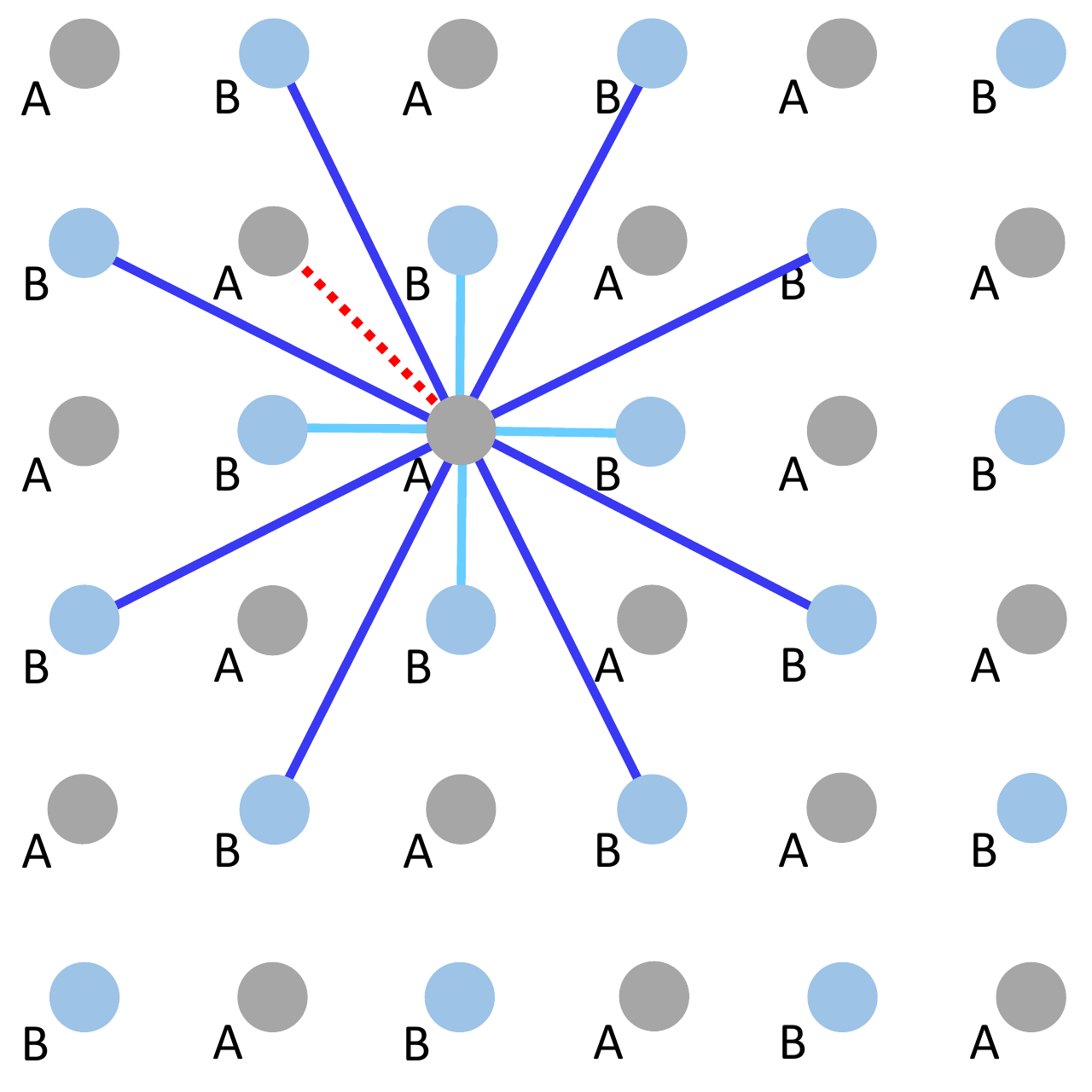}}
    	\quad
    	\subfigure[]{ \label{c} \includegraphics[scale=0.14]{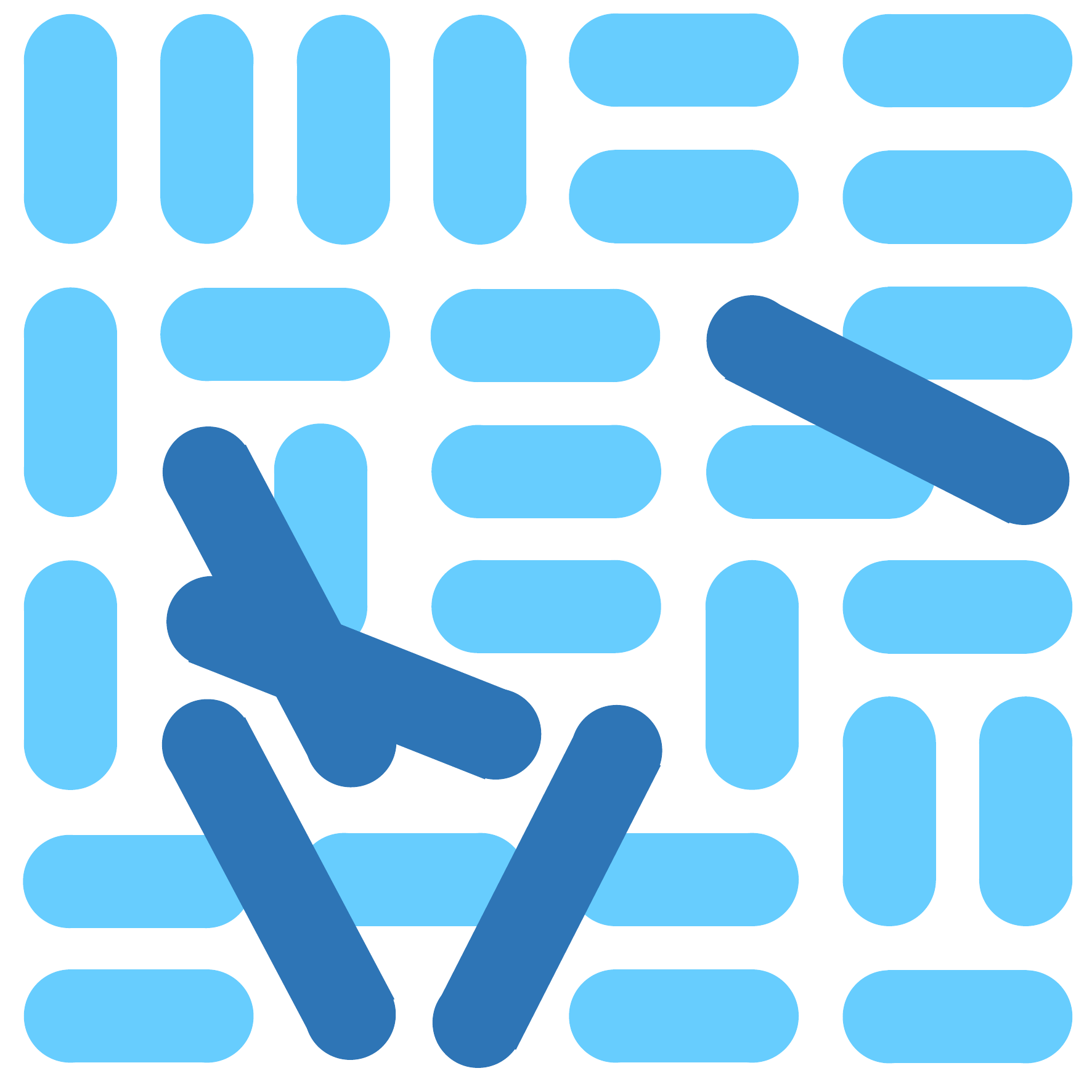}}
    	\quad
    	\subfigure[]{ \label{d} \includegraphics[scale=0.096]{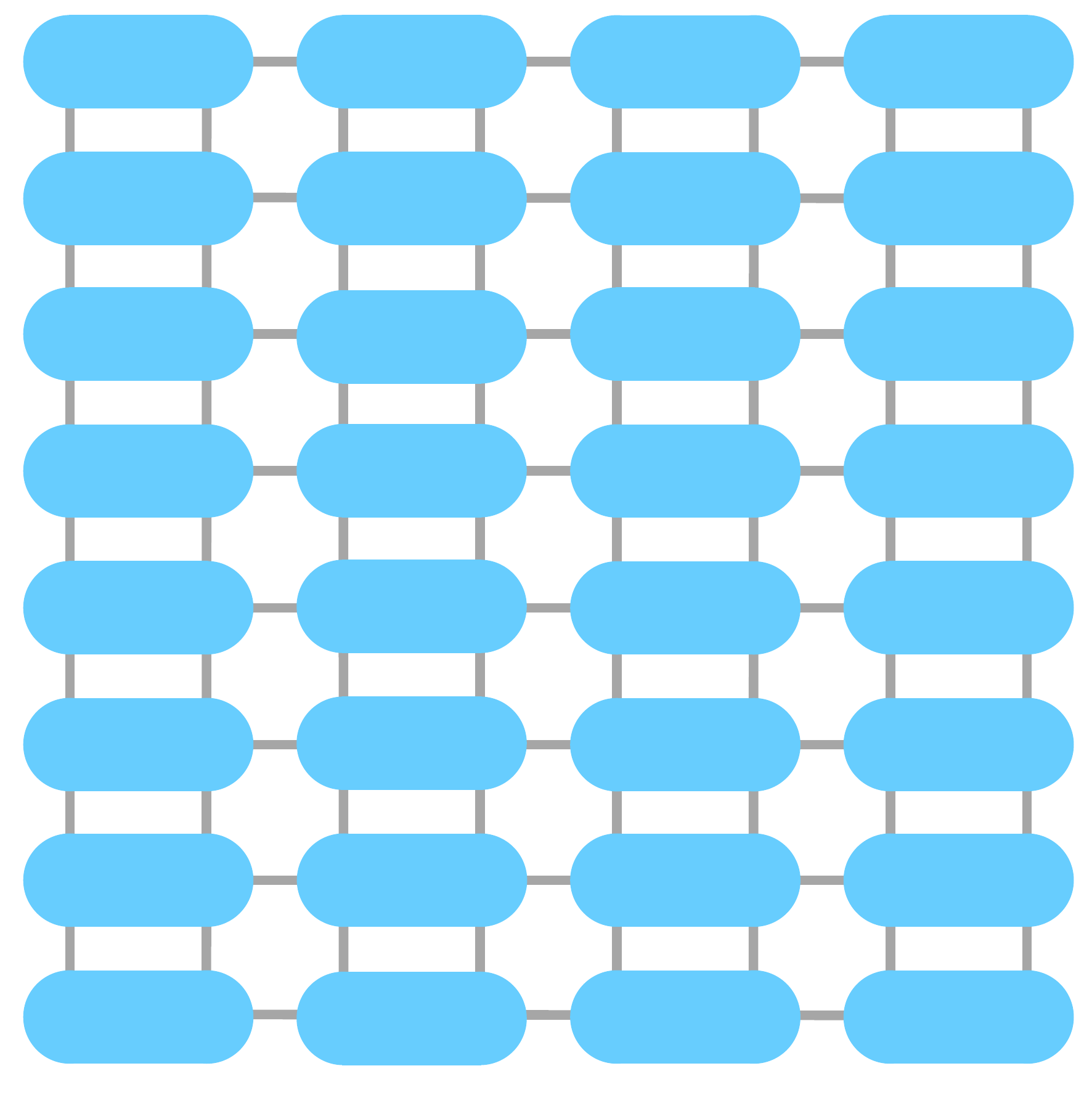}}
    	\quad
    	\subfigure[]{ \label{e} \includegraphics[scale=0.096]{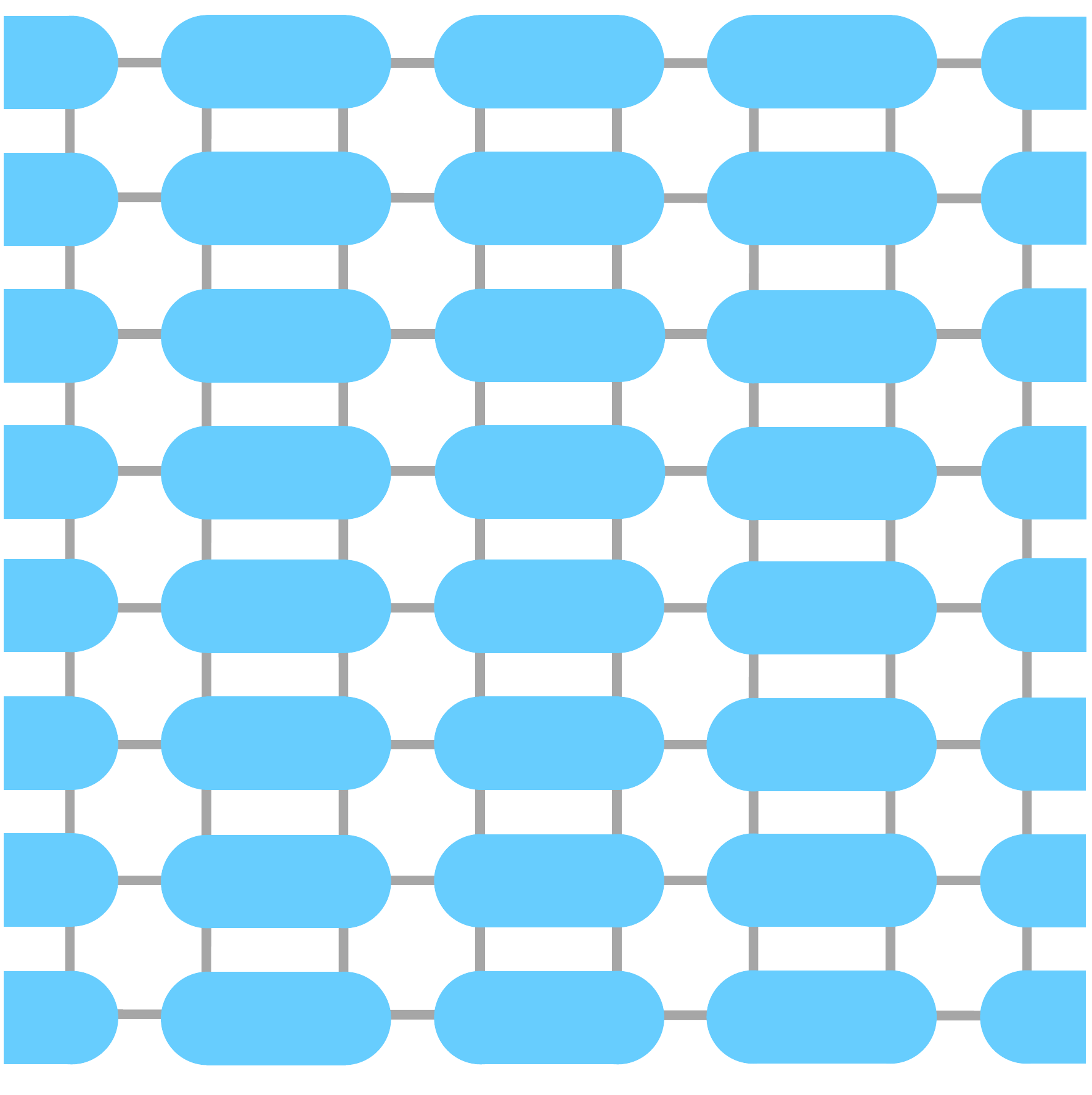}}
    	\quad
    	\subfigure[]{ \label{f} \includegraphics[scale=0.096]{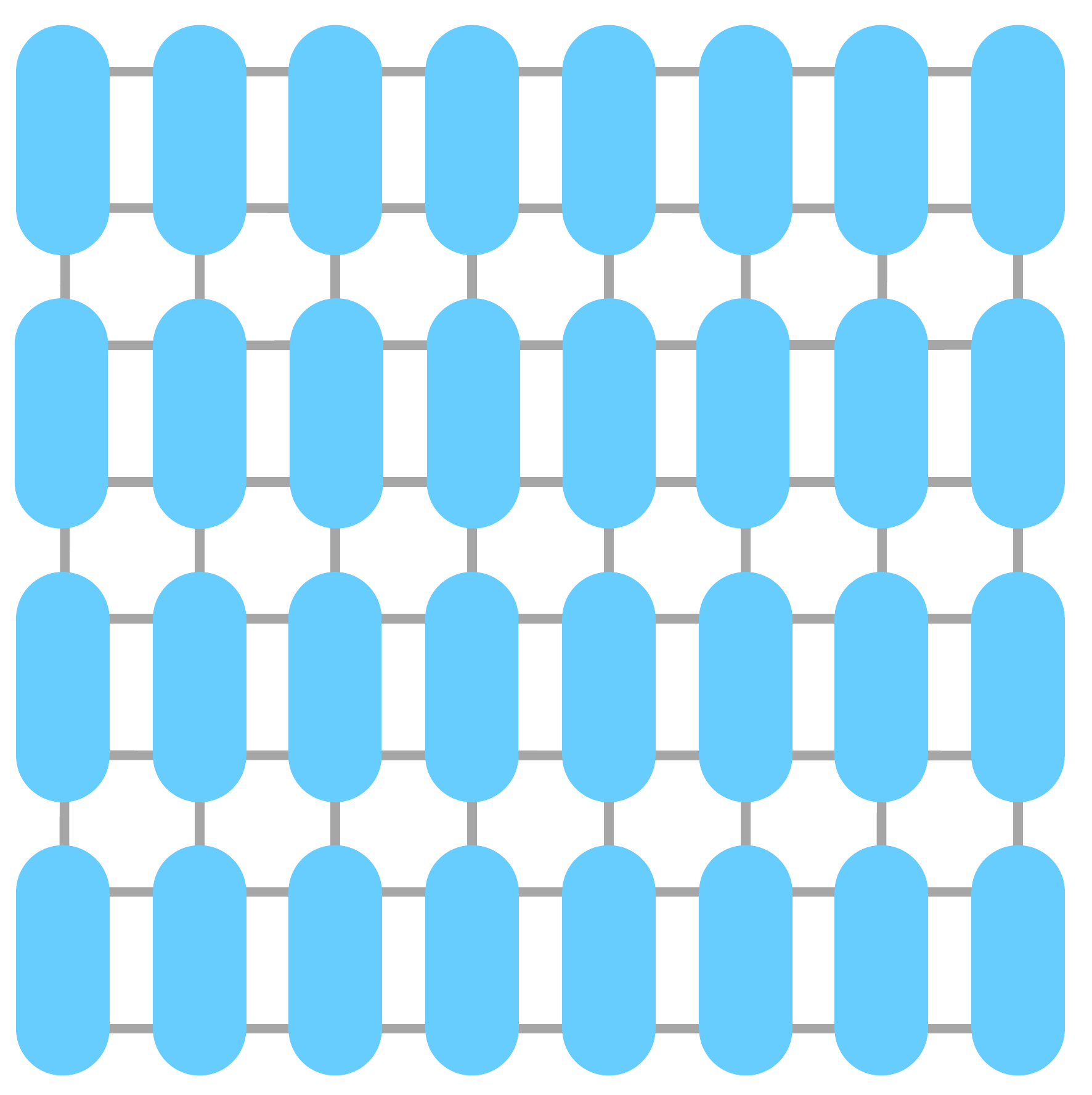}}
    	\quad
    	\subfigure[]{ \label{g} \includegraphics[scale=0.096]{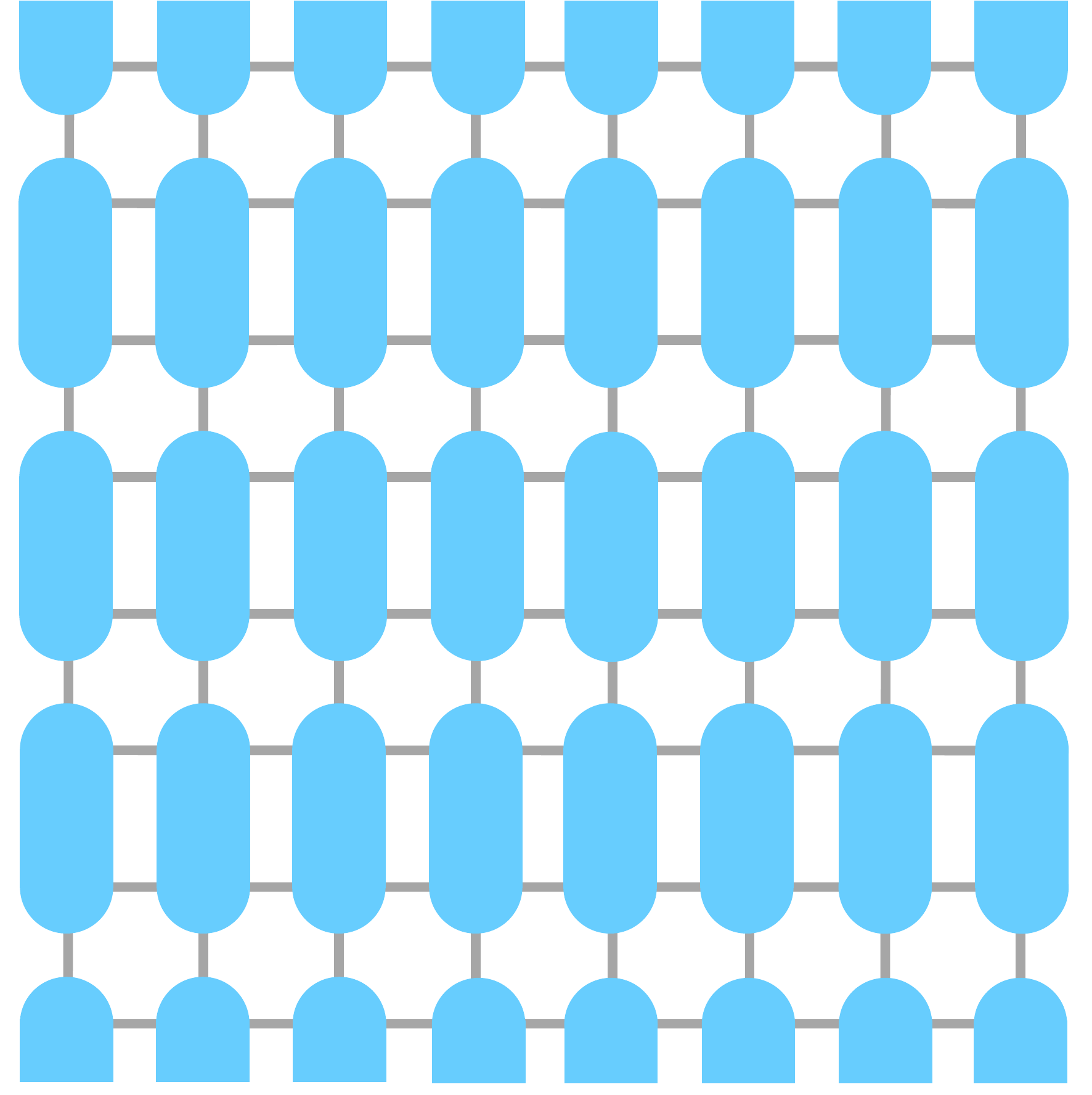}}
		\caption{(a): classical dimer model and plaquette RVB (gray line). (b): geometric constraints. (c): softed dimer model. (d)(e)(f)(g): fourfold degenerate states when T=0(columnar states). }
	\end{figure}

	The model studied by us is points on the square lattice bond to another nearest one and every point could just bond once. Each of bond is called a dimer (see Fig.\ref{a} and tint line in Fig.\ref{b}). As the temperature increases from $T=0$ to infinity, bonds in the model will change. Different temperatures correspond to different confinements with respect to a relatively stable state at a unique temperature. Specially, when $T=0$, dimer model will stay at ordered ground states denoted by \textbf{columnar states} shown in Fig.\ref{d}, \ref{e}, \ref{f} and \ref{g}. One mentionable point is our ground states are fourfold degenerate due to the whole system is double lattice (an atom A is the nearest neighbor to an atom B alternately). These columnar phases break translational and rotational symmetry. According to the motivation of RVB and interaction in Heisenberg model referred in the introduction, a pair of parallel dimers in a plaquette will be regarded as a unit of energy. Considering the geometric structure, we define the partition function
	\begin{equation}
		Z=\sum_{state}exp[-\frac{k}{T}(N(\mathop{\rule[0.1pt]{0.23cm}{0.05cm}}^{\rule[-0.1pt]{0.23cm}{0.05cm}})+N(\rule[-0.6pt]{0.05cm}{0.23cm}\kern0.08cm\rule[-0.6pt]{0.05cm}{0.23cm}))]
	\end{equation}
	where $k$ is Boltzmann factor and $T$ is temperature. Without loss of generality, we can set up $k=-1/ln2$. $N(\pmb{=})$ and $N(\rule[-0.6pt]{0.05cm}{0.23cm}\kern0.08cm\rule[-0.6pt]{0.05cm}{0.23cm})$ are numbers of plaquette dimer pairs in a state and we can assign a unit of energy $k$.

	Unlike conventional order parameter used before, we propose a new 2-dimentional order parameter could distinguish fourfold degenerate states,
	\begin{equation}
		\pmb{\mu} = (\epsilon_1N(\rule[2.5pt]{0.23cm}{0.05cm}),\epsilon_2N(\rule[0pt]{0.05cm}{0.23cm}))   
	\end{equation}
	Where $N(\rule[2.5pt]{0.23cm}{0.05cm})$ and $N(\rule[0pt]{0.05cm}{0.23cm})$ are the numbers of horizontal and vertical dimers, $\epsilon_1$ and $\epsilon_2$ are odd/even factors. If the bond is A-B, $\epsilon_i$=1; otherwise B-A, $\epsilon_i$=-1.

	We now introduce the breaking of the geometric constraint in our model. From the nearest neighbor model and based on the hypothesis of double lattice, we connect A (or B) with the next nearest neighbor B (or A) shown in Fig.\ref{b} (deep line) and we call the new bond \textbf{snAB bond}(second-nearest A-B bond). The reason why we don’t connect the diagonal line of a plaquette is if the bond is formed by two same atoms, frustrations will be introduced to the system(see Fig.\ref{b} dashed line). Meanwhile, in superconductivity, atom A bonding to atom B is a springboard such as valence bond in $La_{2-x}Sr_xCuO_4$ between metal and the oxygen. New constrained system could be a simplification of doped superconductivity. We let the snAB bond is just the entanglement of atom A (or B) and the next nearest atom B (or A). That is, whether a singlet snAB bond or bond-bond pairs are without energy. For the softed dimer model(see Fig.\ref{c}), we consider its ground states are the same as our fourfold degenerate phases above. Because the energy of every formed snAB bond is 0, more than the energy $k$ of a plaquette of dimers. Both two models are set by periodic boundary conditions in this article and we use Metropolis sampling method \cite{1953Equation} cephalocaudally.	
\subsection{Algorithm achievement}
\subsubsection{Simulation process}
	The foundamental method in our simulation process is Monte Carlo. From fourfold degenerate columnar states at T=0, we increase temperature at regular intervals. At each temperature, if we implement updates until physical quantities are convergent, we can derive statistical averages as quantities what we need.  At finite temperatures, we can finally derive the thermodynamic evolution processes and the proporties of transition of our model.
\subsubsection{State presentation}
	Normally, lattice used in the physics field can be represented by graph. However, due to the supremely unique geometric constraints condition, we produce a more streamlined representation defined as \textbf{H.J.J. matrix} (called by the names of three authors of this article). The order of \textbf{H.J.J.} is $L$ and every point on the lattice can be mapped to a single matrix element whose value reflects the angle of the dimer coordinating in the point. For instance, ``\rule[2.5pt]{0.23cm}{0.05cm}" can be denoted by $\begin{bmatrix}0&\pi\end{bmatrix}$, and "\rule[0pt]{0.05cm}{0.23cm}" can be denoted by $\begin{bmatrix}\frac{3\pi}{2}\\\frac{\pi}{2}\end{bmatrix}$. As a matter of fact, corresponding to the graph, \textbf{H.J.J.} is a simplified form of the collection of adjacency lists.
\subsubsection{Directed loop algorithm}
\begin{figure}[htbp]
		\centering
		\subfigure[]{ \label{3a} \includegraphics[scale=0.45]{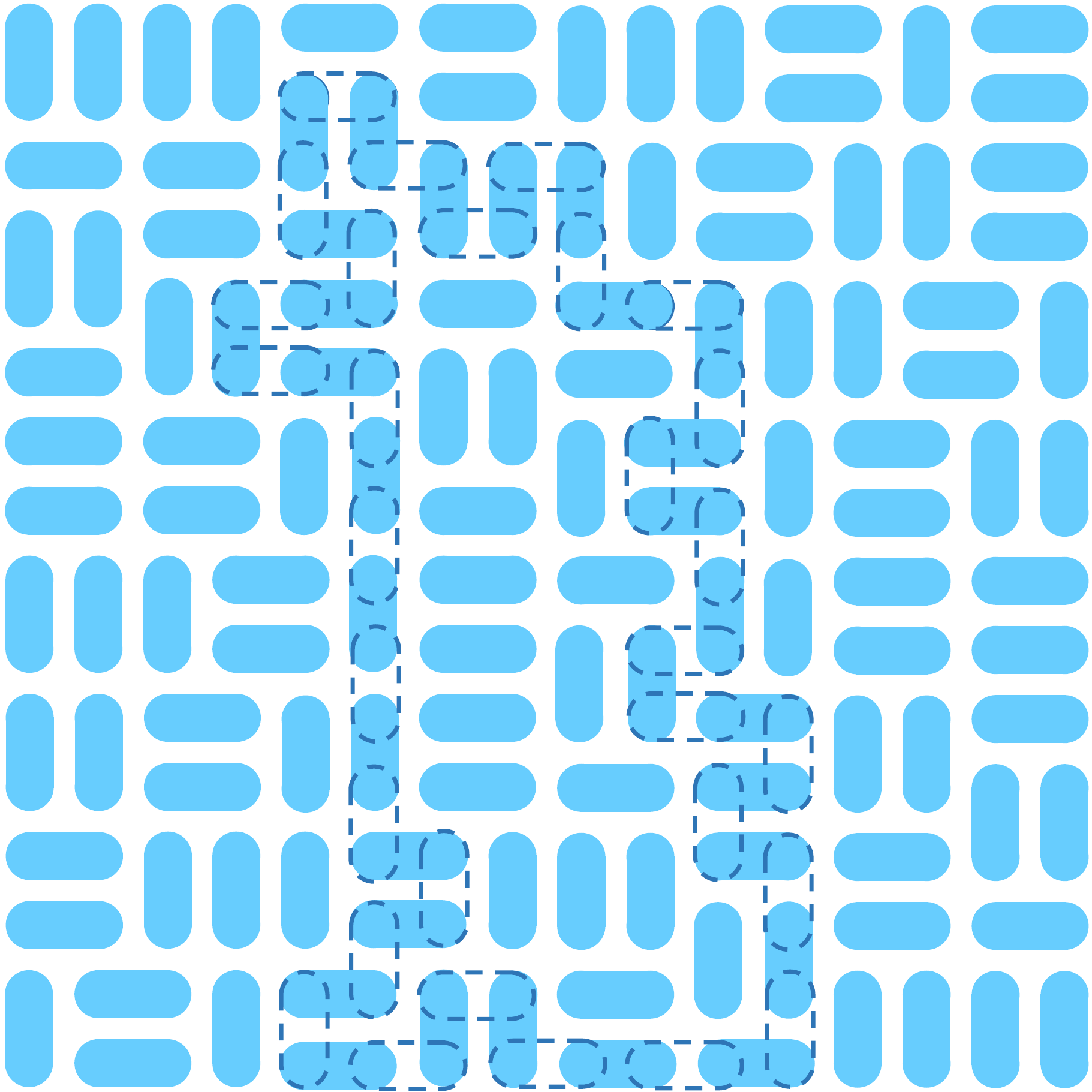}}
		\quad
		\subfigure[]{ \label{3b} \includegraphics[scale=0.45]{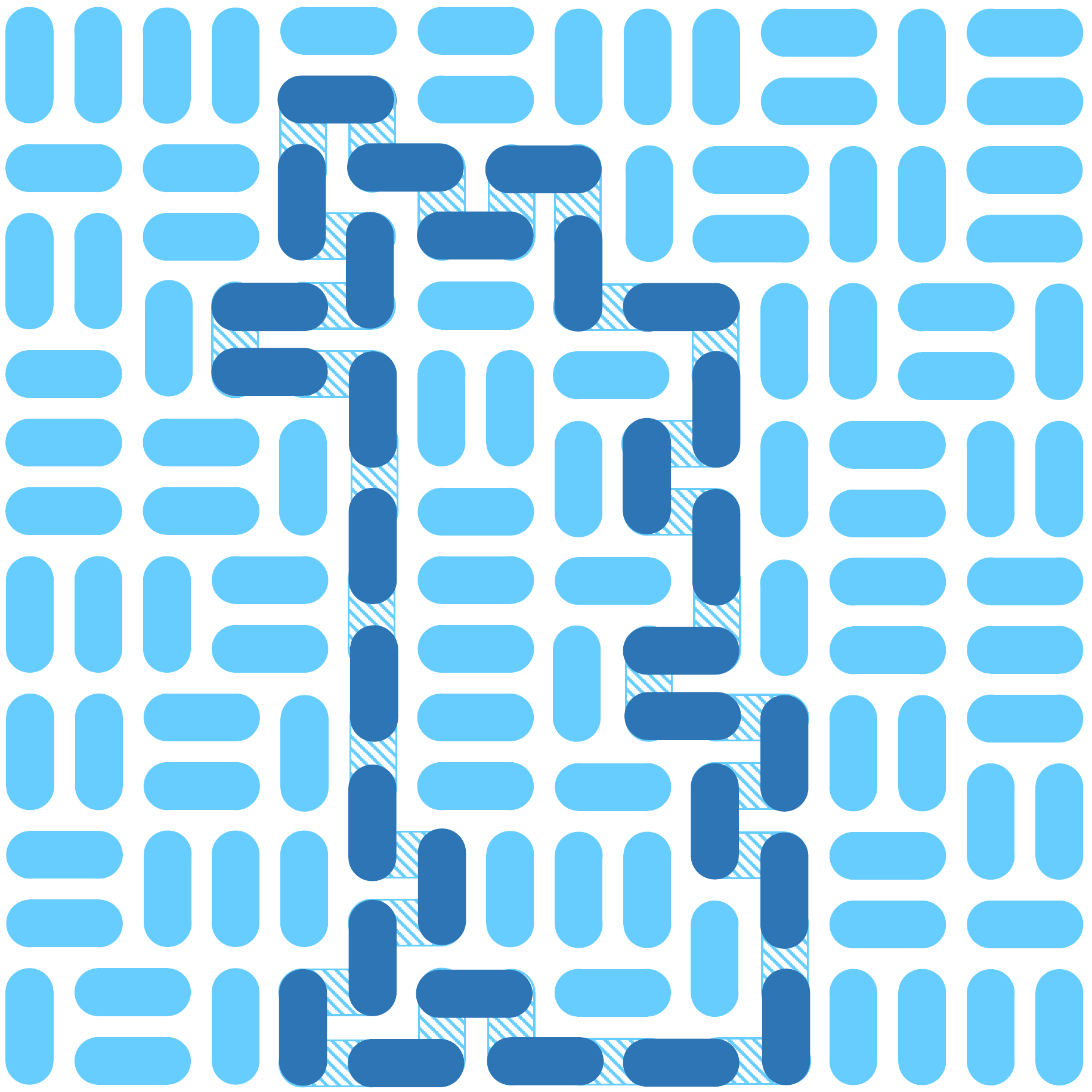}}
		\caption{(a): the chosen process of directed loop algorithm. Former dimers in the loop will vanish while new dimers(dimers with dashed line in the figure) are going to form; (b): finishment of update using the (a) loop. }
	\end{figure}

	We use directed loop algorithm \cite{2003The} in the classical dimer model and softed dimer model. For convenient description, the following dimer is directed from the forward points to the subsequent points. For a given state in the Monte Carlo process, we can choose a point on the lattice randomly. Then we can construct a line through the dimer occupied on this point and define the ending point of the dimer as our new starting point. There are 3 points (11, if softed dimer model) near this new point which means the line has three possible directions to go and we choose one of them randomly. Repeating the operation until the line forms a loop(see Fig.\ref{3a}), we could shift all dimers to all nearest pockets along the loop direction. Compute the whole energy change and utilize the Metropolis algorithm to judge whether this update is accepted or not (see Fig.\ref{3b}).
\subsubsection{Edged cluster algorithm}
	Edged cluster algorithm introduced by us can form long loop and traverse topological sections uniformly. It is an improved algorithm from the fundamental of pocket cluster algorithm \cite{Werner2003Pocket}. The utilization of the symmetry axis and the introduction of the seed are two keys of our algorithm. The formation process is in the following:
	\begin{figure}[htbp]
		\centering
		\subfigure[]{ \label{4x} \includegraphics[scale=0.22]{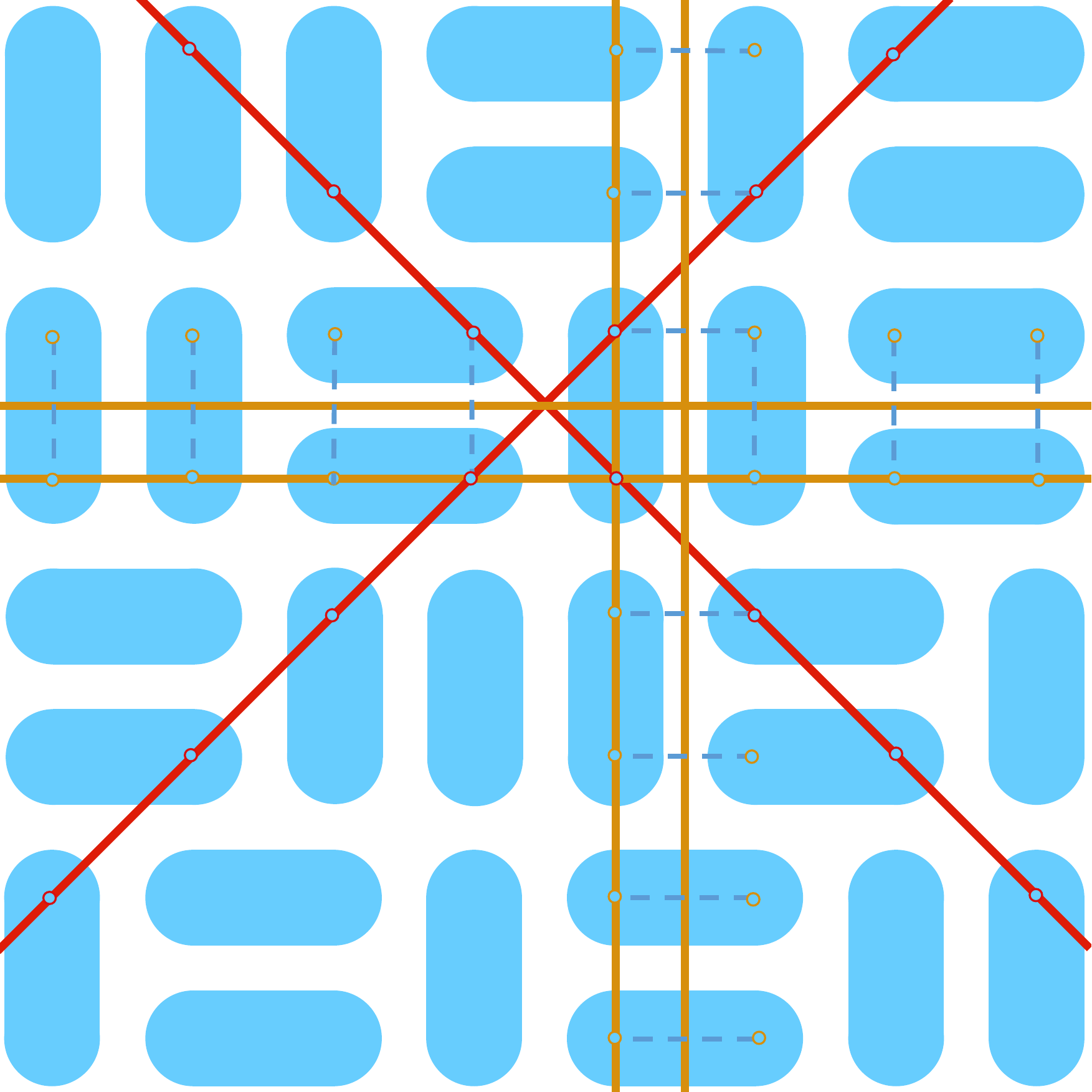}}
		\quad		
		\subfigure[]{ \label{4a} \includegraphics[scale=0.225]{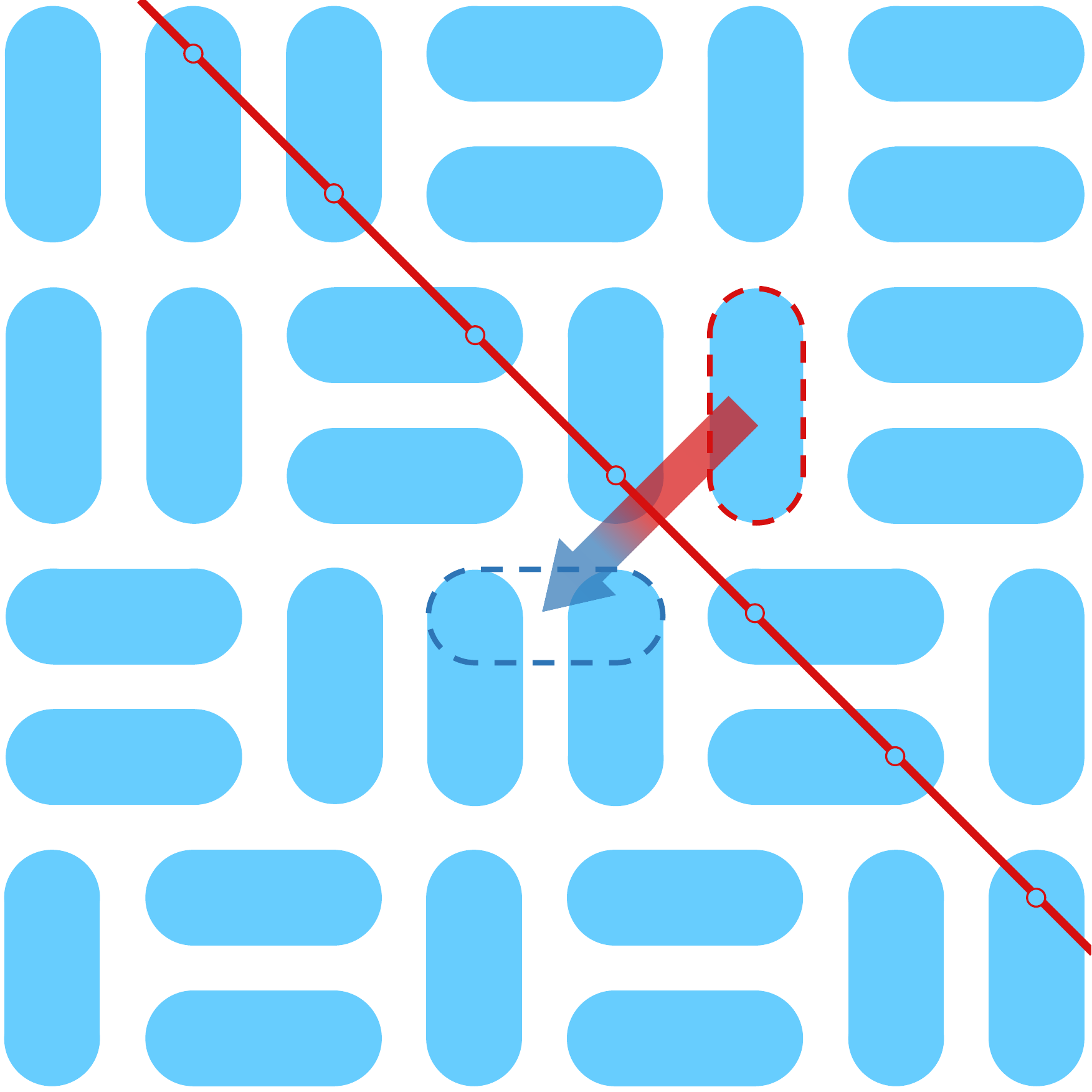}}
		\quad	
		\subfigure[]{ \label{4b} \includegraphics[scale=0.215]{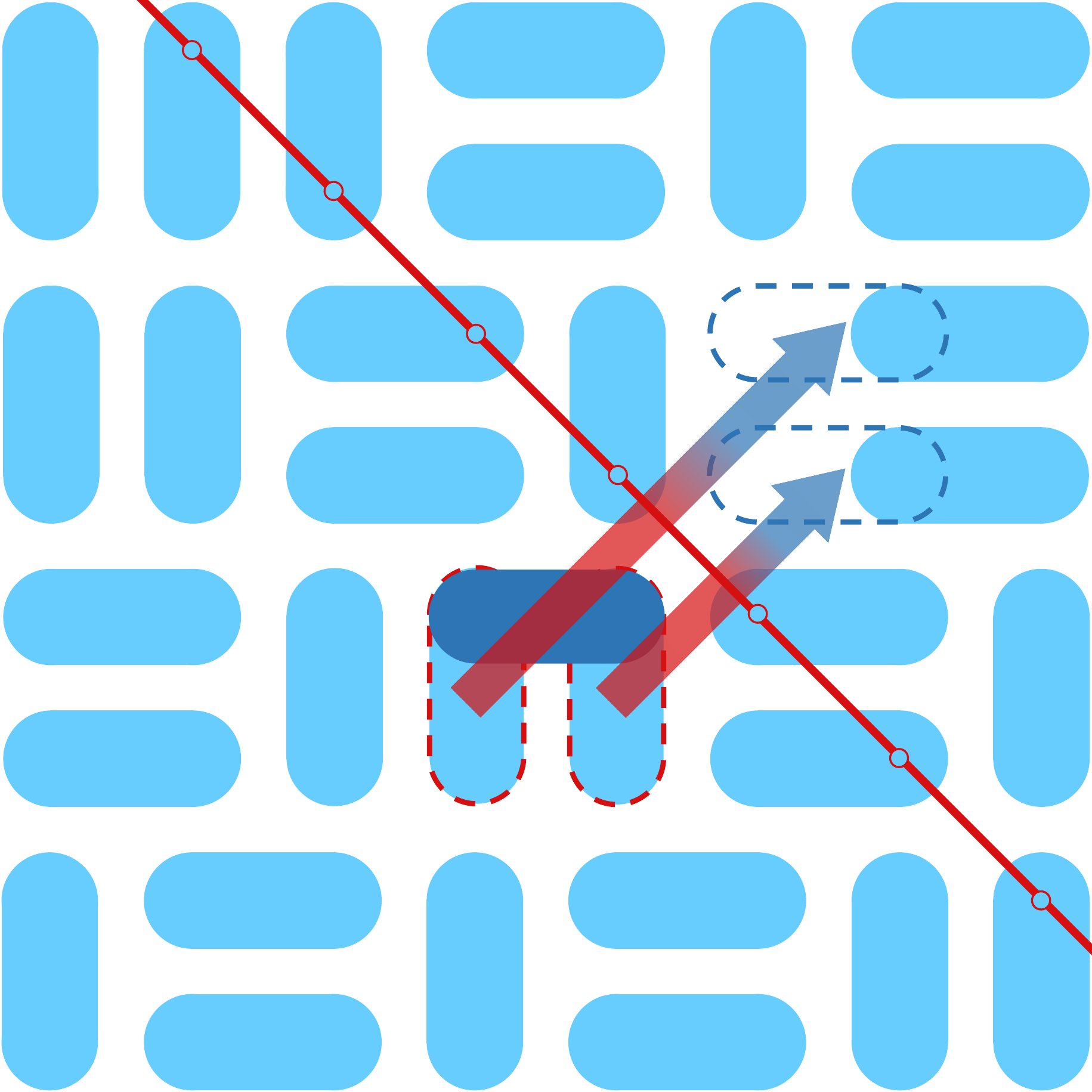}}
		\quad
		\subfigure[]{ \label{4y} \includegraphics[scale=0.22]{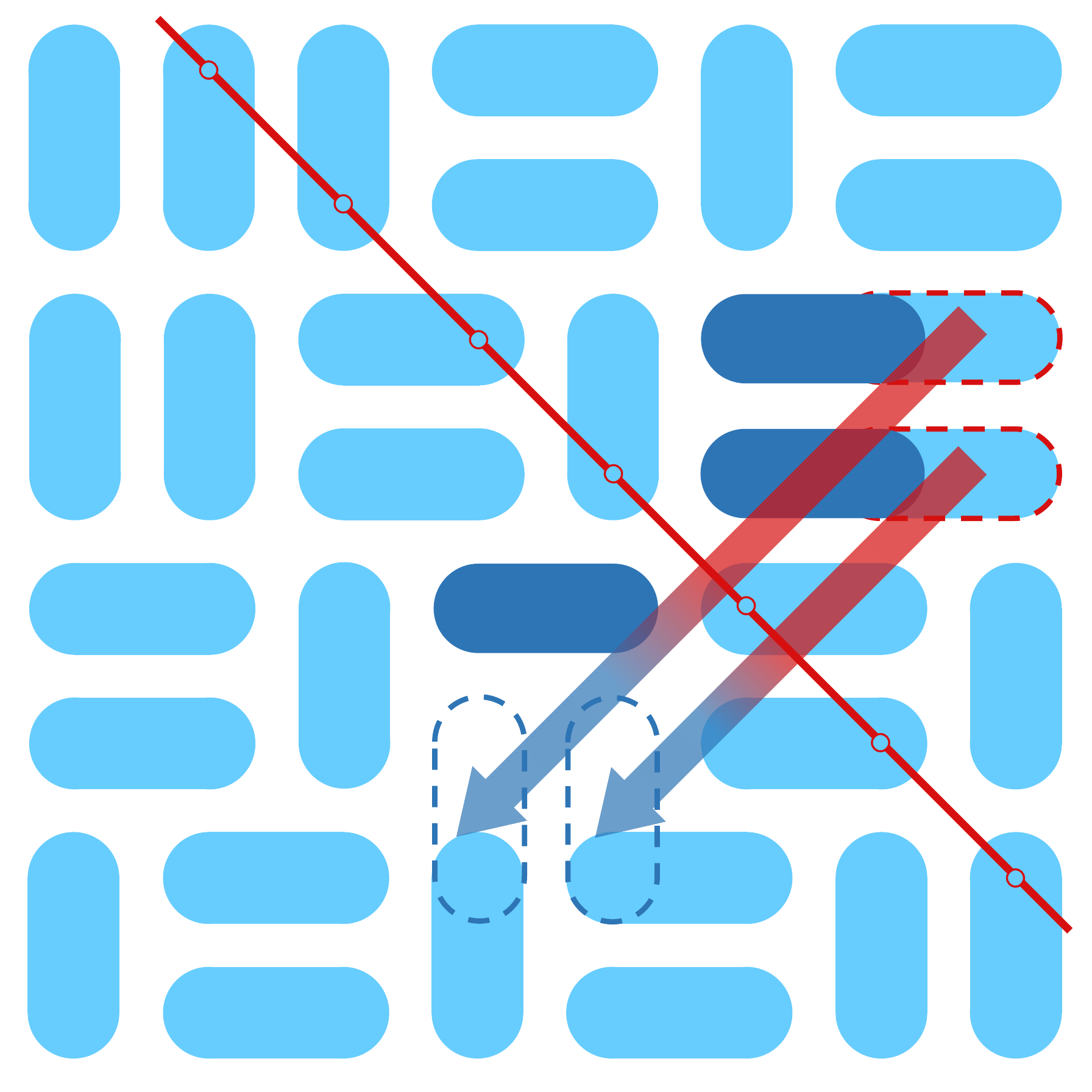}}
		\quad		
		\subfigure[]{ \label{4c} \includegraphics[scale=0.47]{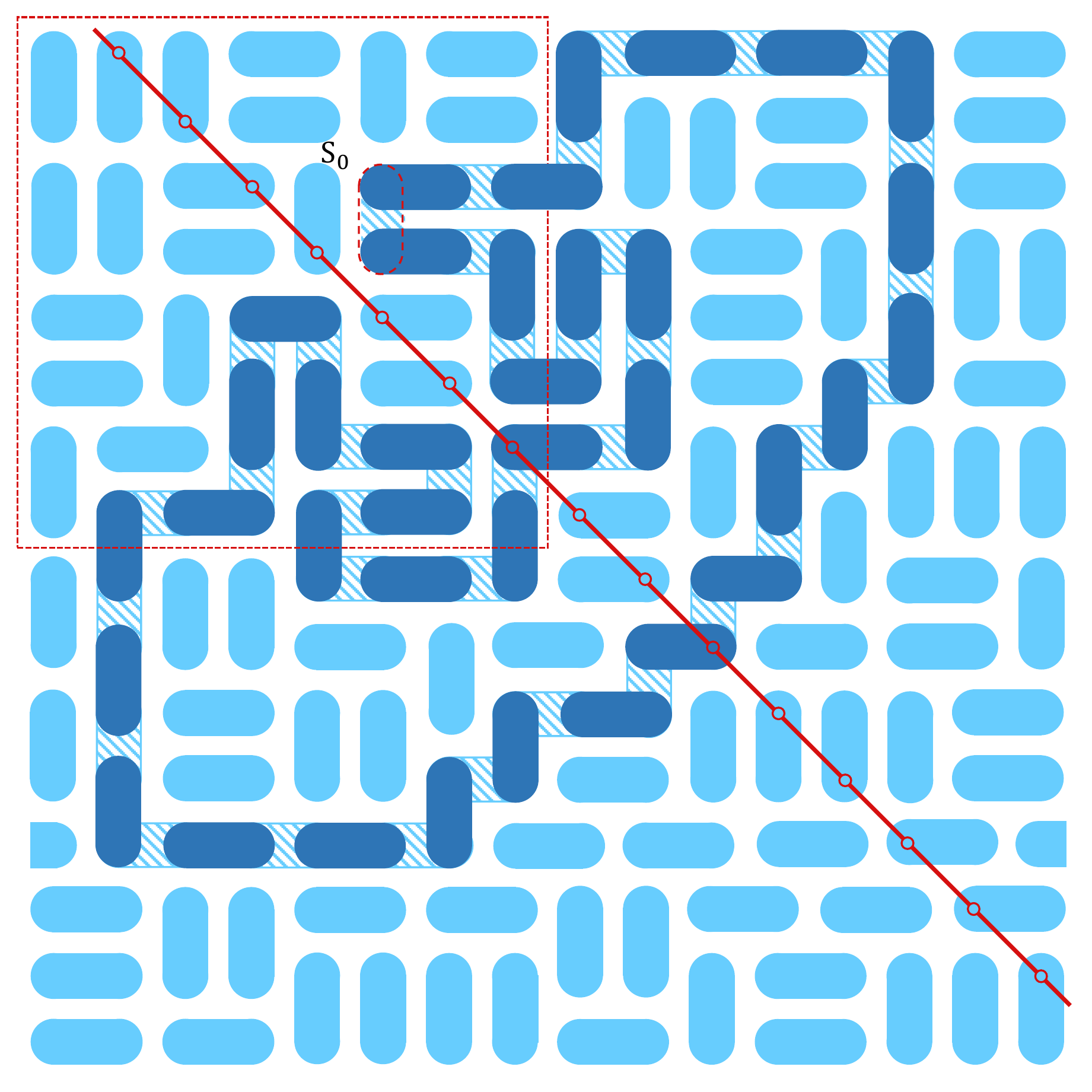}}
		\caption{(a): two kinds of axes can be chosen (vertical and horizontal axes are marked by the tint lines and diagonal axes are marked by deep lines); (b)(c)(d): the formation process of edged loop where red line is the selected axis. We wrap seed dimers by red dashed lines and wrap holes by blue dashed lines; (e): the whole loop formed by edged cluster algorithm (deep marks are new dimers and tint marks are original dimers).}
	\end{figure}

	1). We choose an axis in classical dimer model. Two kinds of axes can be chosen here. One is vertical or horizontal axes shown in Fig.\ref{4x} tint lines. This kind of axes are parallel to grid lines. At the same time, axes must be along points of the square lattice or bisect the line segment between two points. The other is diagonal axes tilting at 45° along the points of the square lattice shown in Fig.\ref{4x} tint lines.

	2). We choose a dimer as a seed randomly (we wrap the chosen dimer with the red dashed lines in Fig.\ref{4a}). 

	3). We reflect the chosen dimer with the selected symmetry axis. The original chosen dimer will leave a hole on the lattice (we wrap the hole with the blue dashed lines in Fig.\ref{4a}). If the reflected new dimer overlaps dimers, overlapped dimers will be chosen as new seeds and the reflected one will stay as the new dimer (reflected dimer is emphasized by deep color and overlapped dimers are marked by red dashed lines in Fig.\ref{4b}).

	4). The seeds reflects and overlaps repeatedly until the final seed are reflected to itself, so we derive a new loop shown in Fig.\ref{4c}.
\section{Numerical Simulation Results}
\subsection{The properties of square lattice}
\begin{figure*}[htbp]
		\subfigure[]{ \label{2a}  \includegraphics[width=12.8cm,height=8.4cm]{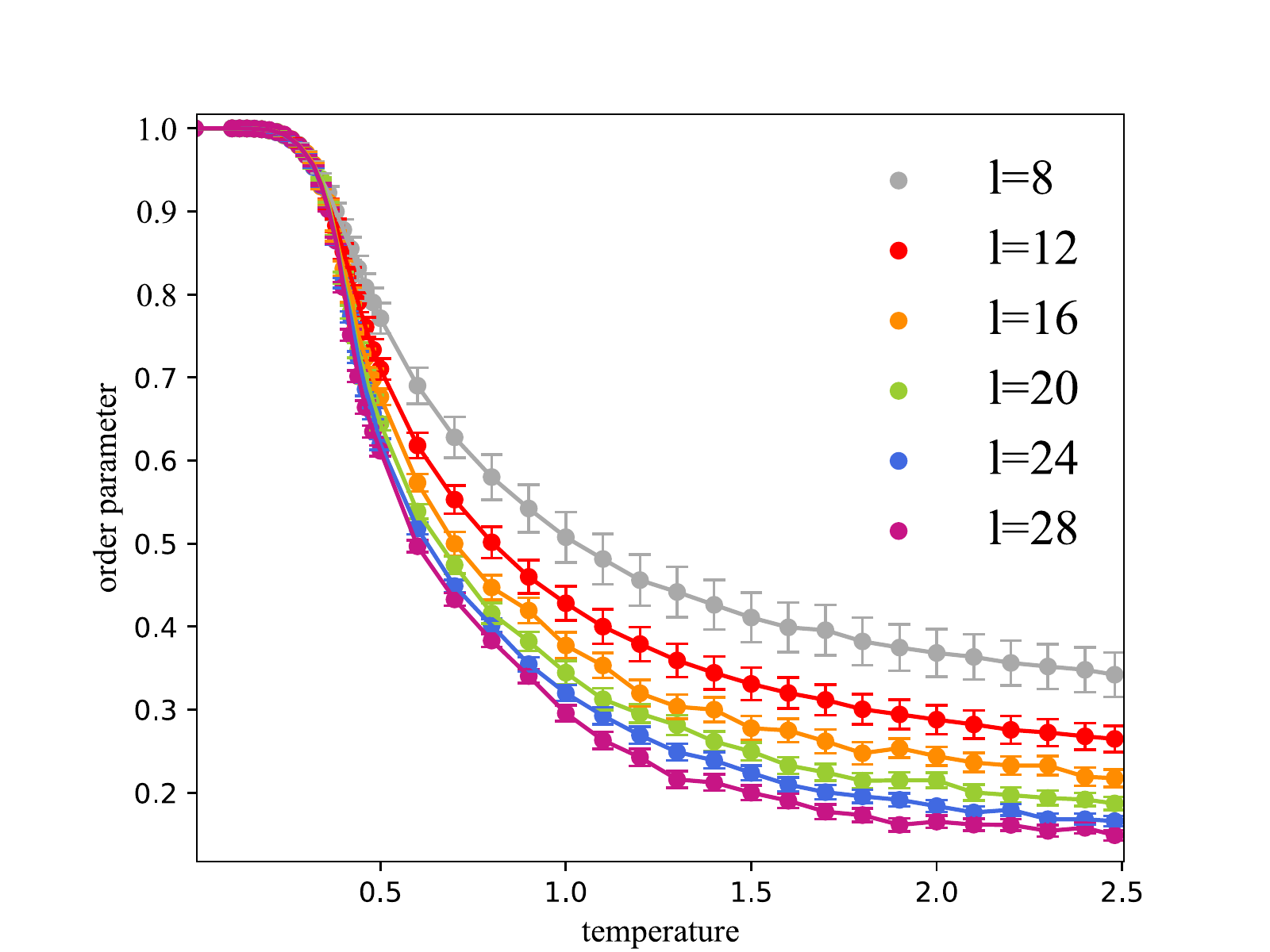}}
		\quad
		\subfigure[]{ \label{2b} \includegraphics[scale=0.45]{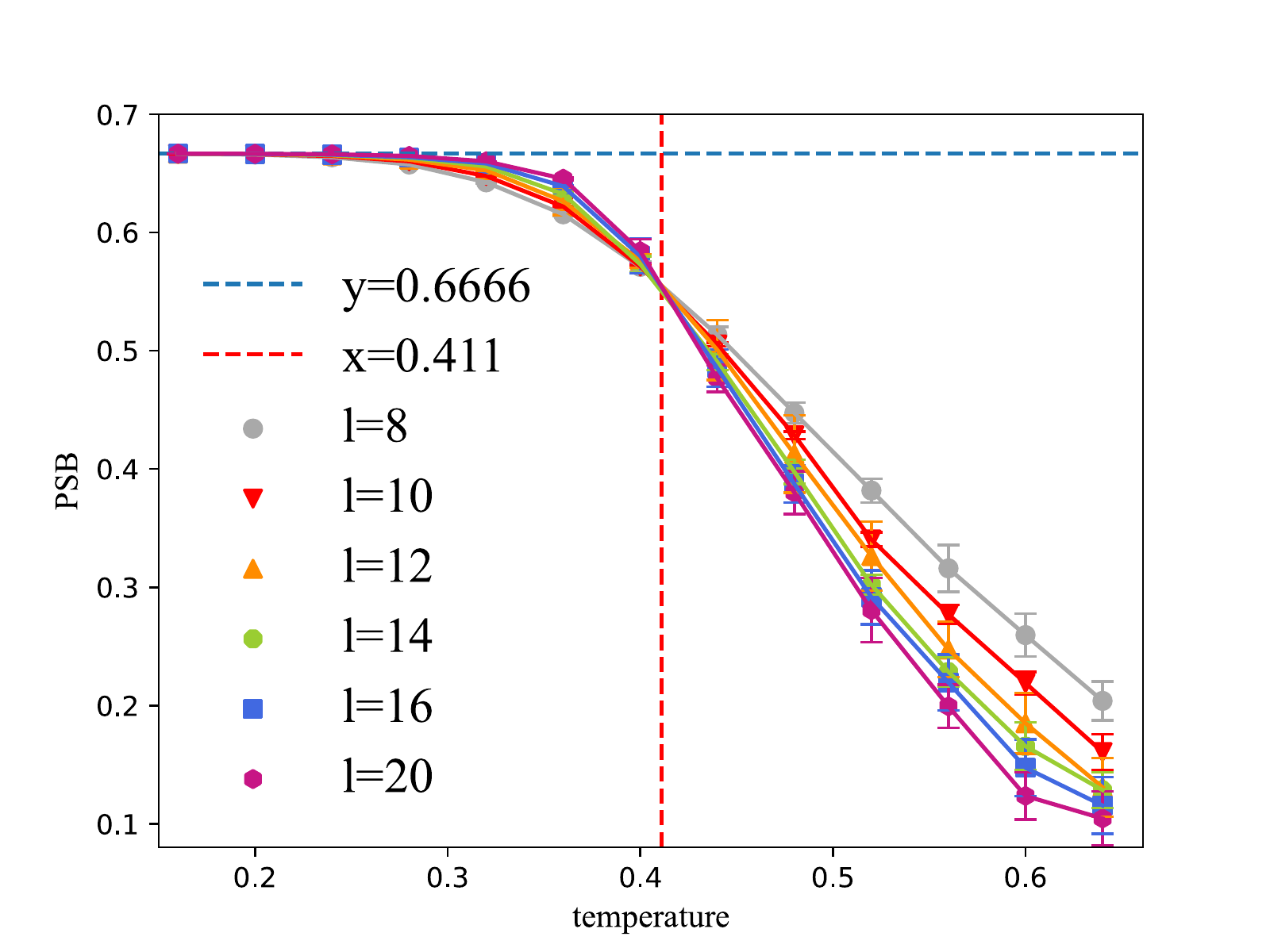}}
		\quad
		\subfigure[]{ \label{2c} \includegraphics[scale=0.45]{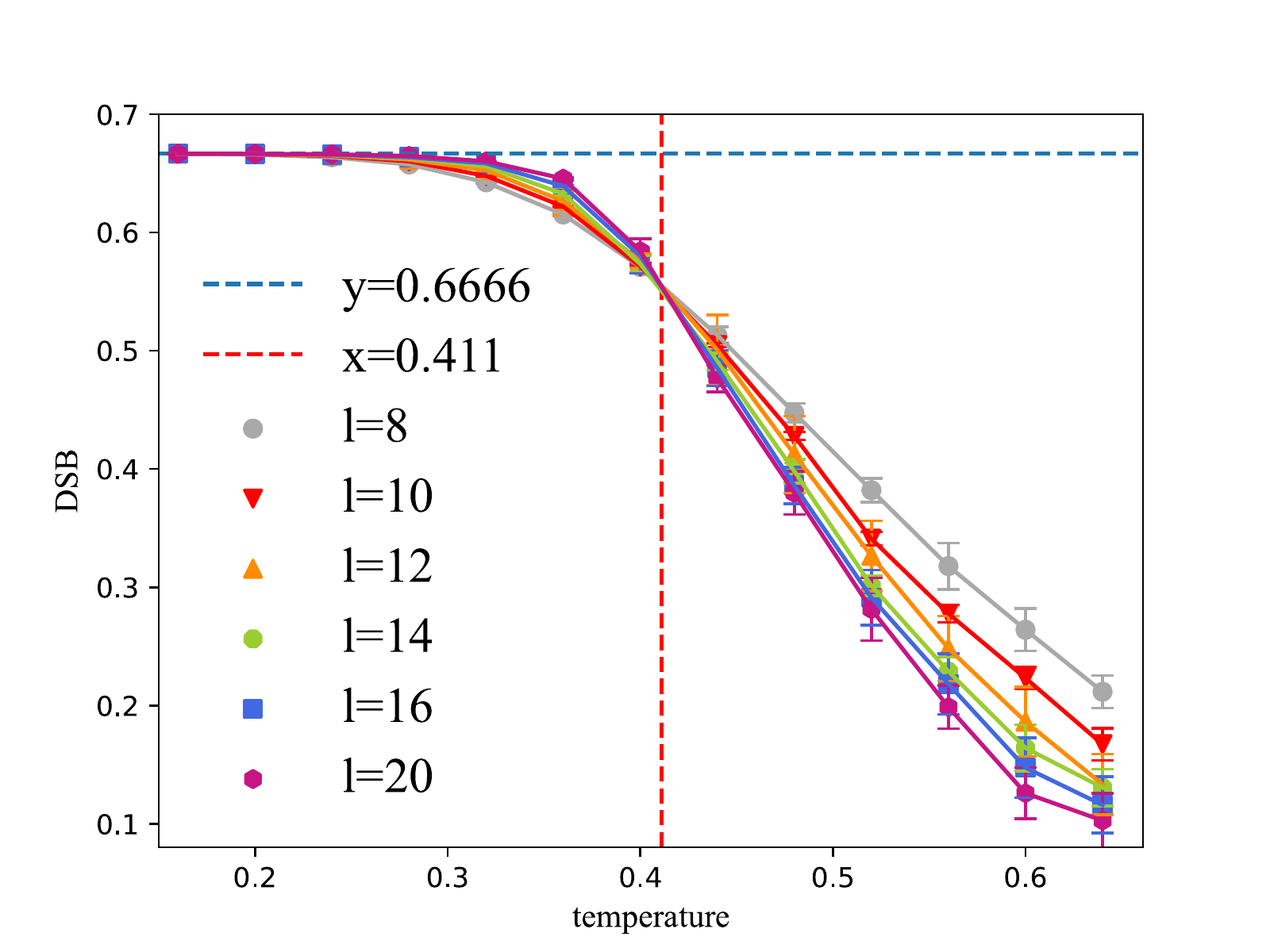}}
		\quad
		\subfigure[]{ \label{2d} \includegraphics[width=12.8cm,height=8.4cm]{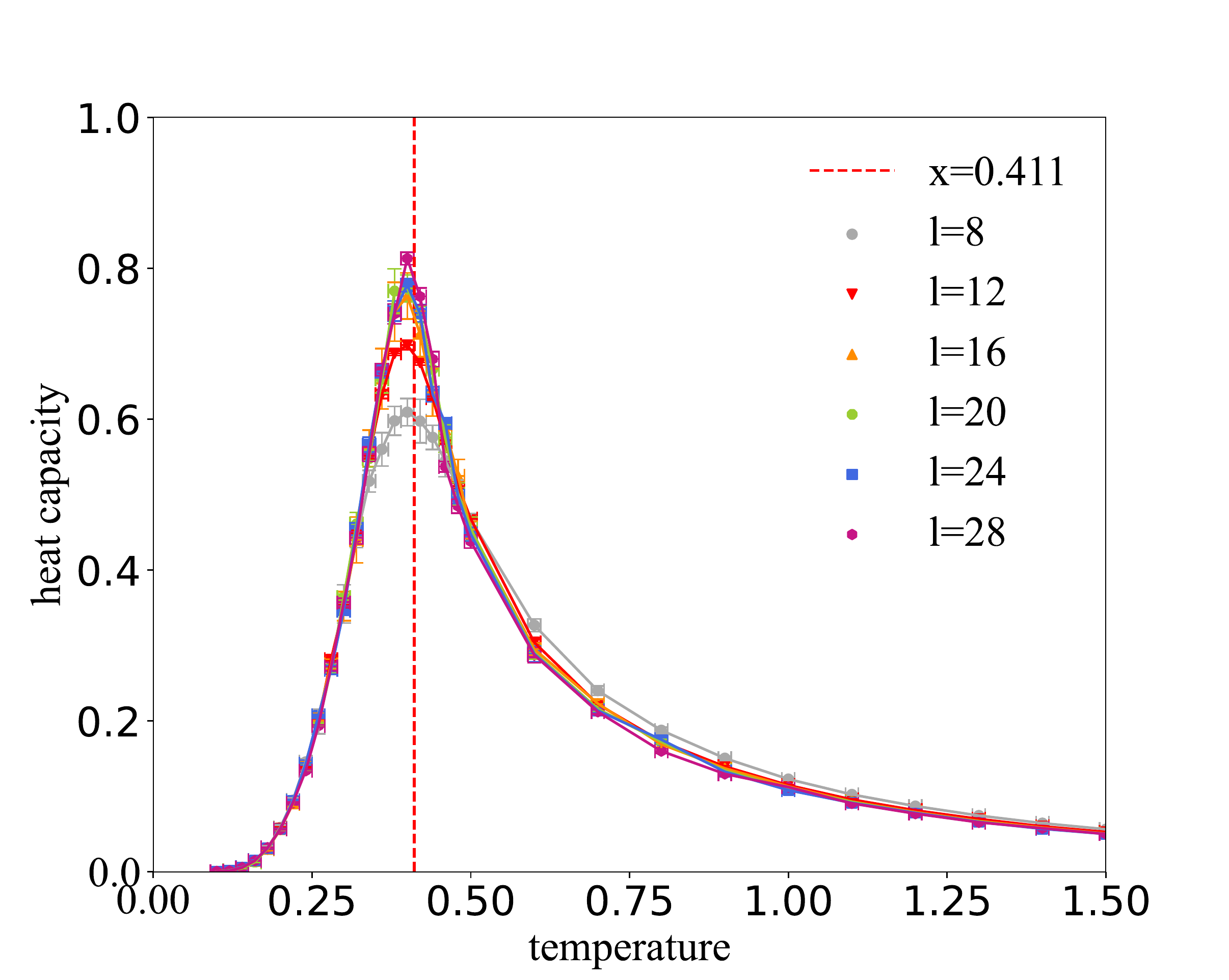}}
		\caption{(a): order parameter; (b)(c): PSB and DSB Binder ratio; (d): specific heat capacity.}
	\end{figure*}

	Using the directed loop algorithm, simulations are made on $N=L\times L$ lattices with periodic boundary condition. According to our proposal above, we compute the norm of every geometric order parameter and normalize them. Dual to the Heisenberg antiferromagnetic system, the transition of the classical dimers will happen at finite temperature (see Fig.\ref{2a}). Noting that the temperature between 0.4 and 0.46 (here what we must remind is  that thermometric scale is depended on the choice of the value of $k$ above and all of our temperatures in our article are not Kelvin standard but depended on our set of $k=-1/ln2$ without losing of generality), we could see an obvious order parameter transition happen. At high temperatures, order parameters will approach to zero for disordered phases if the lattice is infinite theoretically. However, we could see order parameters are not zero. Thus, we argue it is an elementary hint that the transition happened there is not a regular thermodynamic transition and there could still exist some quasi long-range correlations. 

	To strengthen our argument, we simulate the Binder ratio \cite{K1981Finite}, heat capacity and the correlation function immediately. For Binder ratio, we follow the definition of the QDM \cite{1996Columnar}
	\begin{equation}
		DSB = N^{-1}|N(\rule[2.5pt]{0.23cm}{0.05cm})-N(\rule[0pt]{0.05cm}{0.23cm})|
	\end{equation}
	\begin{equation}
		PSB =N^{-1}|N(\mathop{\rule[0.1pt]{0.23cm}{0.05cm}}^{\rule[-0.1pt]{0.23cm}{0.05cm}})-N(\rule[-0.6pt]{0.05cm}{0.23cm}\kern0.08cm\rule[-0.6pt]{0.05cm}{0.23cm})|
	\end{equation}
	Where $DSB$ is the dimer rotational symmetry breaking and $PSB$ is the pair rotational symmetry breaking. Both two quantities are  long-range order in every configuration shown in low-temperature ordered phase. But they will vanish in high-temperature disordered phases. We can use Binder ratio to locate the transition with high precision. Defined by
	\begin{equation}
		B_{DSB} = 1-\left \langle DSB^4  \right \rangle/(3\left \langle DSB^2  \right \rangle^2)  
	\end{equation}
	and
	\begin{equation}
		B_{PSB} = 1-\left \langle PSB^4  \right \rangle/(3\left \langle PSB^2  \right \rangle^2)  
	\end{equation}
	they will start from 2/3 in a long range ordered phase and decrease as the temperature increases. But these curves for different $L$ all cross at a unique temperature $T$ if their sizes are finite (see Fig.\ref{2b} and \ref{2c}). In our simulation, we can derive two basilic factos. The first is the crossing of curves shows that the transition process is a second-order transition or a BKT transition. The second is curves will not be convergent to zero gives us a keynote that the quasi long-range order still exists when the temperature climbs to the infinity. The phenomenon of the curves crossing in a sole point is the same as the conclusions derived in the 2-dimensional XY system which has been proven that it possesses coupled chiral vortexes by early researchers \cite{1999Binder}.  Meanwhile, their data also show the Binder ratio will not go to zero as the temperature increases. At the end of our curves, we can see they still remain the downward trend on the lattices of $L=8,10,12$ because of the effect of finite size. If $L$ goes to 14, 16, 20, curves turn to flatness and are higher than zero whether how big the temperature is. Another thing deserves to be emphasized is we identify the transition temperature $T_c$=0.411 precisely. 

	The specific heat capacity per site $c_v$ defined as
	\begin{equation}
		c_v=\frac{C_v}{N}=\frac{\left \langle E^2  \right \rangle-\left \langle E  \right \rangle^2}{NkT^2}
	\end{equation}
	displays a peak in our simulation and it does not diverge in the thermodynamic limit (Fig.\ref{2d}). It also shows that there exists a second-order transition or a BKT transition. The critical exponent $\alpha$  can be measured in the following Table \ref{tab1}. All of specific heat capacity $c_v$ in our simulation are less than 0.9 and we can see all curves of specific heat capacity $c_v$ will overlap no matter how the system size increases. Likewise the Binder ratio measurement above, the deviation of the lattice $L=8,12$ are owing to the effect of finite size. Besides, the peak of $c_v$ is located slightly below the value of $T_c$ determined above. Both the shift appeared here and the astringency argued before strongly indicate the transition existed in the model is really a BKT transition.
	\begin{table}[h]
		\centering
		\begin{tabular}{ p{0.6cm} p{1.2cm} p{1.2cm} p{1.2cm} p{1.2cm} p{1.2cm} p{1.2cm} }
		\hline
		\hline
		&L=8 &L=12 &L=16 &L=20 &L=24 &L=28\\
		\hline
		&2.262 &2.546 &2.792 &2.822 &2.817 &2.971\\
		$\alpha_{0+}$&$\sim$&$\sim$&$\sim$&$\sim$&$\sim$&$\sim$\\
		&3.250 &3.472 &3.593 &3.635 &3.600 &3.650\\
		\hline	
		&-1.925 &-2.146 &-2.275 &-2.399 &-2.397 &-2.567\\	
		$\alpha_{0-}$&$\sim$&$\sim$&$\sim$&$\sim$&$\sim$&$\sim$\\	
		&-1.790 &-2.035 &-2.098 &-2.136 &-2.125 &-2.138\\
		\hline
		\hline
		\end{tabular}
		\caption{critical exponents of classical dimer model transition.}\label{tab1}
	\end{table}

	To synchronize explanation, we calculate two correlation functions: dimer-dimer correlation function $D(x)$ and monomer-monomer correlation function $M(x)$.
	\begin{equation}
		D(x)=\frac{\left \langle n^x_-(r)n^x_-(r+x)  \right \rangle+\left \langle n^y_|(r)n^y_|(r+y)  \right \rangle}{2}
	\end{equation}
	\begin{equation}
		M(x)=\frac{\left \langle n_=^x(r)n_=^x(r+x)  \right \rangle+\left \langle n^y_{||}(r)n^y_{||}(r+y)\right \rangle}{2}
	\end{equation}

\begin{figure}[h]
		\centering
		\subfigure[]{ \label{5a} \includegraphics[width=5.8cm,height=5cm]{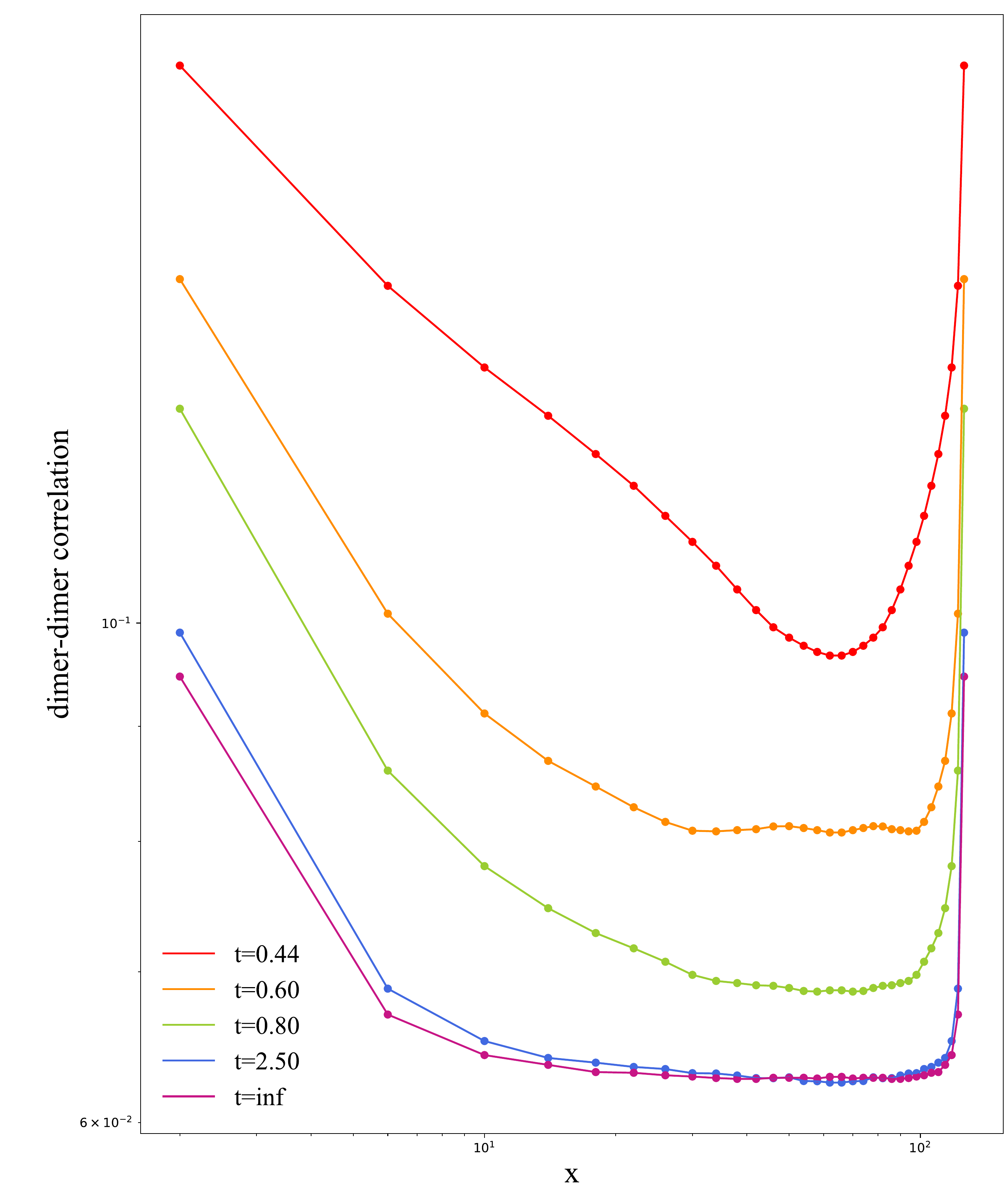}}
		\quad
		\subfigure[]{ \label{5b} \includegraphics[width=6.4cm,height=5cm]{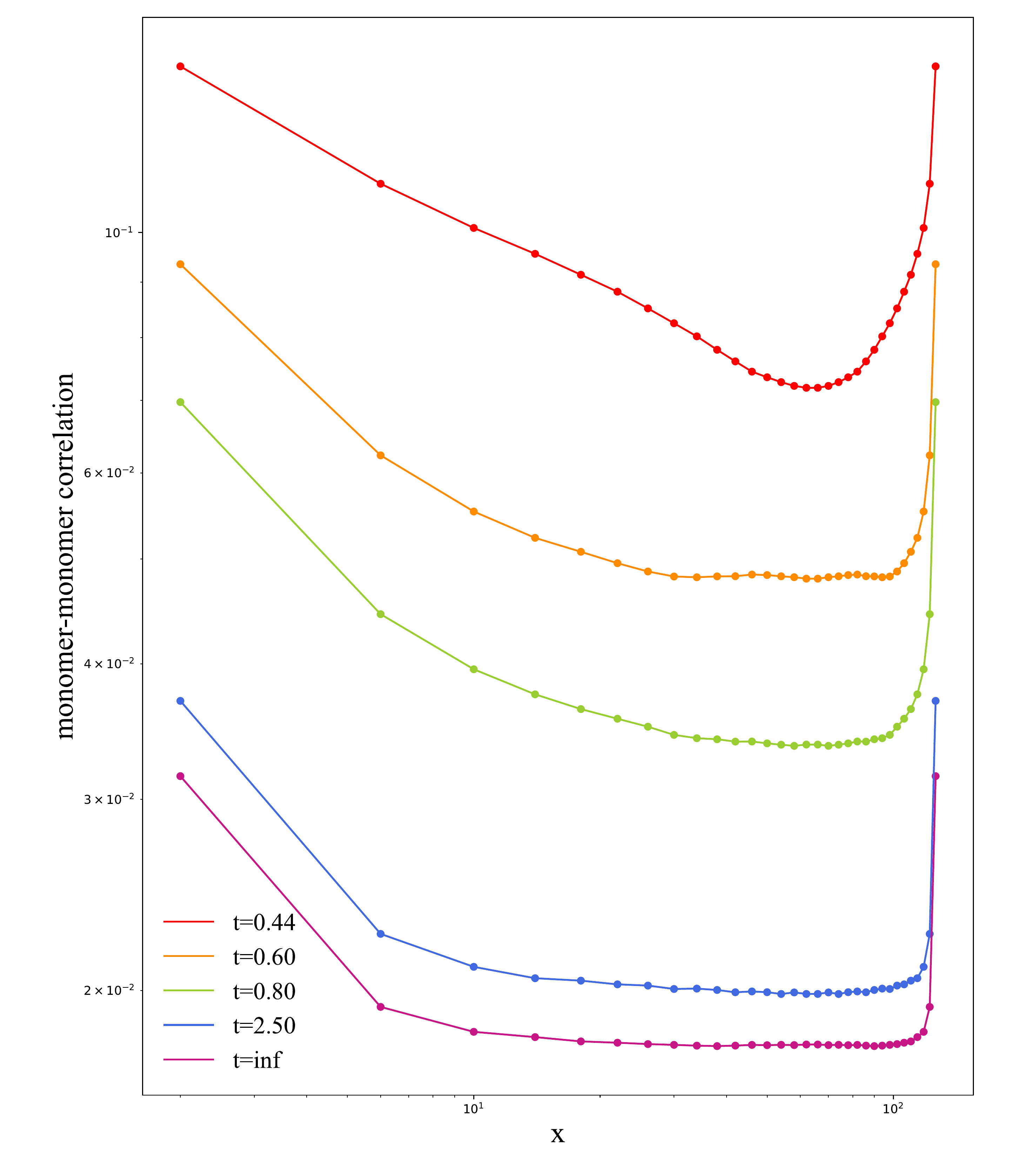}}
		\caption{(a): dimer-dimer correlation functions versus distance $x$ in a log-log scale for various temperatures; (b) monomer-monomer correlation functions versus distance $x$ in a log-log scale for various temperatures}	
	\end{figure}

	Where $n_-^x (r)=1$ ($n_=^x (r)=1$) for horizontal dimer (monomer) at site $r$, or 0 otherwise. We finally derive algebraic decay of these correlations for $T>Tc$, and flat line for $T<Tc$. After fitting analysis, we find both correlators are power law decay and the critical exponents $\delta ^d$ and $\delta ^m$ vary continuously with temperature, where $\delta ^d$ is defined as $D(x)\sim(-1)^x x^{-\delta^d}$ and $\delta ^m$ is defined as $M(x)\sim(-1)^x x^{-\delta^m}$. In our simulations, $\delta ^d$=0.1949(24)  and $\delta ^m$=0.2062(26) at $T=0.44$ (this temperature goes off the transition temperature slightly). Factor (-1) appearing here is due to the geometric constraints of dimers.  If the point bonds, it can not bond to another, which means correlators will bounce from the odd index to the even index. Without loss of sense, we can neglect odd terms shown in the Fig.\ref{5a} and \ref{5b}. The correlation functions distinctly reveal the quasi long-range order in the transition of the classical dimer model on the 2-dimensional square lattice and powerfully demonstrate the transition is actually a BKT transition at finite temperature.

	\begin{figure}[htbp]
		\centering
		\subfigure[]{ \label{Ya} \includegraphics[scale=0.5]{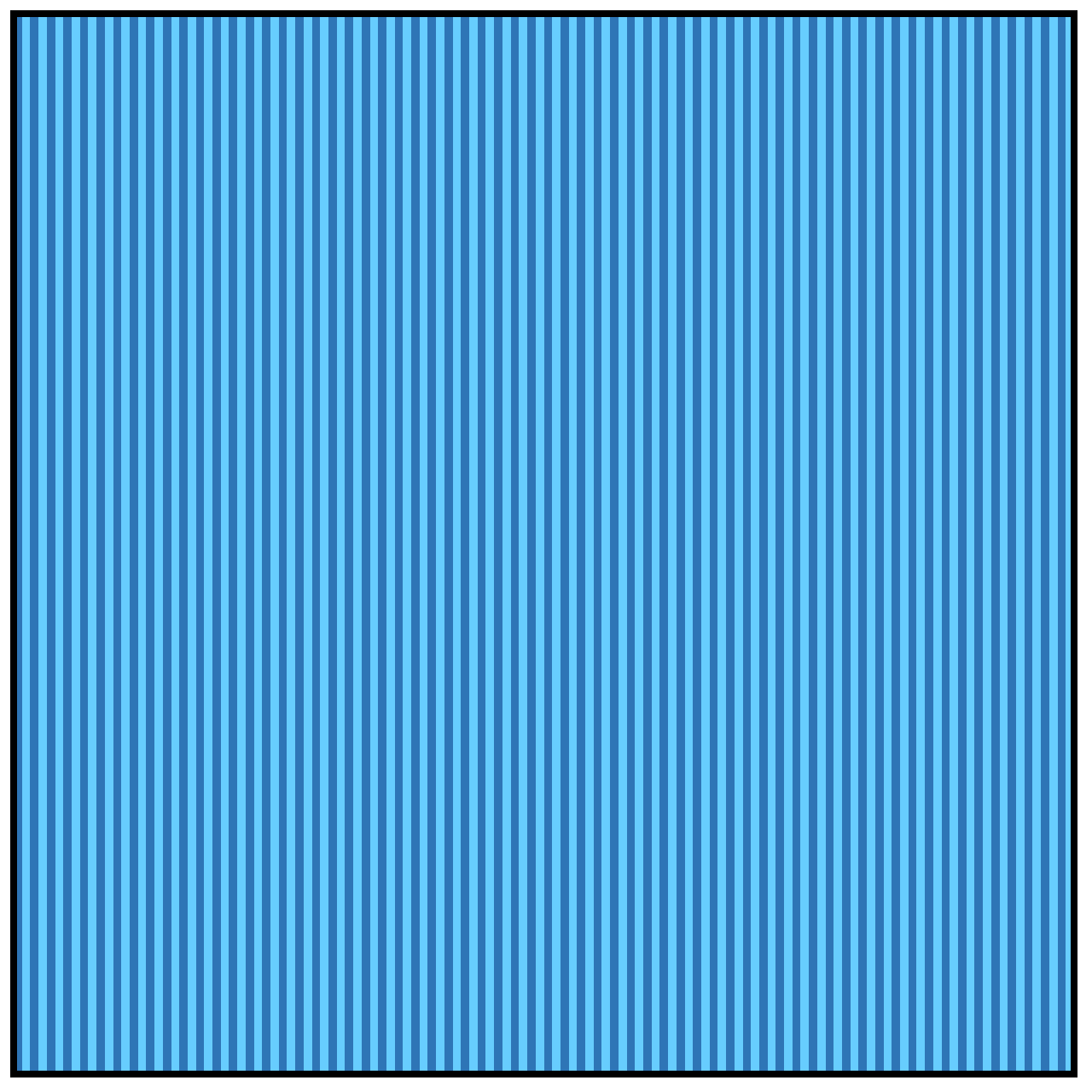}}
		\quad
		\subfigure[]{ \label{Yb} \includegraphics[scale=0.5]{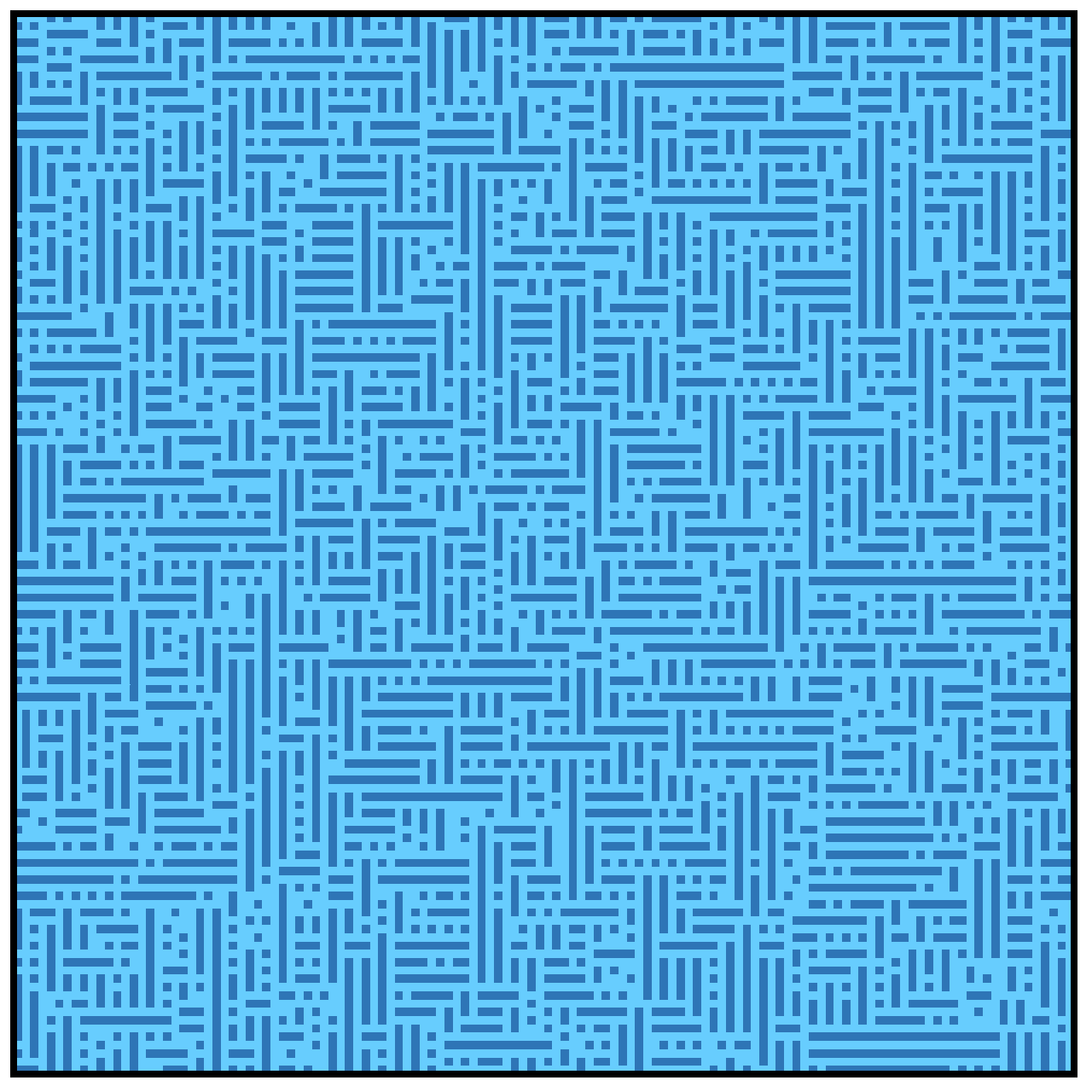}}
		\quad
		\subfigure[]{ \label{Yc} \includegraphics[scale=0.5]{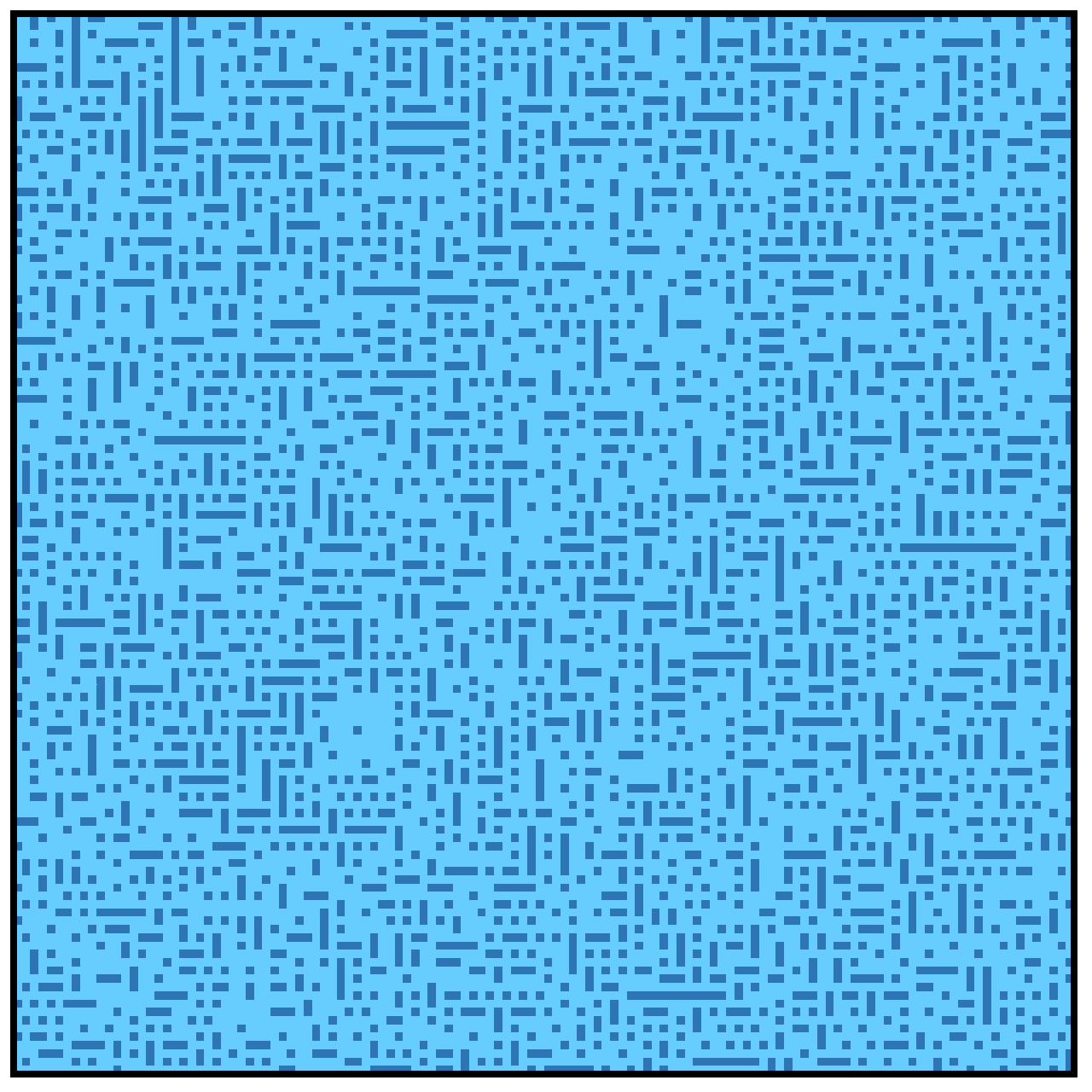}}
		\caption{Long-range order distribution (a): At zero temperature, the model is occupied by long-range order completely; (b): At $T=0.6$, there still exits quasi long-range order; (c): At the infinite temperature, long-range order disappears and there are just short-range bonds.}
	\end{figure}
	To represent the quasi long-range order clearly, we choose three states at different temperatures and draw their displays (see Fig.6). In low temperature, the system is in one of fourfold degenerate Columnar states, which means the system is occupied by long-range order completely. Theoretically, if the system undergoes a transition, the long-range order will disappear suddenly. However, at $T=0.6$ shown in Fig.\ref{Yb}, we could see there still exits quasi long bond. Although there are no bonds like a thread through the whole graph, quasi long bonds still exit. There is a BKT transition that has been proven by the bias of transition point of specific heat capacity and the correlators can not be trivial at $T=0.6$. In classical dimer model, BKT transition produces these quasi long bonds and causes the existence of the long-range order. As the temperature increases, quasi long-range order will disappear in strong heat fluctuations. Among these, the BKT transition happens and eliminates these topological long bonds. We could see in Fig.\ref{Yc}, there are no quasi long-range order at the infinite temperature. These three states at different temperatures sufficiently indicate the existence of topological transition and represents the appearance/disappearance process of the quasi long-range order. 

\begin{figure}[htbp]
		\centering
		\subfigure[]{ \label{Wa} \includegraphics[scale=0.22]{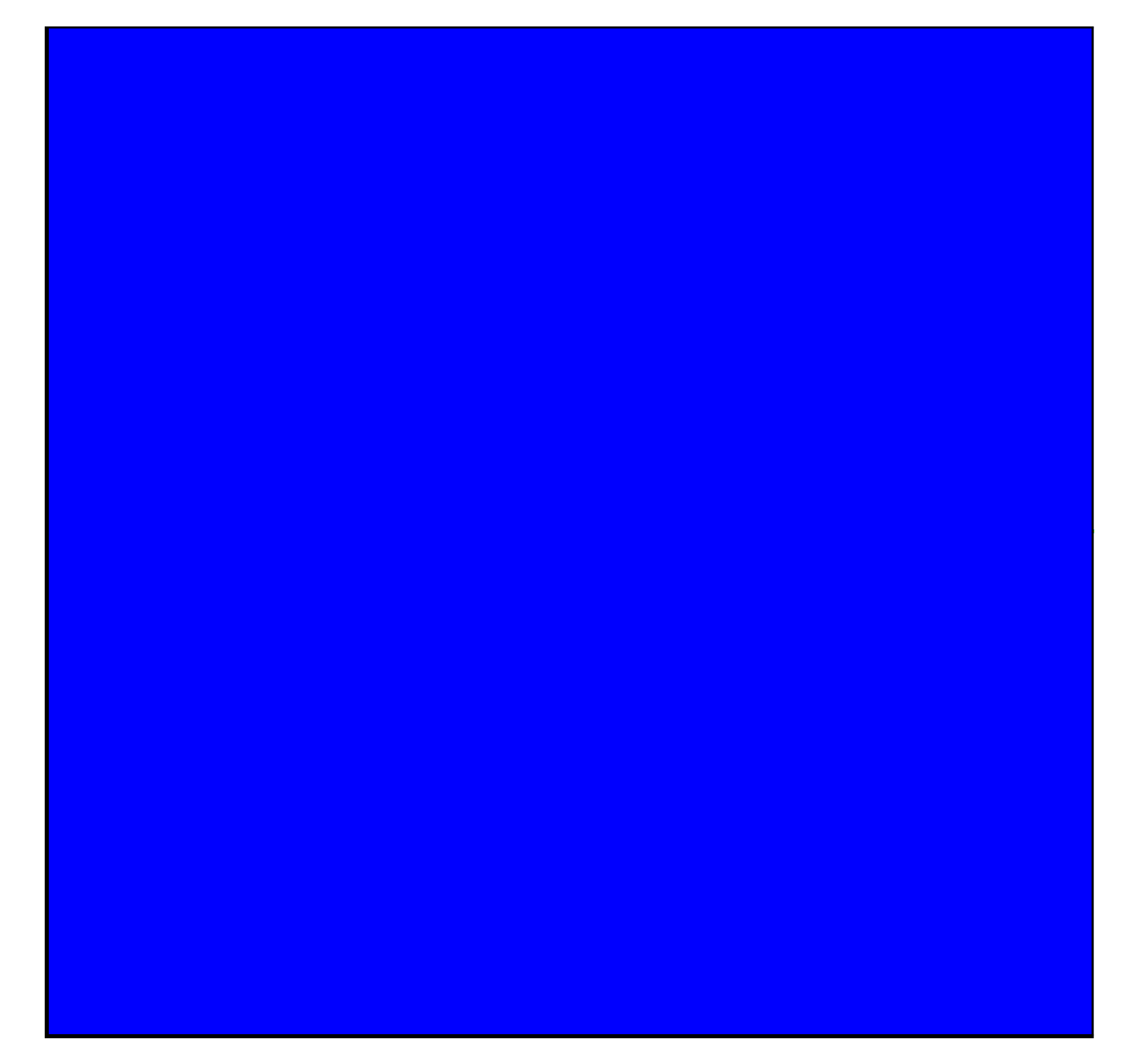}}
		\quad
		\subfigure[]{ \label{Wb} \includegraphics[scale=0.22]{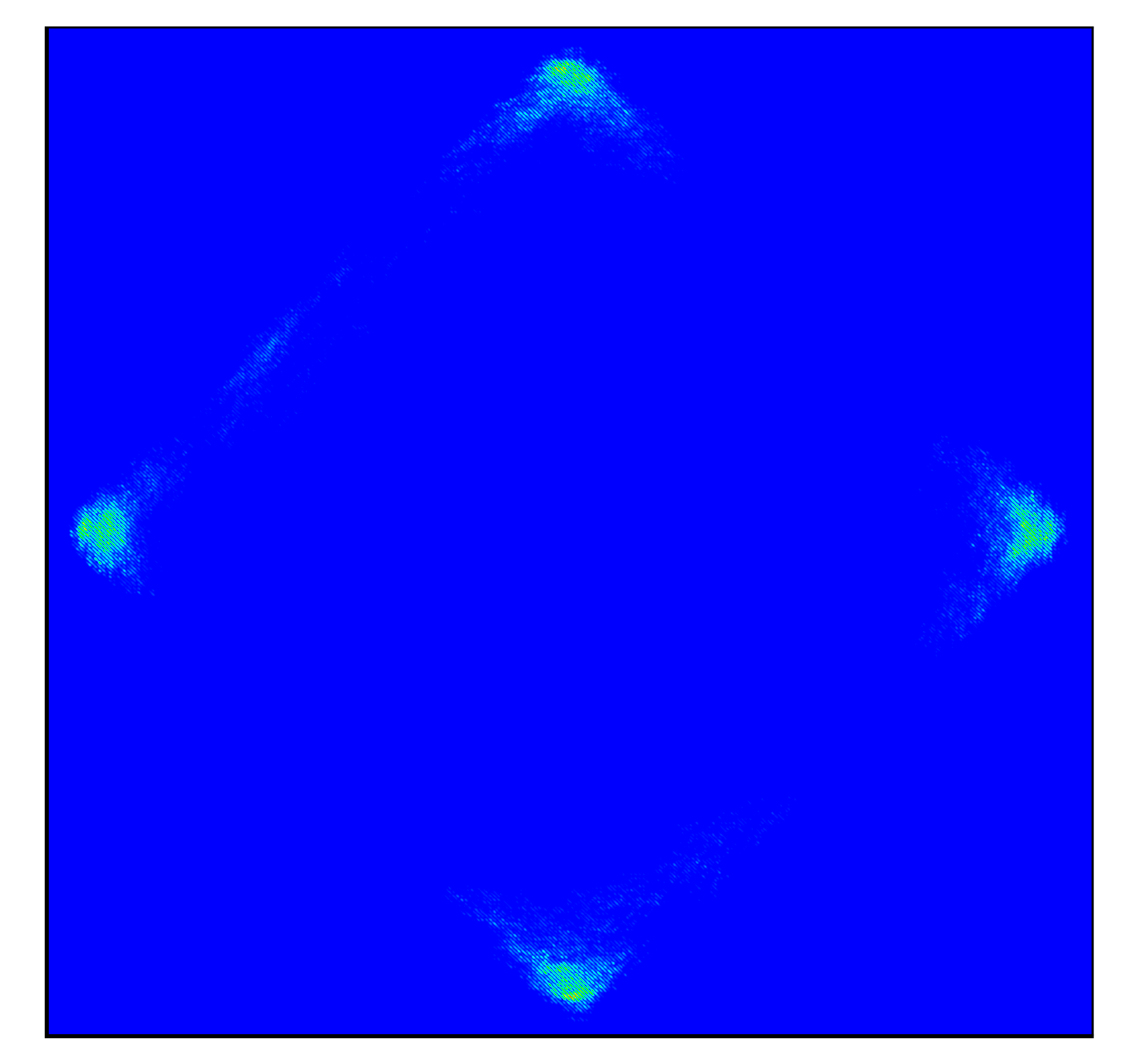}}
		\quad
		\subfigure[]{ \label{Wc} \includegraphics[scale=0.22]{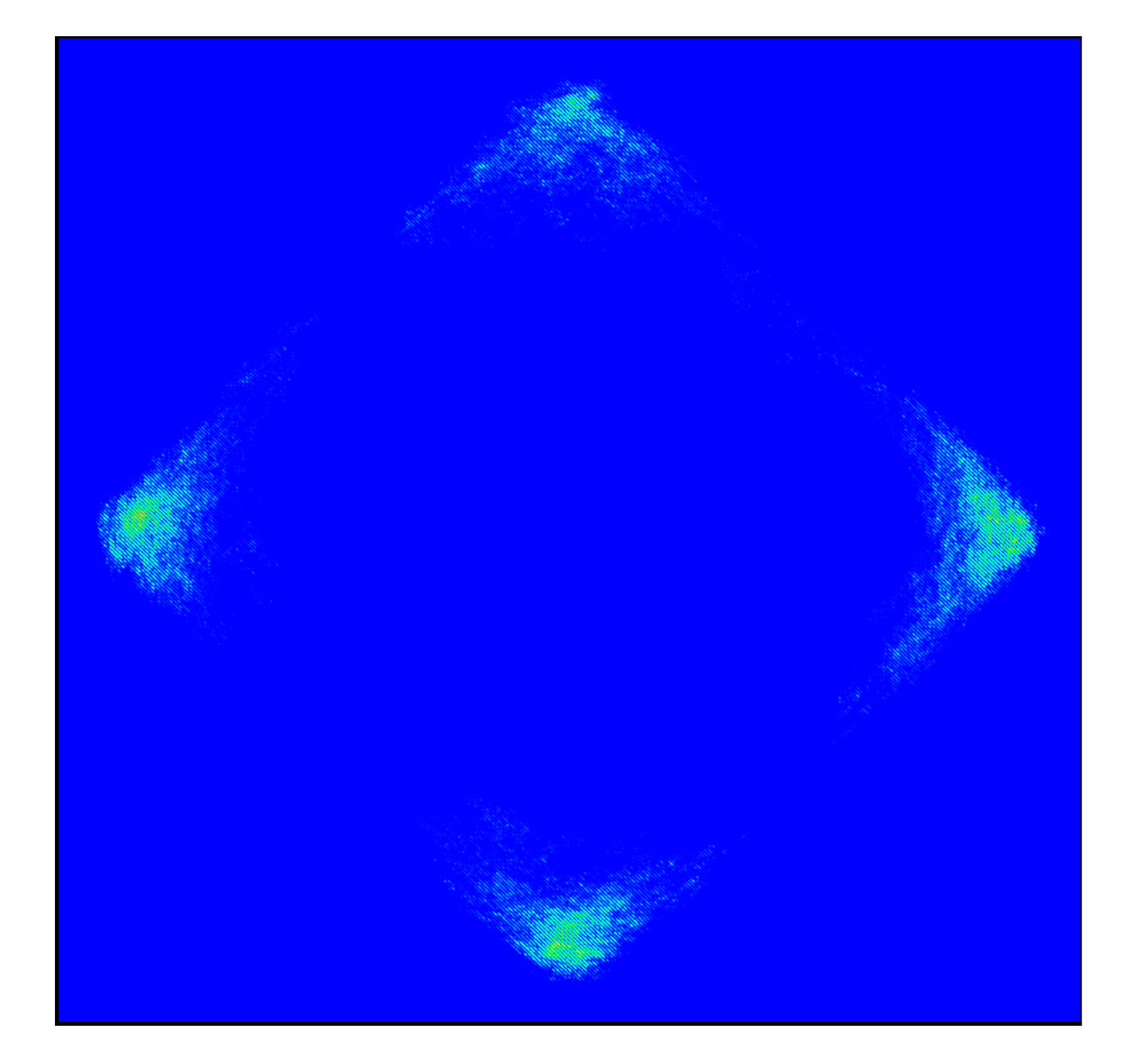}}
		\quad
		\subfigure[]{ \label{Wd} \includegraphics[scale=0.22]{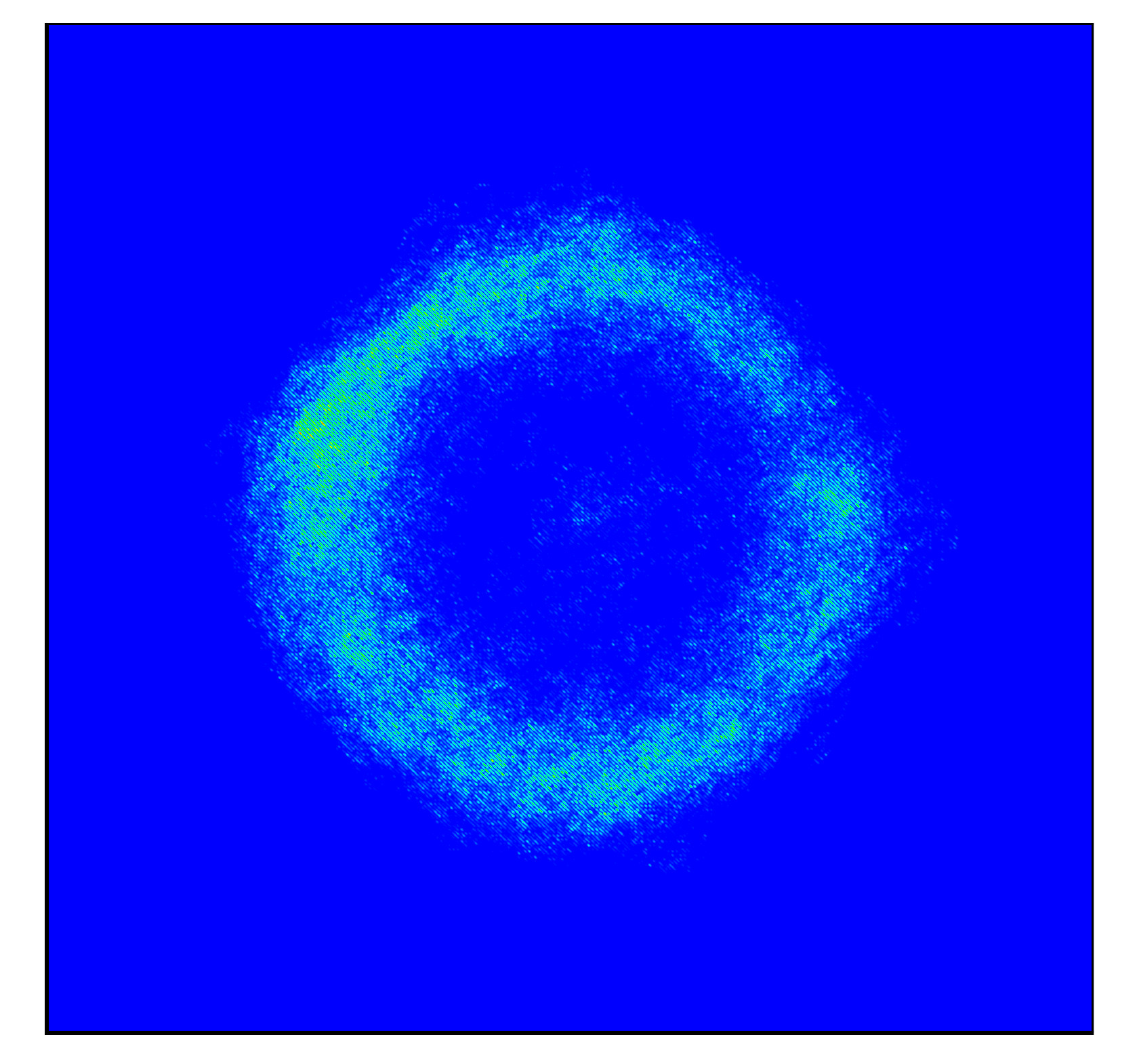}}
		\quad
		\subfigure[]{ \label{We} \includegraphics[scale=0.22]{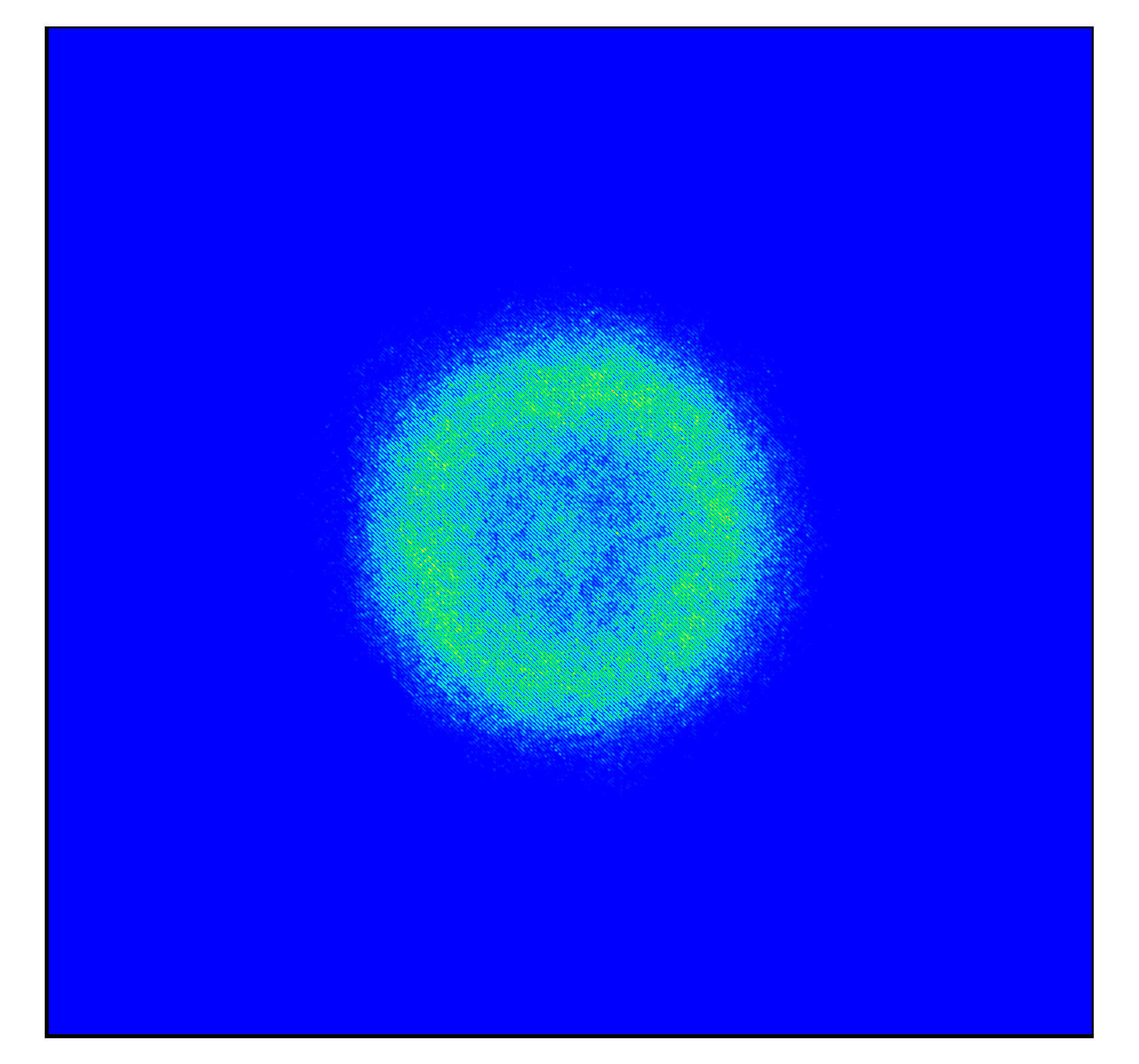}}
		\quad
		\subfigure[]{ \label{Wf} \includegraphics[scale=0.22]{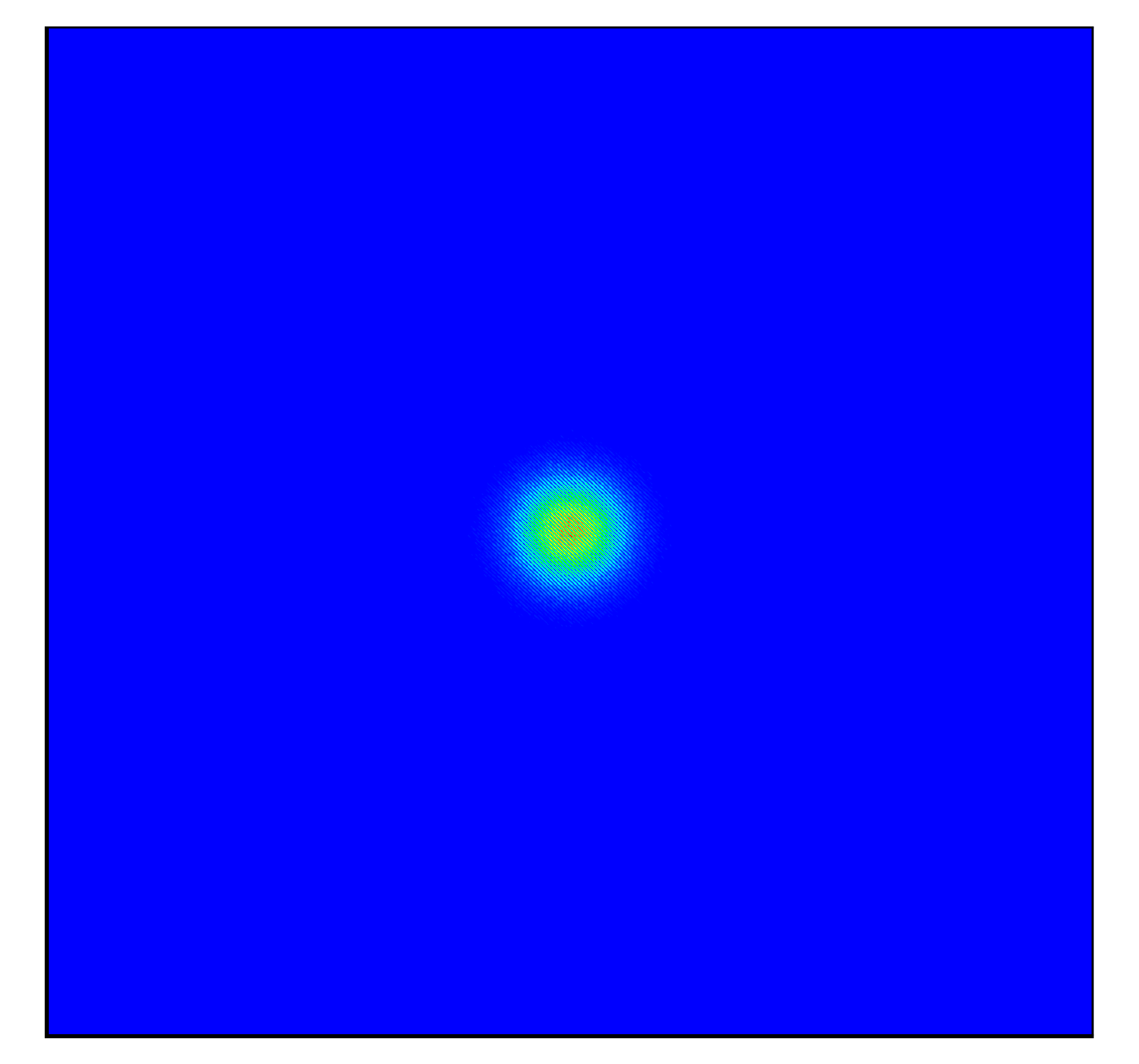}}
		\caption{Distribution of order parameters. (a): $T=0$ (fourfold degenerate states); (b): $T=0.38$; (c): $T=0.411$ at the transition temperature; (d): $T=0.6$, symmetry can not restore completely; (e): $T=1.0$, in high temperature zone; (f): $T=infinite$, the system recaptures $U(1)$ symmetry.}
	\end{figure}

	We want to study the symmetry properties behind the transition process and indicate our conclusion in another visualized way. As mentioned above, our system is in one of fourfold degenerate Columnar states (see the distribution in Fig.\ref{Wa}). According to our 2-dimensional order parameter defined before, we can represent the distribution of order parameters in the complex plain. Order parameters of microstates appeared at low temperatures are distributed around these four states shown in Fig.\ref{Wb}. As the temperature increases, order parameter will start to approach to zero. Before the transition, the symmetry of the system is broken. It just stays at the $S(4)$ symmetry. However, if the system undergoes the transition process, it will recapture wider symmetry partly shown in Fig.\ref{Wc} and \ref{Wd}. Different from the regular thermodynamics transition, in a certain size zone of temperatures behind the transition temperature, the symmetry of the system cannot restore completely. It shows local breaking of the symmetry. The phenomenon confirms there still exits a BKT transition again. At the infinite temperature, the symmetry of classical dimer restores eventually. Strong heat fluctuations at high temperatures will lead the system to enter the $U(1)$ symmetry shown in Fig.\ref{We}.
\subsection{Edged cluster algorithm in the square lattice}
	Directed loop algorithm could be a passable method to solve the question on the small sized lattice ($L\le$30). However, in our simulation, if lattices scale up ($L\ge$30), our simulation results will be impacted by the metastable states. It is possible to form a long loop such that we could flip enough dimers theoretically. But actually, the average length of loop is too short to refrain from metastable states [Table \ref{tab2}]. We can see even if at infinite temperature, on the lattice of $L=48$, max length of loop $l_{max}=44\ll48\times48/2=1152$. Thus, loop algorithm is not appropriate for the large system obviously.
	\begin{table}[h]
		\centering
		\begin{tabular}{ p{0.5cm} p{0.95cm} p{0.8cm} p{0.95cm} p{0.8cm} p{0.95cm} p{0.8cm} p{0.95cm} p{0.8cm} }
		\hline
		\hline
		&$L=$ &08&$L=$ &16&$L=$ &32&$L=$&48\\
		\hline
		$T$&Average &Max &Average &Max &Average &Max &Average &Max \\
		&length &length&length&length&length&length&length&length\\
		\hline
		0.3&2.62 &24&2.47 &30&2.47 &26&2.46&32\\
	    0.4&2.71 &24&2.54 &34&2.53 &28&2.53&38\\
	    1.0&3.17 &25&2.82 &39&2.78 &39&2.78&34\\
		inf&3.78 &27&3.19 &50&3.09 &60&3.09&44\\	
		\hline
		\hline
		\end{tabular}
		\caption{Loop average length and max length in directed loop algorithm updates.}\label{tab2}
	\end{table}

	It encourages us to develop an algorithm which could flip larger size of cluster of dimers. Inspired by the seed selection of Wolff algorithm, we try to introduce the concept of seed to dimer model. Based on the pocket algorithm \cite{Werner2003Pocket}, we introduce the edged cluster algorithm to get rid of the impact of metastable states.

\begin{figure}[h]
		\centering
		\subfigure[]{ \label{8a} \includegraphics[width=6.3cm,height=3.6cm]{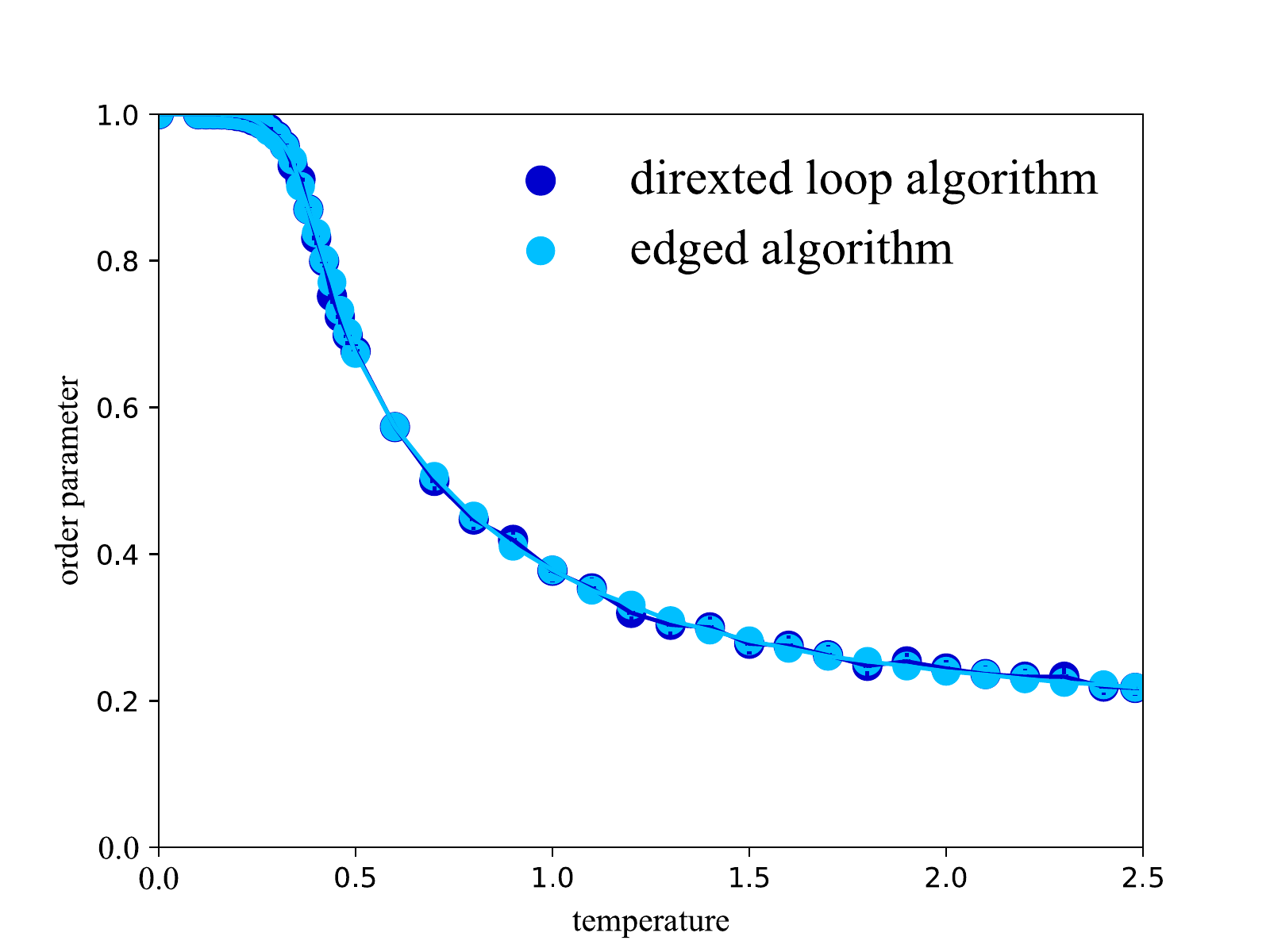}}
		\quad
		\subfigure[]{ \label{8b} \includegraphics[width=6.3cm,height=3.6cm]{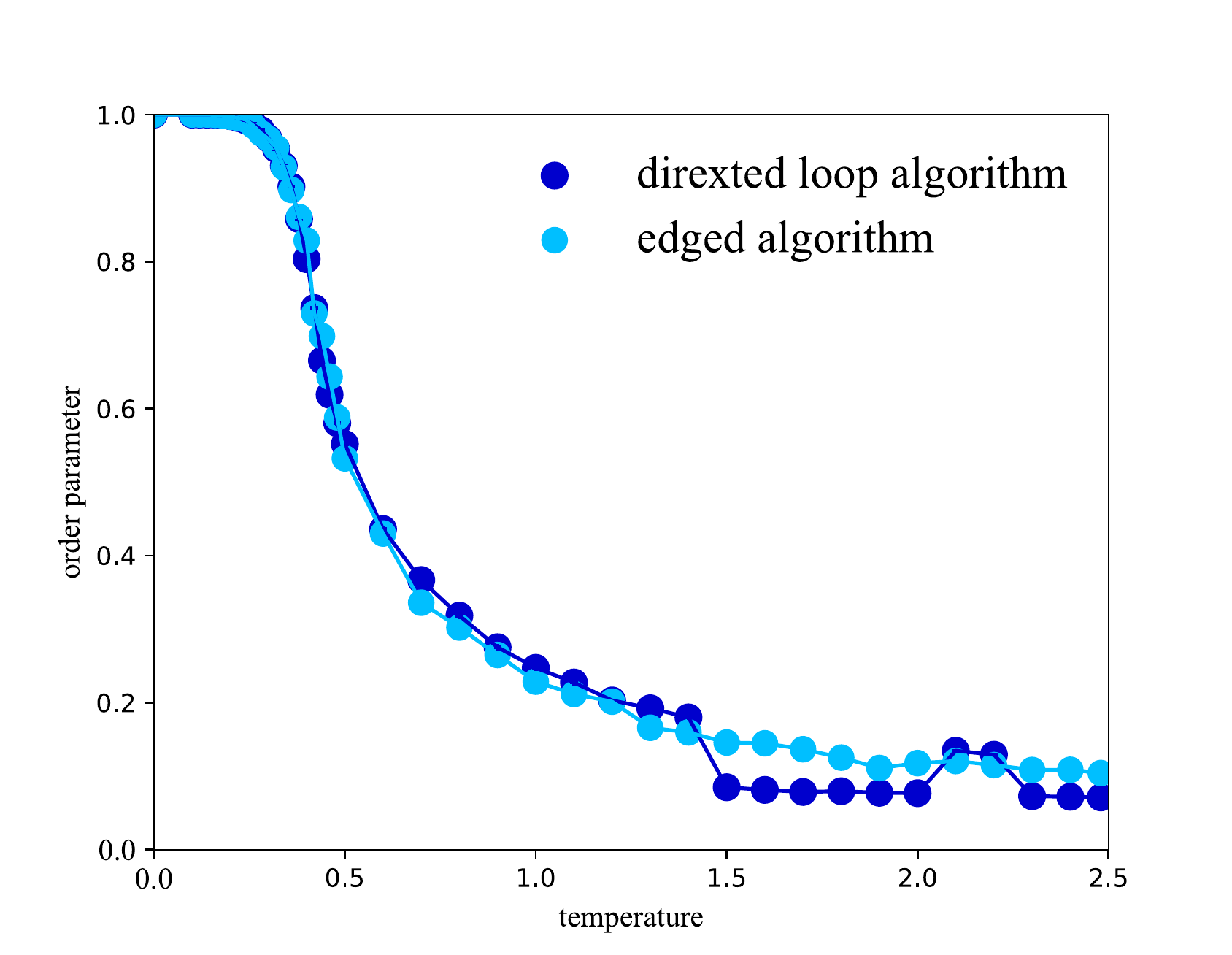}}
		\subfigure[]{ \label{8c} \includegraphics[width=6.3cm,height=3.6cm]{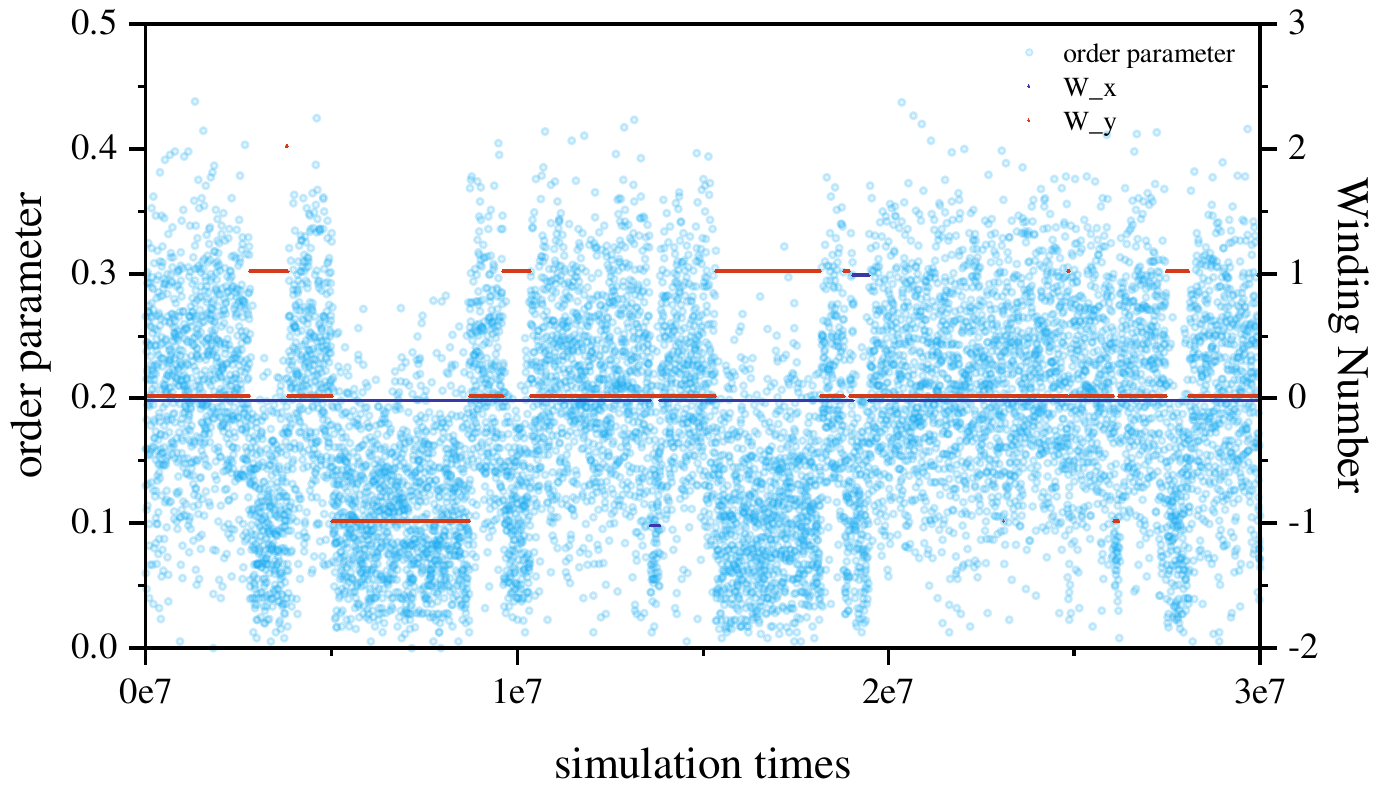}}
		\subfigure[]{ \label{8d} \includegraphics[width=6.3cm,height=3.6cm]{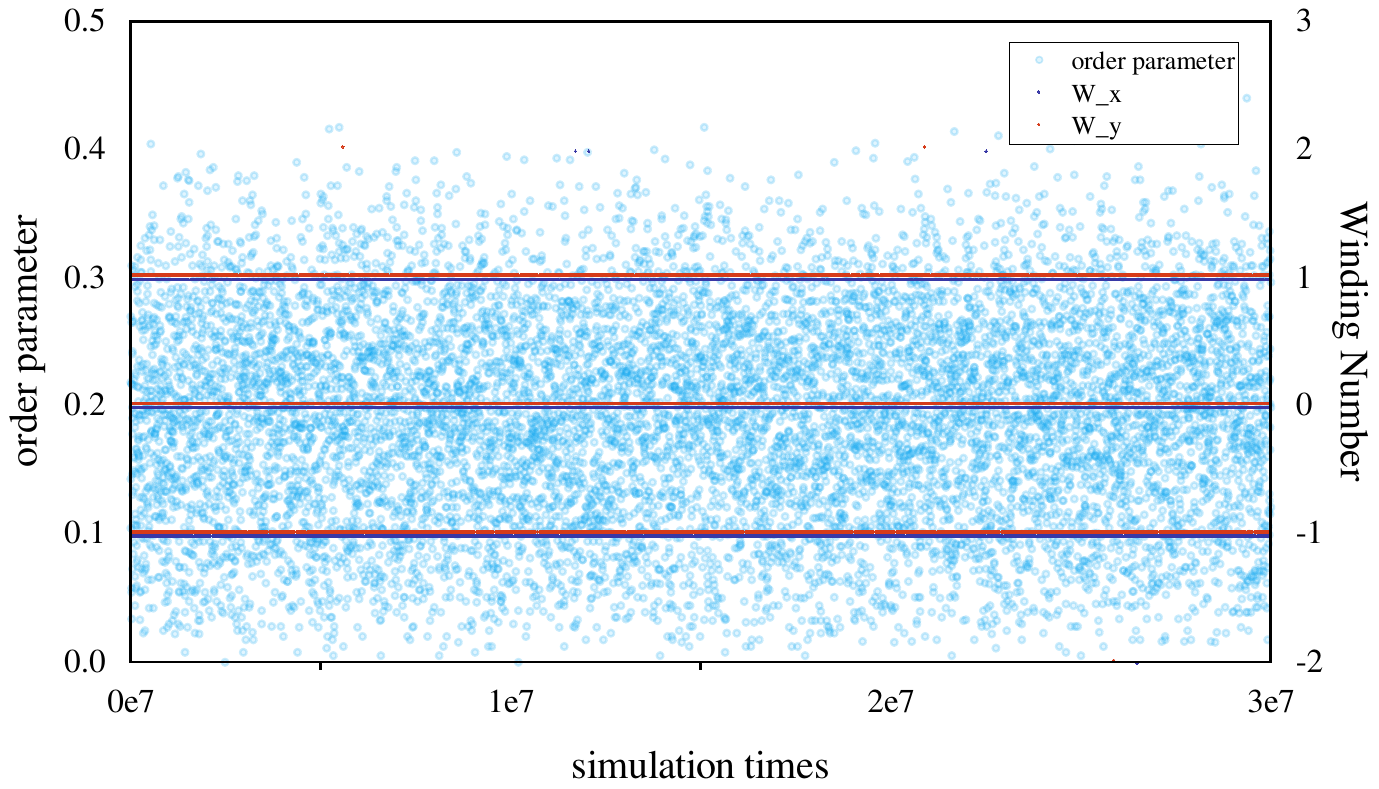}}
		\caption{(a): order parameter of lattice $L=16$ (deep line shows directed loop algorithm and tint line shows edged cluster algorithm); (b): order parameter of lattice $L=32$ (deep line shows directed loop algorithm and tint line shows edged cluster algorithm); (c): the use of directed loop algorithm (d): the use of edged cluster algorithm.}	
	\end{figure}

	Edged cluster algorithm satisfies detailed balance condition because the choice of axis and seed are the same weight. But if we use it to solve the softed dimer model, question is not so easy. In this article, we do not plan to argue this question. To check the correctness of our algorithm, we simulate order parameter change at finite temperatures again and we derive the same conclusion as previous sector (see Fig.\ref{8a}).
 
	Using our new algorithm, the length of loop at this time could reach a high level compared with the data derived before [Table \ref{tab21}]. For example, the average length of loop $l_{ave}\sim30$ and $l_{max}\sim300$ on the lattice $L=32$ at the infinite temperature. 

	\begin{table}[h]
		\centering
		\begin{tabular}{ p{0.5cm} p{0.95cm} p{0.8cm} p{0.95cm} p{0.8cm} p{0.95cm} p{0.8cm} p{0.95cm} p{0.8cm} }
		\hline
		\hline
		&$L=$ &08&$L=$ &16&$L=$ &32&$L=$&48\\
		\hline
		$T$&Average &Max &Average &Max &Average &Max &Average &Max \\
		&length &length&length&length&length&length&length&length\\
		\hline
		0.3&2.41 &26&3.07 &52&4.33 &97&5.55&152\\
		0.4&2.96 &26&4.44 &106&6.09 &298&8.21&296\\
		1.0&5.02 &32&9.97 &98&20.02 &310&29.87&554\\
		inf&6.19 &32&13.08 &100&26.98 &304&40.84&542\\	
		\hline
		\hline
		\end{tabular}
		\caption{Loop average length and max length in edged cluster algorithm updates.}\label{tab21}
	\end{table}
	As for metastable states, we simulate order parameter change at the limit of temperature on the lattice of $L=32$. We could see simulation implemented by directed loop algorithm exist some sub-equilibrium states, which cause the curve fluctuates near and after transition process (see Fig.\ref{8b} deep line). However, the results derived from the algorithm produced by us could derive smooth curve shown in Fig.\ref{8b} tint line.

	Ulteriorly, we note that the fluctuation in Fig.\ref{8b} deep line, staggered characterization could be found. If we compute topological winding number $(W_x,W_y)$ of each simulation microstate, we spectacularly discover these fluctuations are locked in different topological sections actually. Because the average loop is so short that it can hardly pass through the boundary of lattice, traversing the topological sections is too difficult to achieve. According to our preliminary statistic, if we use directed loop algorithm to solve large size lattice question, almost billions of simulations could counteract the fluctuation generated by different topological sections. States which are measured by us are just the tip of the iceberg. Decided by current topological section, our updates are among part of the whole ensemble of states. 

	We measure topological winding numbers as the simulation proceeds on the lattice of $L=32$ in the classical dimer model. We could see if we use directed loop algorithm, order parameter will fluctuate as a lot of packets. Because directed loop algorithm could hardly traverse topological sections and once winding numbers have been changed, configurations will be locked in this section. That is the reason why we can see fluctuations at high temperatures in the Fig.\ref{8b}.

	However, if we use edged cluster algorithm, topological winding numbers will change uniformly. Due to the long length of the loop, winding numbers can shift among different topological sections swiftly, which could make us derive relatively smooth curve of order parameter without fluctuations. We can see the distribution of the order parameter is more uniform in the Fig.\ref{8d}. The blue line and the red line are winding number $W_x$ and $W_y$. Fig.\ref{8c} shows the winding number is intermittent because of the lock of the topological sections and in Fig.\ref{8d}, we can see the continuous traverse among every topological sections of $W_x$ and $W_y$.  
\subsection{The square lattice with the breaking of the geometric constraint}
	Now let us break the geometric constraint of classical dimers (the new model is called softed dimer model). In this article, we introduce snAB bonds to the lattice. There exists a BKT transition from classical dimer model we studied before, but the introduction of snAB bonds will cause the change of universality class of thermodynamic transition. To study the changing process and the properties of the new transition, we measure different weight of two kinds of bonds on the lattice to explore the impact of the emergence of snAB bond. The algorithm used here is still directed loop algorithm, and we must compute the detailed balance in the updating process because of different weight introduced here (see Appendix \ref{app2}).

\begin{figure*}[htbp]
		\centering	
		\subfigure[]{ \label{7a} 
		\includegraphics[scale=0.45]{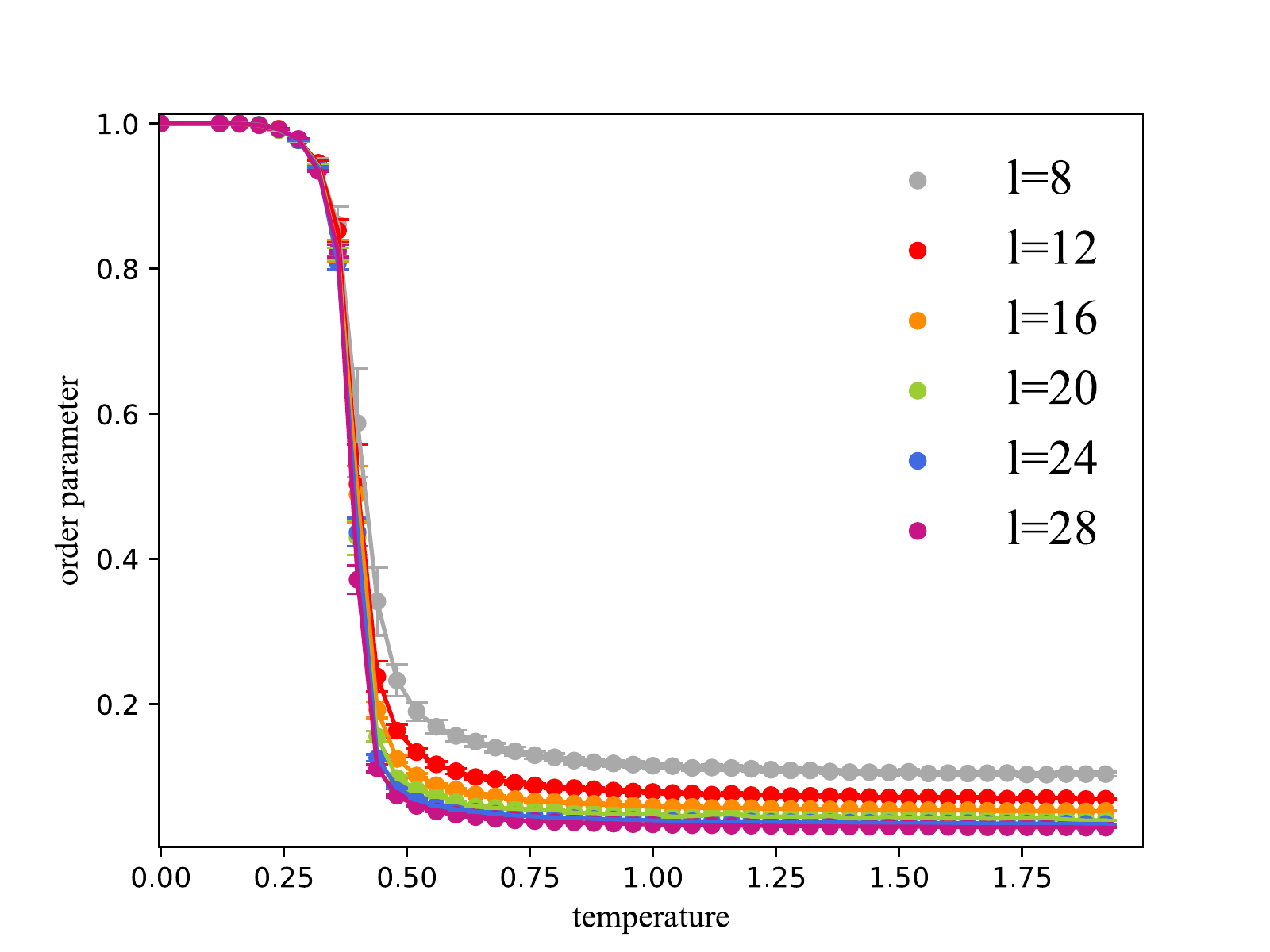}}
		\subfigure[]{ \label{7b}
		\includegraphics[width=5cm,height=5cm]{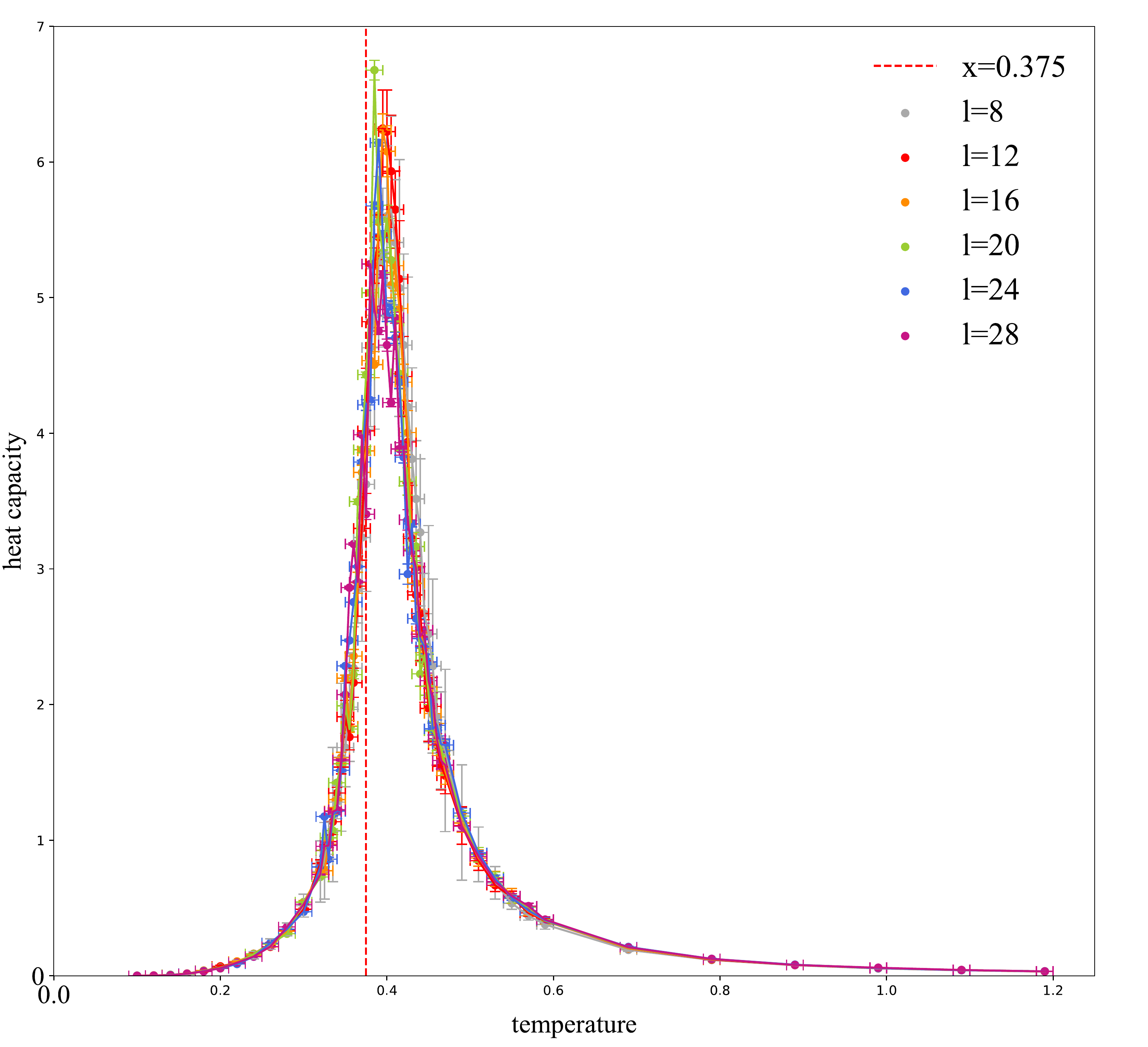}} 
		\quad	
		\subfigure[]{ \label{7c} \includegraphics[scale=0.35]{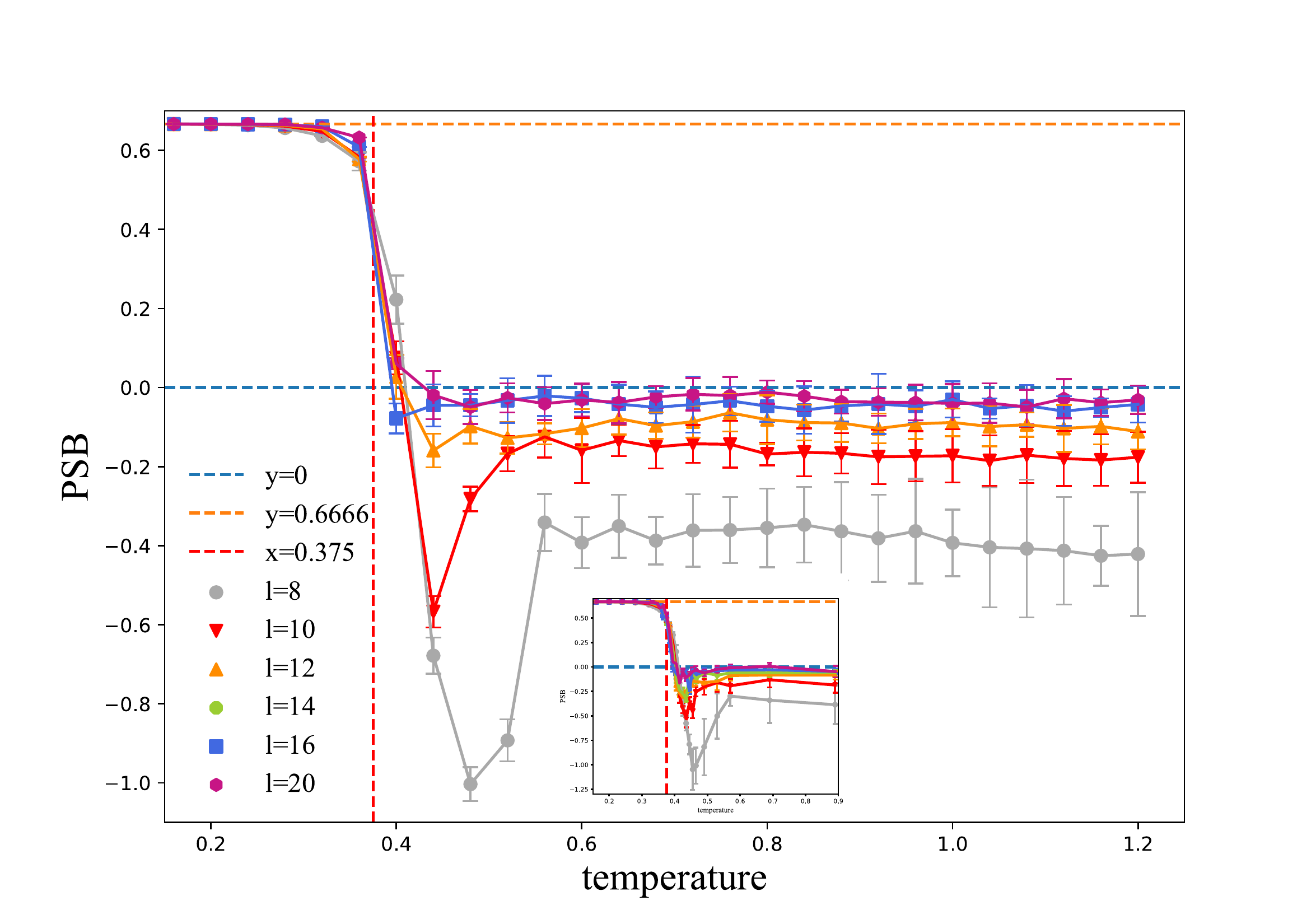}}
		\quad
		\subfigure[]{ \label{7d} \includegraphics[scale=0.32]{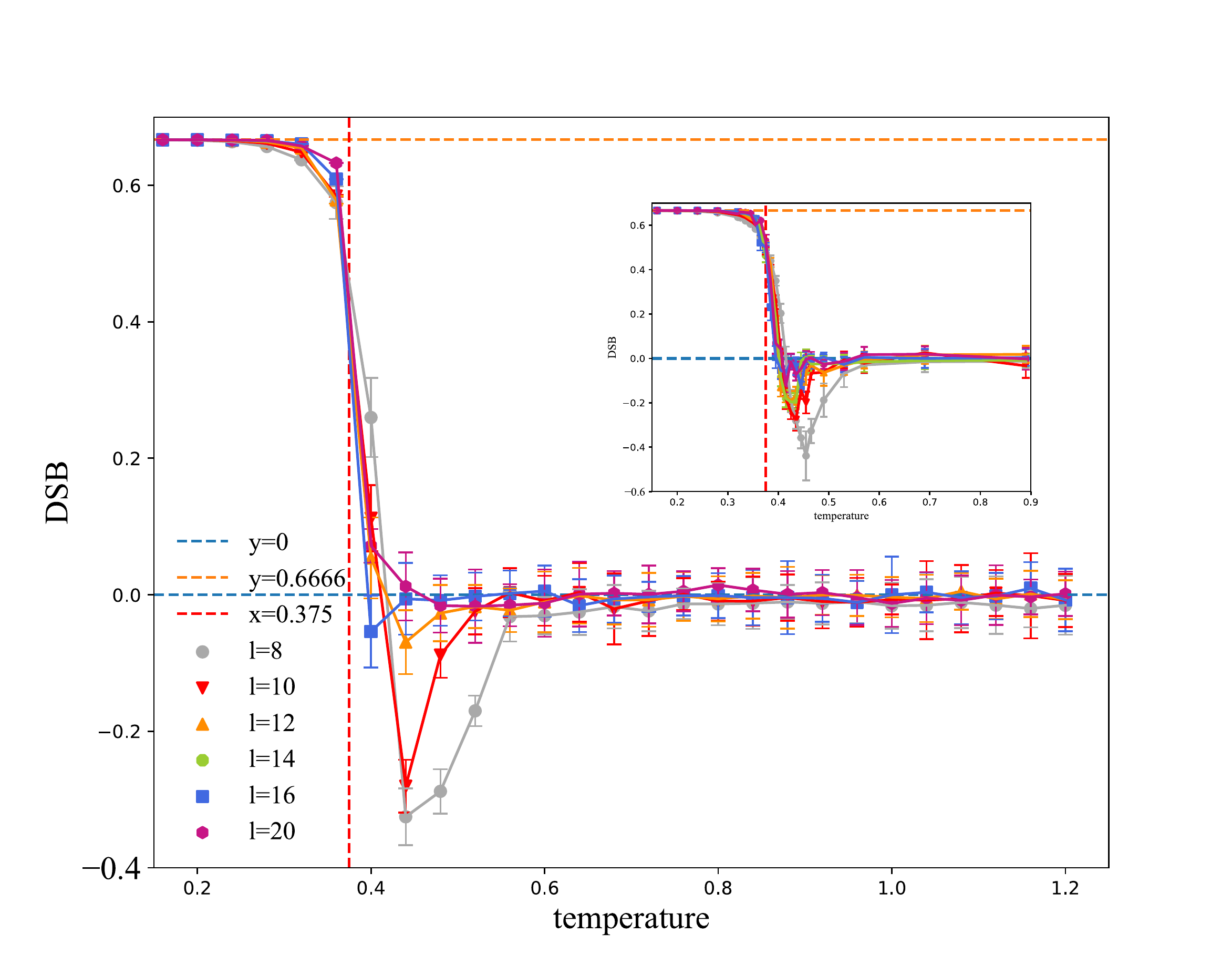}}
		\caption{(a): order parameter; (b): special heat capacity ; (c) (d):PSB and DSB Binder ratio (small figures in Fig.\ref{7c} and \ref{7d} are simulation process with more resolutions but more fluctuations in the transition).}		
	\end{figure*}

	In softed dimer model, we inspect order parameter on lattices of different size tentatively. Differing from the conclusion of classical dimer model studied by us before, order parameter at the transition temperature will decay sharply. At the limit of high temperature, we find the value of order parameter will approach to zero shown in the Fig.\ref{7a}. The simulation curves reveal there exists a BKT transition here probably and the class of the transition could turn to be a one-order or two-order thermodynamic transition.

	Then we compute the special heat capacity, Binder ratio and correlators. Specific heat capacity $c_v$ here shows great divergence at the transition temperature (see Fig.\ref{7b}). Compared with the specific heat capacity in classical dimer model, the value of $c_v$>1 at $T$=0.375 but early value $c_v<1.0$ no matter how big the size is. The peak of $c_v$ presents turbulent situation and we guess here $c_v$ could be infinity. It is restricted by the resolution of points of temperatures, that is $c_v$ may be divergent value so that the slope is too cliffy to be described by discrete points. Then, we measure the critical exponents of softed dimer model [Table \ref{tab3}]. Comparing with the critical exponents of classical dimer model, $\alpha_{0+}$ and $\alpha_{0-}$ are obviously bigger than the exponents above [Table \ref{tab1}]. On the lattice $L=20$, the exponent reaches to 12.180$\sim$14.810, which demonstrates cruel divergent trend.

	\begin{table}[h]
		\centering
		\begin{tabular}{ p{0.6cm} p{1.2cm} p{1.2cm} p{1.2cm} p{1.2cm} p{1.2cm} p{1.2cm} }
		\hline
		\hline
		&L=8 &L=12 &L=16 &L=20 &L=24 &L=28\\
		\hline
		&6.859 &9.640 &9.236 
		&12.180 &9.672 &8.882\\
		$\alpha_{0+}$&$\sim$&$\sim$&$\sim$&$\sim$&$\sim$&$\sim$\\
		&9.509 &11.360 &10.840 &14.810 &11.490 &12.660\\
		\hline	
		&-9.006 &-9.973 &-9.167 &-7.652 &-7.669 &-7.236\\	
		$\alpha_{0-}$&$\sim$&$\sim$&$\sim$&$\sim$&$\sim$&$\sim$\\	
		&-8.300 &-8.835&-8.062 &-6.451 &-6.731 &-6.236\\
		\hline
		\hline
		\end{tabular}
		\caption{critical exponents of softed dimer model transition.}\label{tab3}
	\end{table}

	For Binder ratio, whether $B_{DSB}$ or $B_{PSB}$ curves here can hardly find the characteristic cross clearly. And near the transition temperature there exists a drastic drop (see Fig.\ref{7c} and \ref{7d}). For every size of the lattice, both $B_{DSB}$ and $B_{PSB}$ will enter negative values first ($B_{DSB}\sim(-0.11,-0.50)$ and $B_{PSB}\sim(-1.0,-0.13)$), and then they would converge to zero. Zero Binder cumulants sufficiently prove there is no long-range order at the high temperature and indicate when we destroy the geometric constraint, topological relation will disappear, which means the transition is not a BKT transition. Additional explanation of our large error bars of the lattice $L=8$ is caused by the effect of finite size. At this time, Binder ratio reaches to zero difficultly and there are fluctuations near the zero.

\begin{figure}[htbp]
		\centering	
		\subfigure[]{ \label{6a} \includegraphics[scale=0.48]{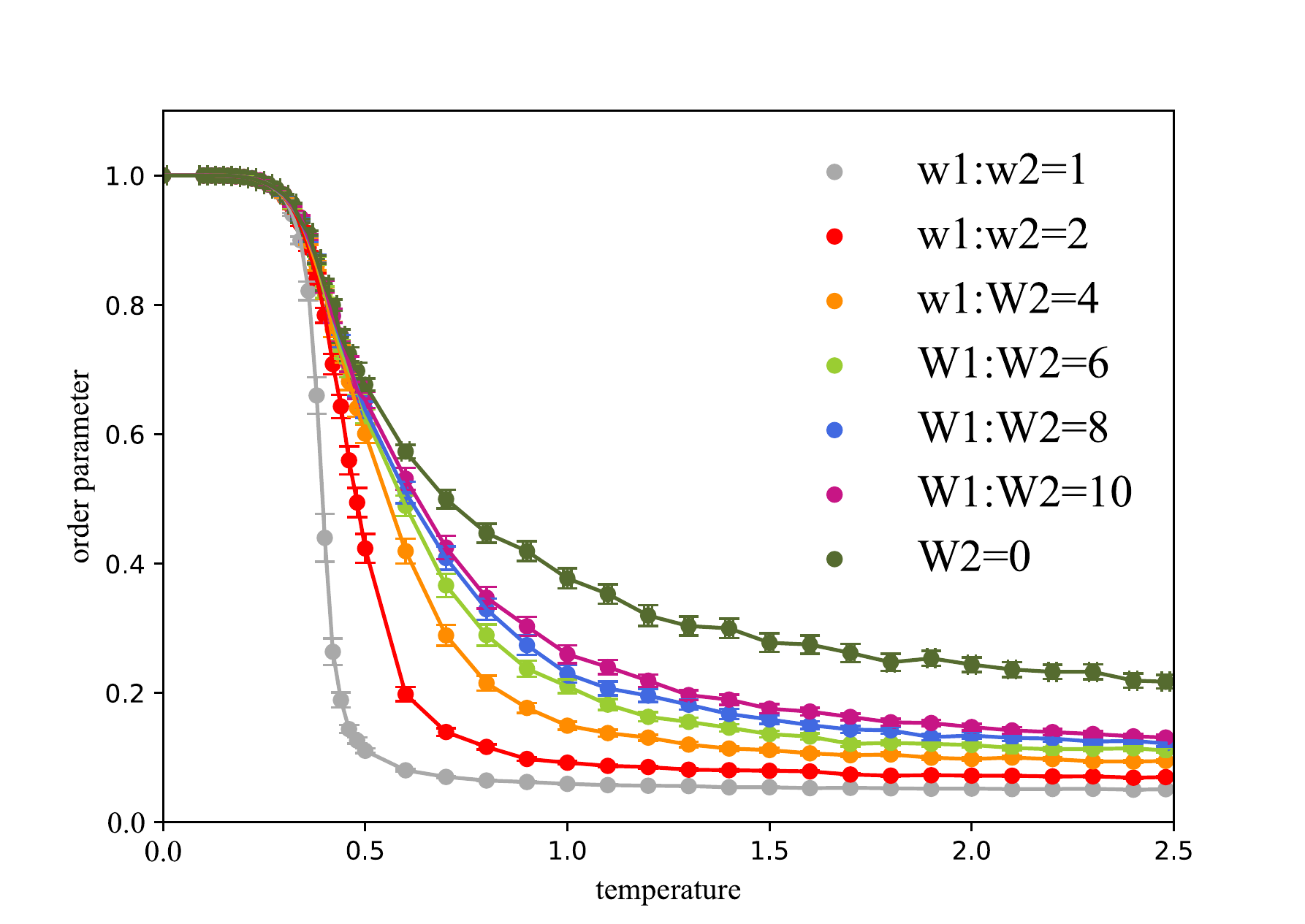}}
		\quad	
		\subfigure[]{ \label{6b} \includegraphics[scale=0.44]{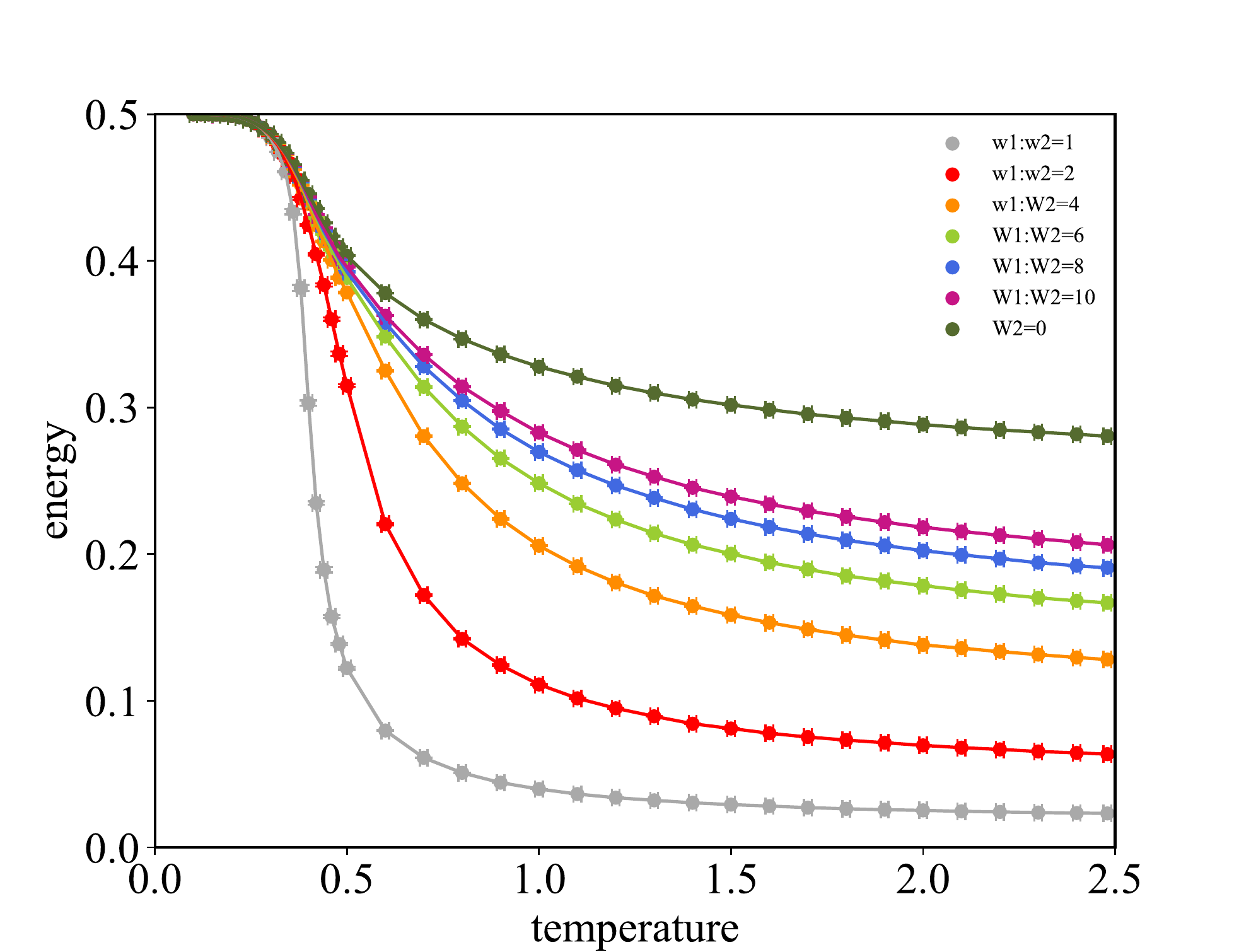}}
		\quad
		\subfigure[]{ \label{6c} \includegraphics[scale=0.44]{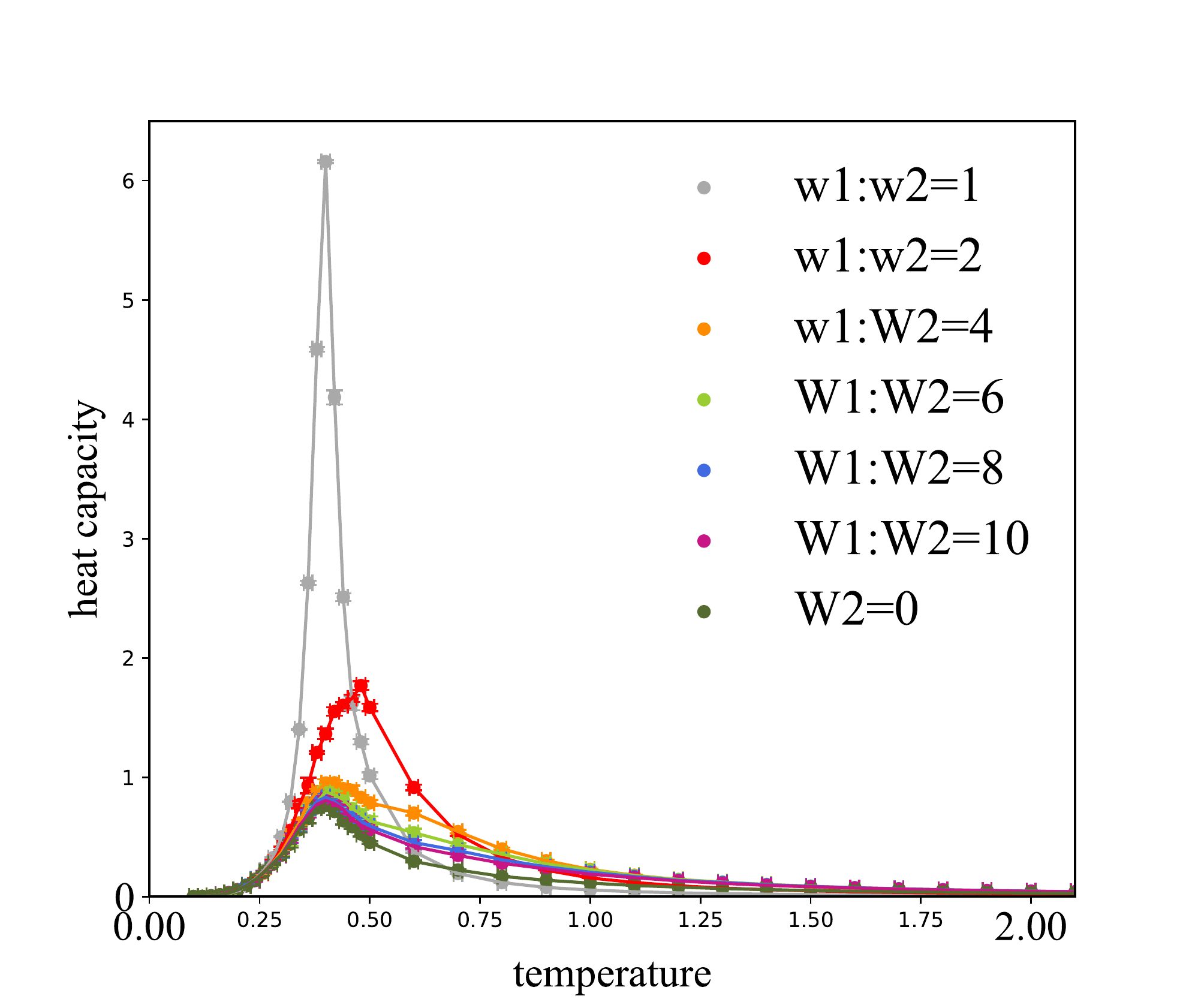}}
		\caption{ (a): order parameter; (b): energy ; (c): heat capacity.}		
	\end{figure}

	For a better look of the impact of introducing the snAB bond, we attempt to test different weight of the distribution of regular dimer bonds and snAB bonds. Consider new detailed balance of the system, we suppose the weight of dimer bond is $w_1$ and the weight of snAB bond is $w_2$. We set $w_1:w_2=1$, $w_1:w_2=2$, $w_1:w_2=4$, $w_1:w_2=6$, $w_1:w_2=8$, $w_1:w_2=10$ and $w_2=0$ seven different weight ratio on the lattice of $L=16$ and simulate the evolvements as the weight ratio change. In our results, order parameter shows a stepwise decay as the weight of $w_2$ increases. When $w_1:w_2=1$, we derive the sharply decaying curve shown in Fig.\ref{6a}. To study what the transition is when we exhaustively destroy the geometric constraint($w_1:w_2=1$), we check the energy change and the specific heat capacity change in finite temperatures (see Fig.\ref{6b} and \ref{6c}). The conclusion is the transition is a one-order transition, as the energy curve toboggans at the transition temperature and consider the analysis of critical exponents we derived before, energy change is stepped and its first-order derivative is divergent.

	Moreover, to represent the transition in softed dimer model, we describe twin peaks coexistence picture(TPCP) shown in Fig.11. Near the transition, one-order transition means two different phases will coexist. We draw TPCP in different weight of dimer bonds and snAB bonds. Initially, in softed dimer model, $w_1:w_2=1$, we can see two peaks clearly. As the weight of $w_2$ decreases, one of the peaks collapses gradually. When the raito of two weight of bonds are bigger than 6, there are no longer two packets. TPCP gives another strong evidence here to indicate the transition of softed dimer model is a one-order transiton.
	\begin{figure}[htbp]
		\centering	
		\subfigure[]{ \label{Xa} \includegraphics[scale=0.43]{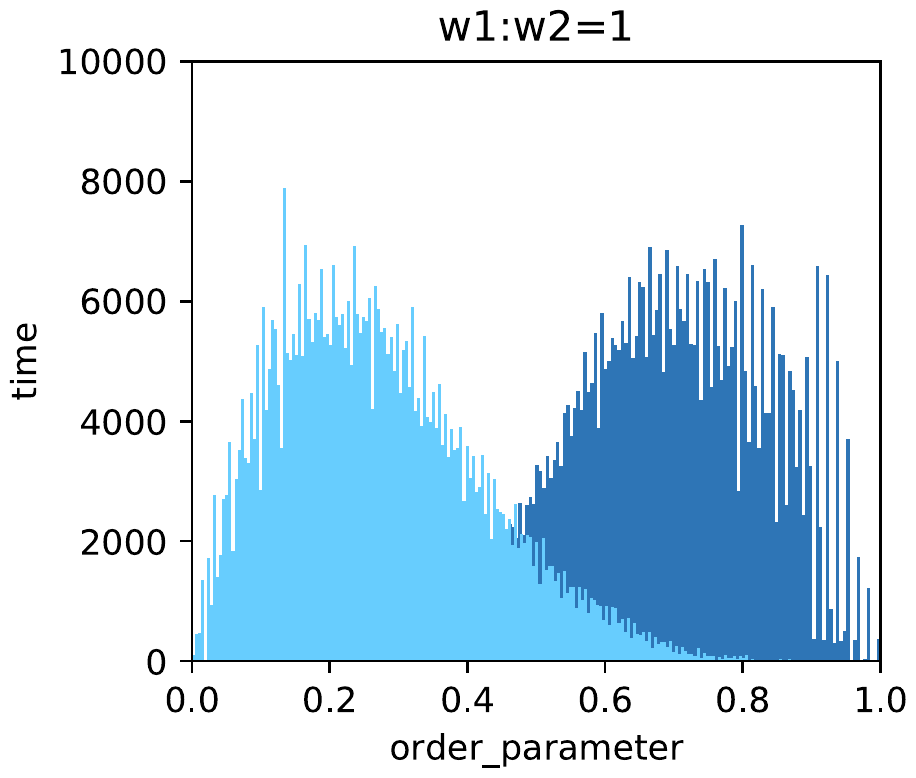}}
		\quad	
		\subfigure[]{ \label{Xb} \includegraphics[scale=0.43]{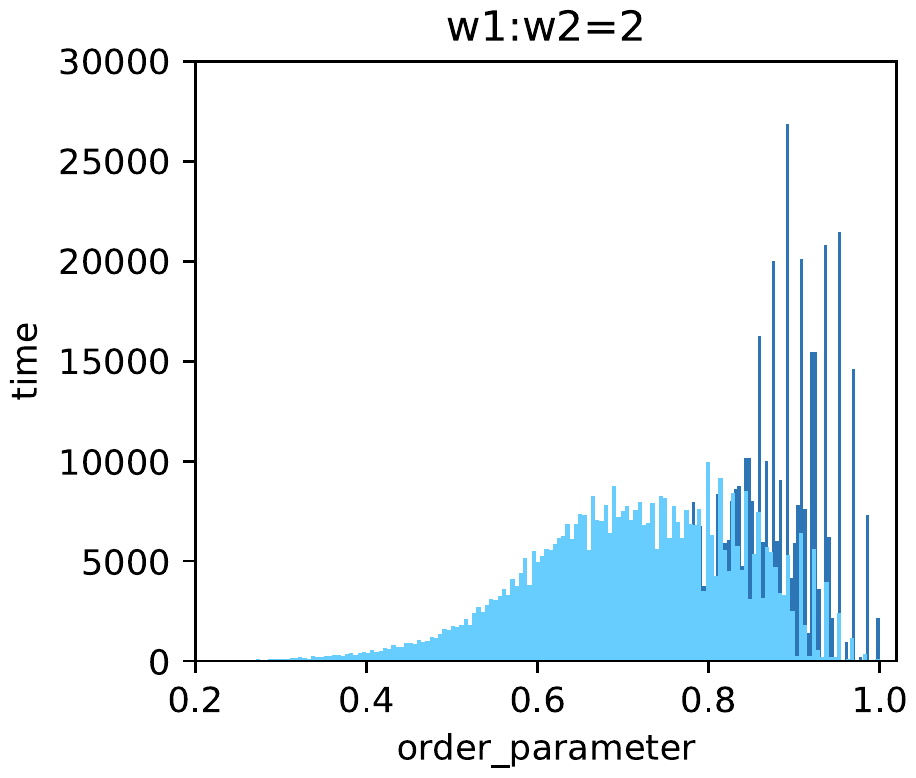}}
		\quad	
		\subfigure[]{ \label{Xc} \includegraphics[scale=0.43]{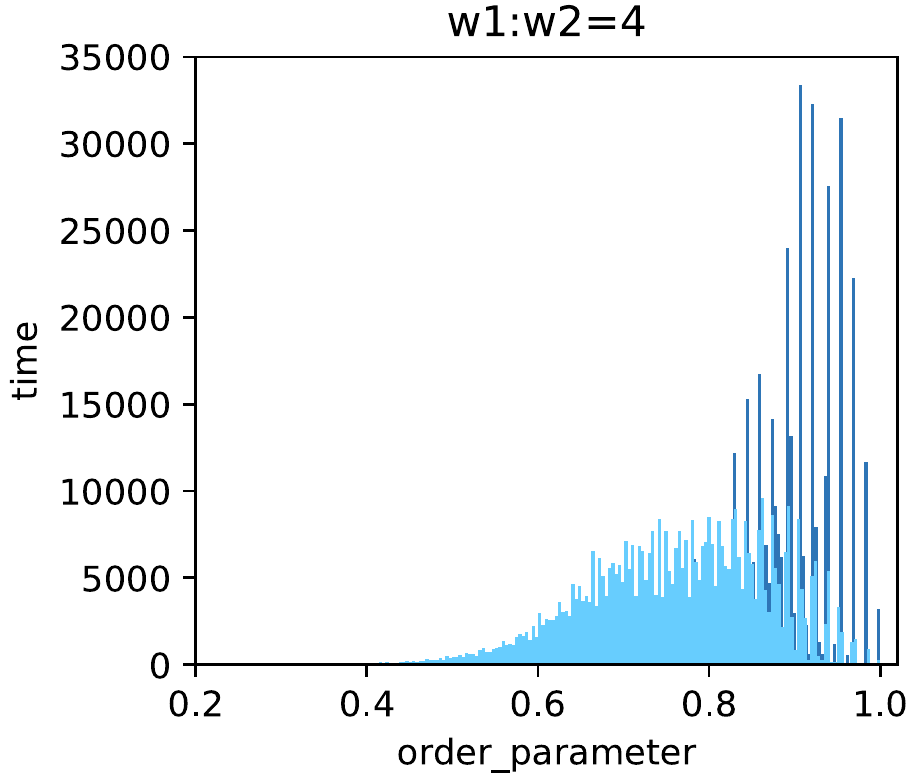}}
		\quad	
		\subfigure[]{ \label{Xd} \includegraphics[scale=0.43]{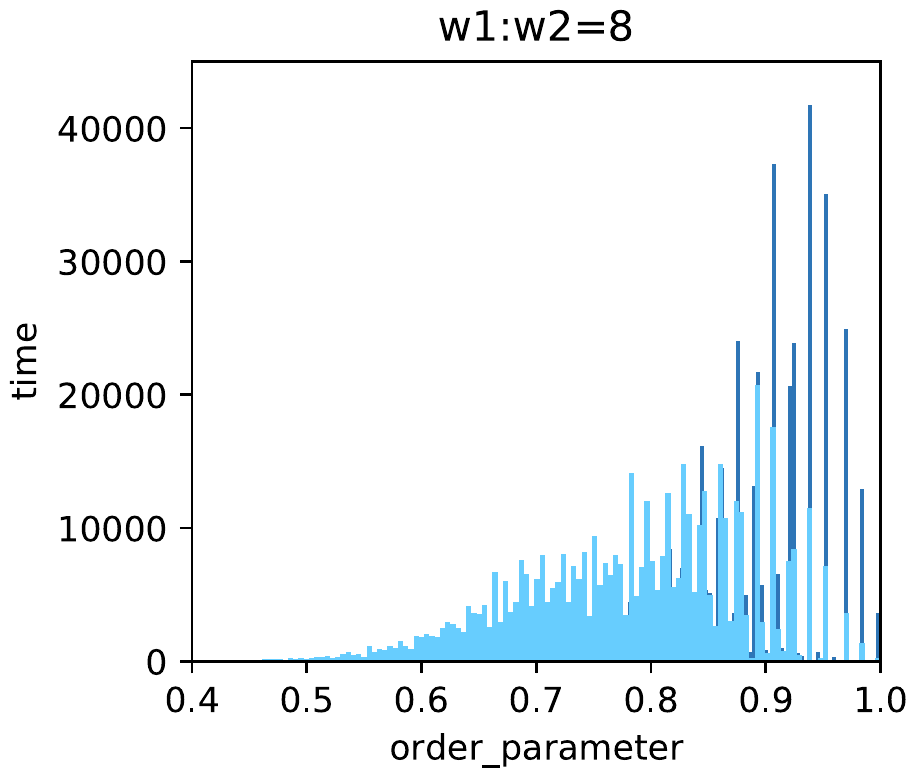}}
		\quad	
		\subfigure[]{ \label{Xe} \includegraphics[scale=0.43]{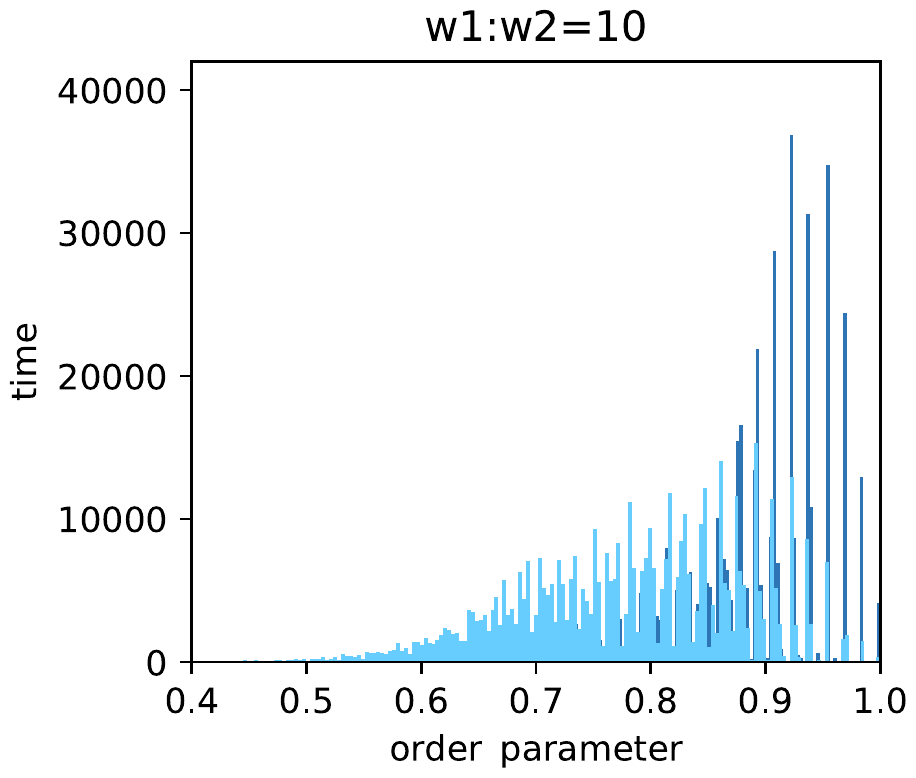}}
		\quad	
		\subfigure[]{ \label{Xf} \includegraphics[scale=0.43]{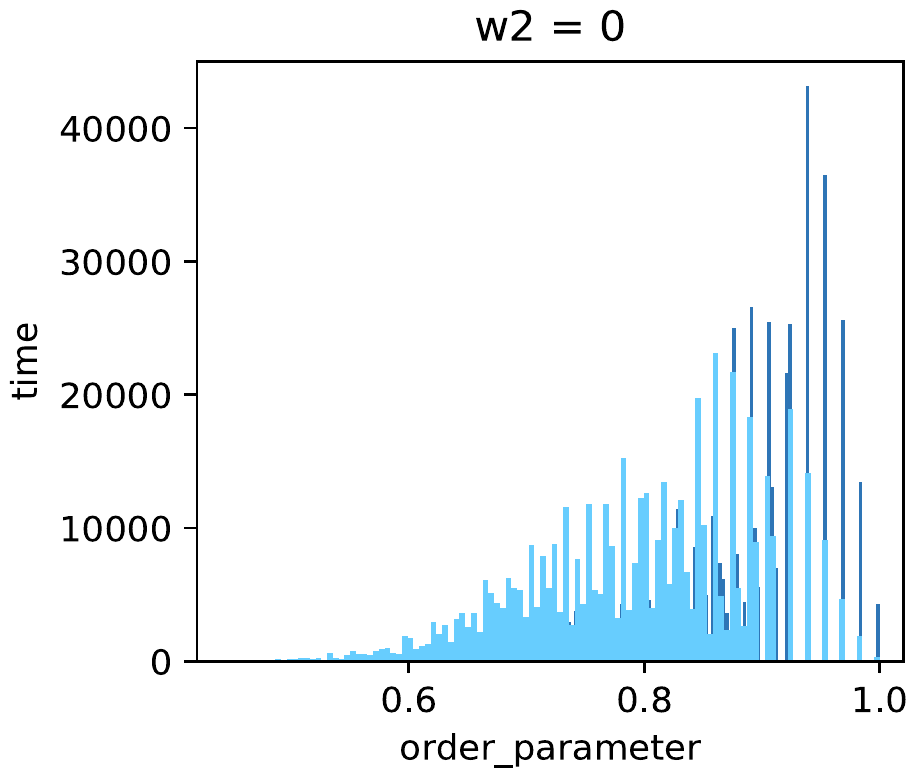}}
		\caption{Twin peaks coexistence picture (TPCP) near the transition temperature in the temperature interval of (0.38,0.40). (a): $w_1:w_2 = 1$, two peaks coexist; (b): $w_1:w_2 = 2$, left peak collapses gradually; (c): $w_1:w_2 = 4$, left peak collapses ulteriorly; (d)(e)(f): $w_1:w_2 = 8$, $w_1:w_2 = 10$, $w_2 =0$, there exits just one peak.}		
	\end{figure}
\section{Conclusion}
	In conclusion, we study classical dimer model on the square lattice, compare widely used directed loop algorithm with edged cluster algorithm introduced by us and destroy the geometric constraint of hard-core dimers to research softed classical dimer model. In classical dimer model, we find and verify the model actually undergoes a BKT transition. We identify the transition temperature $T_c$=0.411 precisely and near the transition process there still exists a quasi long-range order and this order will not disappear thoroughly.

	We introduce a new algorithm in classical dimer model and check the efficiency of it. Edged cluster algorithm can form longer loop than normal directed loop algorithm and more importantly, new algorithm succeeds in rapidly stepping over the winding (topological) sections. That means when we simulate lattice of larger size, the fluctuations generated by the lock of the topological sections could decrease dramatically. Except for that long loop can simulate more efficiently, our algorithm can broaden the ensembles of classical dimers for researchers and offer more possibilities to study lattice of larger size. 

	When breaking the geometric constraint of classical dimer model, we study thermodynamic properties of softed dimer model and make certain the impact of the weight of different bonds. In our softed dimer model, we find the BKT topological transition disappears and the effect of long-range order weaken. We similarly identify the transition temperature $T_c=0.376$. In virtue of that Binder ratio approaches to zero at the infinite temperature and specific heat capacity $c_v$ changes at finite temperatures, we could argue the transition in softed dimer model transfers to a quasi one-order transition. The energy curve sharply decay at the transition temperature and critical exponents forcefully increase ($|\alpha|\sim(7,9)$, previously in the classical dimer model $|\alpha|\sim(1,3)$). To sum over, energy change is stepped and its first-order derivative is divergent, thus the transition here is a  one-order classical thermodynamic transition.
\begin{acknowledgments}		
	``Cruel to be kind!" Firstly, we wish to thank Prof. Yan Chen for giving us this precious chance to finish this work and Prof. Jie Lou and Dr. Zheng Yan for their hard but fruitful discussions with us. We must thank Renfei Gao for offering his computer to support us to complete plenty of calculations. Thank Di Wu and Delong Jiang for their assistance of using showmakers.
	Unforgettably, thank Mr. Big for his endurance of our noisy discussion.
\end{acknowledgments}
\appendix
\section{Topological winding numbers}\label{app1}
\begin{figure}[h]
		\centering
		\subfigure[]{ \label{10a} \includegraphics[scale=0.2]{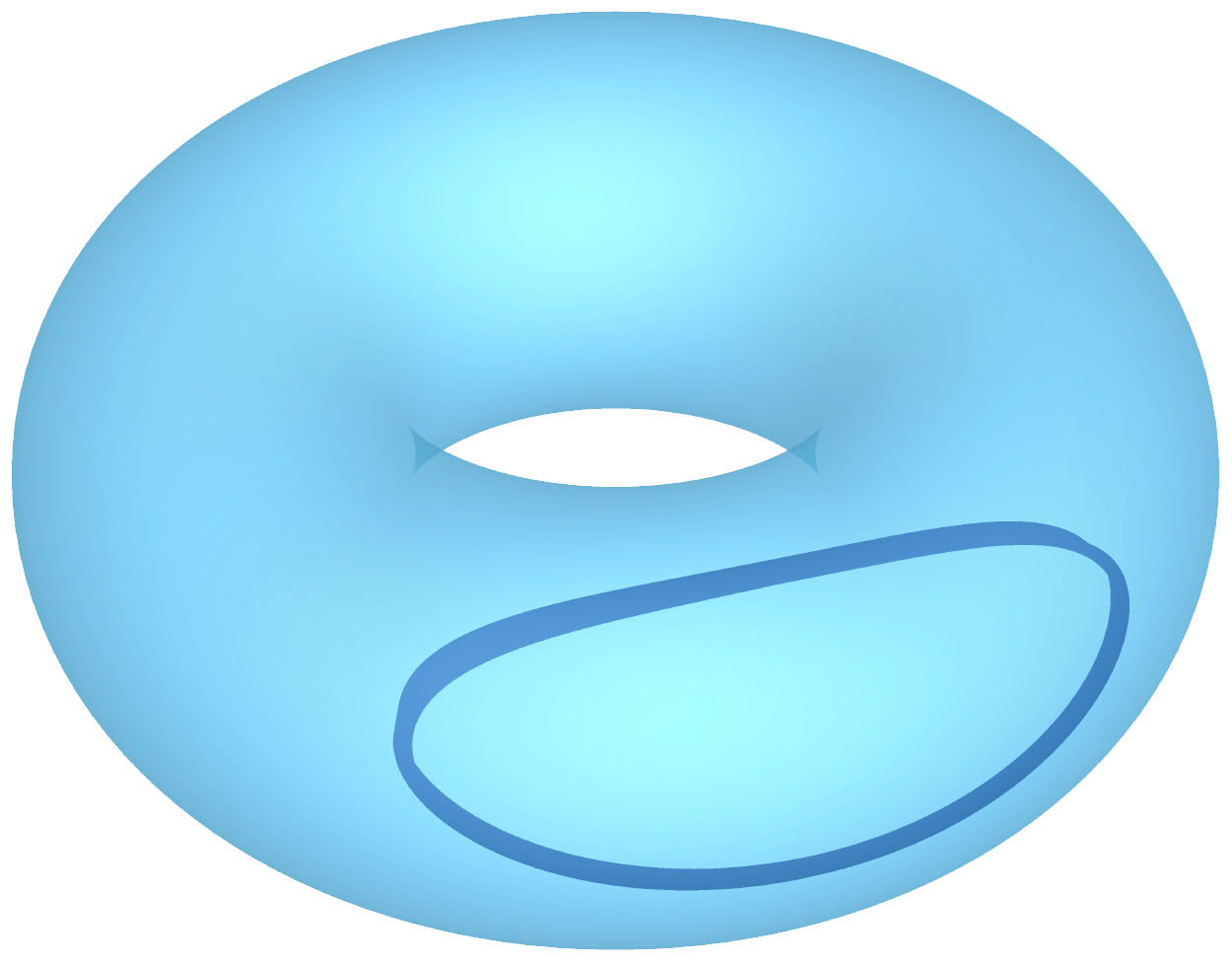}}
		\quad
		\subfigure[]{ \label{10b} \includegraphics[scale=0.2]{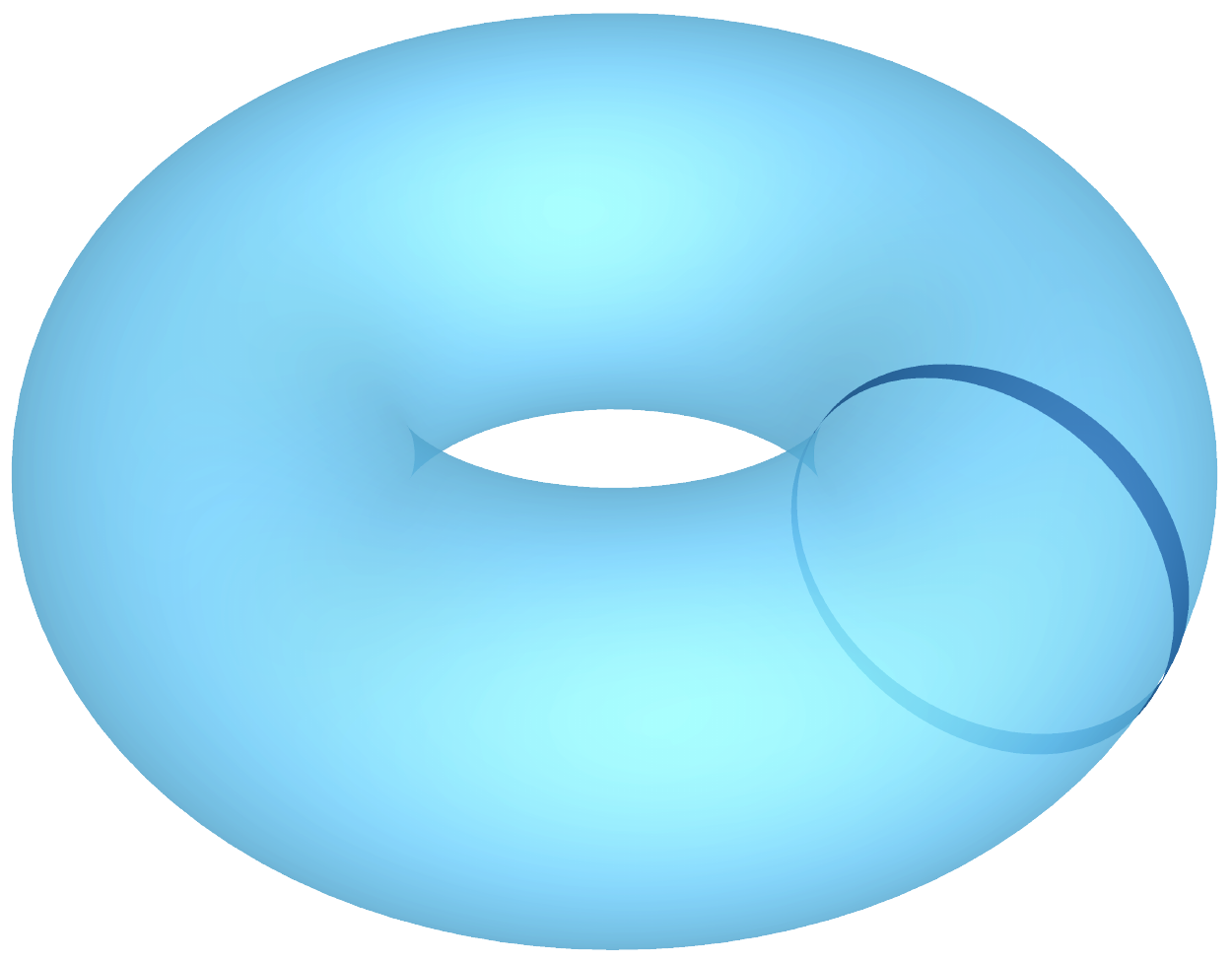}}
		\quad
		\subfigure[]{ \label{10c} \includegraphics[scale=0.2]{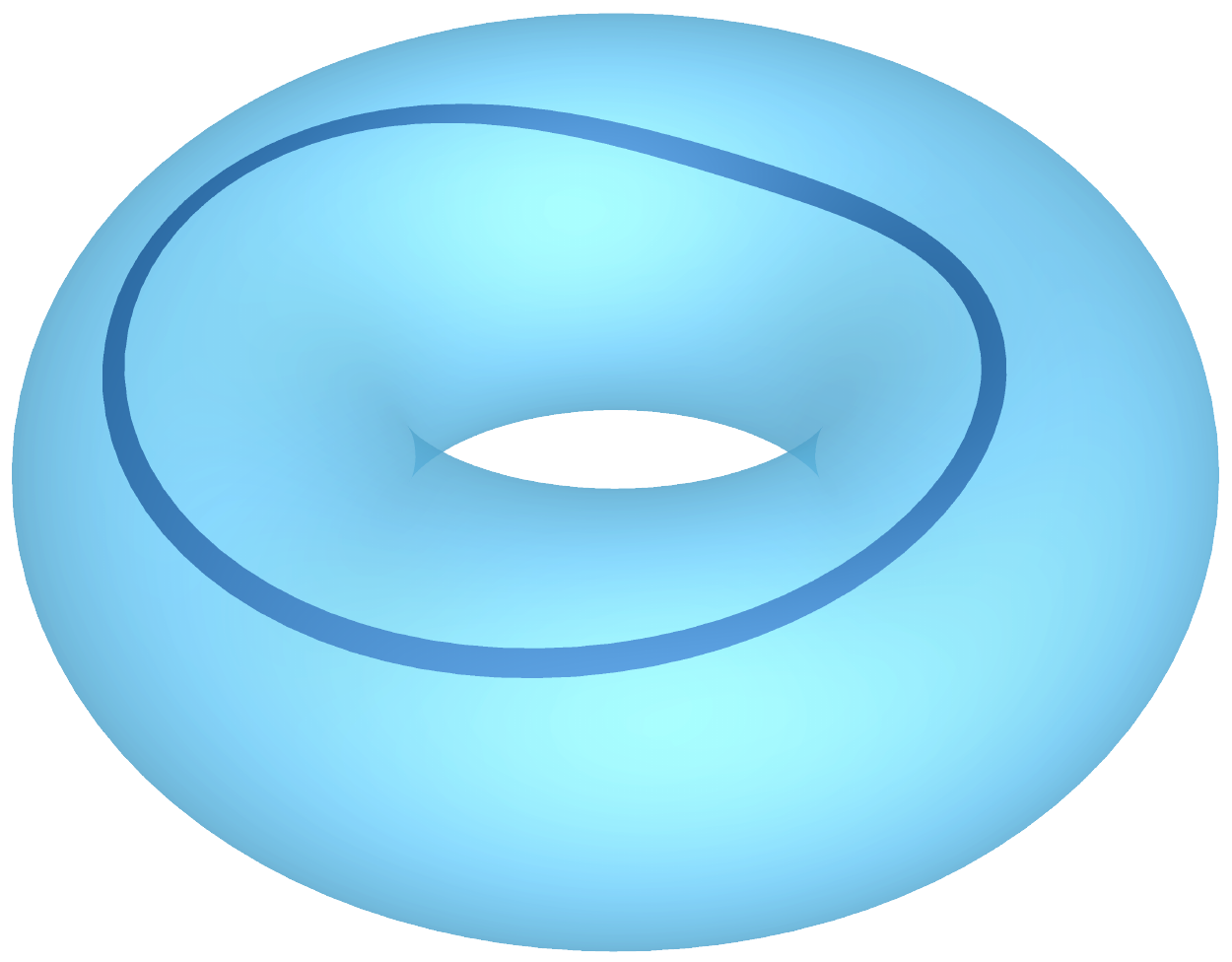}}
		\quad
		\subfigure[]{ \label{10d} \includegraphics[scale=0.2]{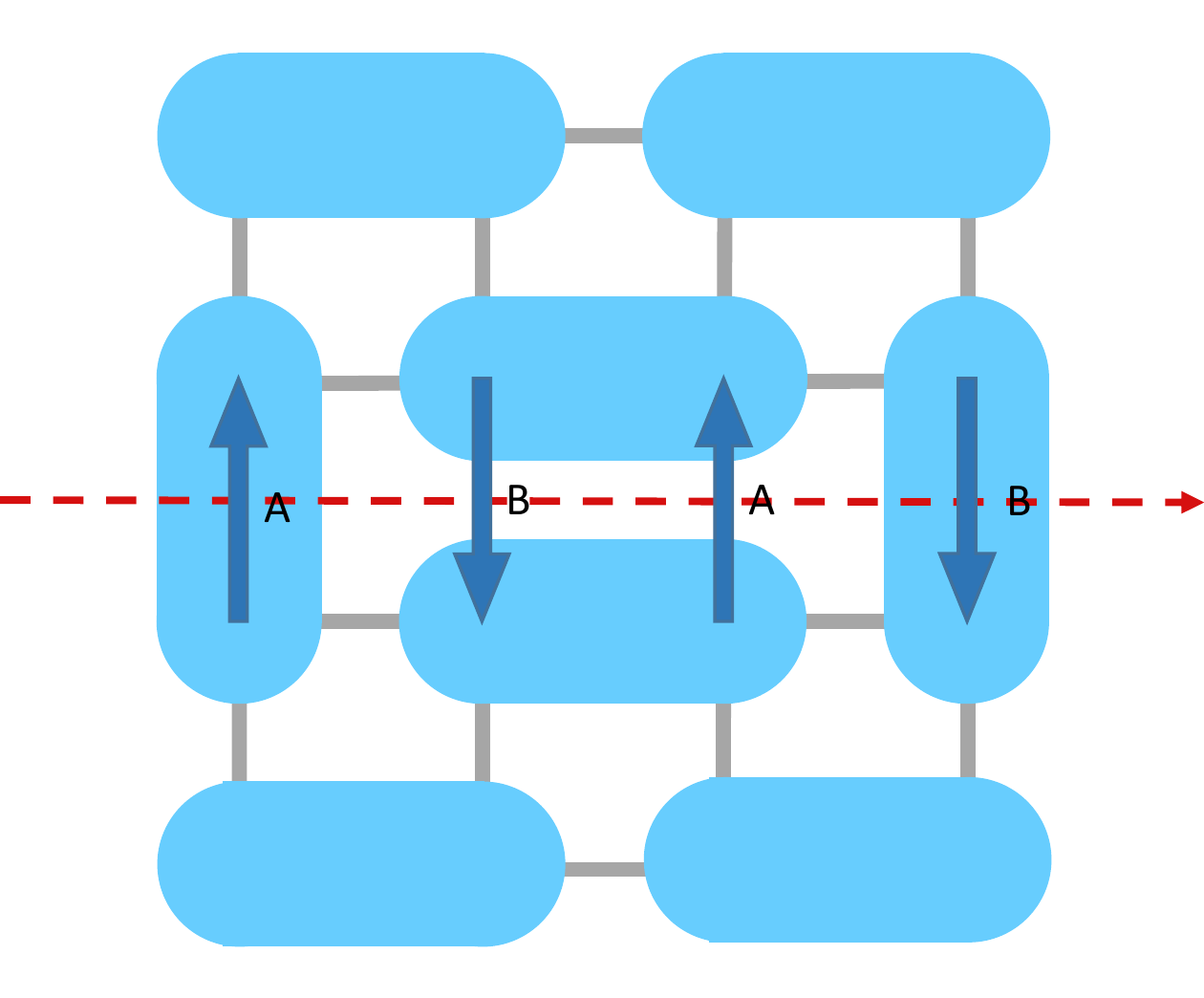}}
		\caption{(a): local loop on the torus and its winding number is (0,0); (b): loop circles around the pipe of the torus and its winding number is (0,1); (c): loop circles around the hole of the torus and its winding number is (1,0). (d): method used to compute the winding number of classical dimer model.}
	\end{figure}

	The start point of topological winding number defined from the algebraic topology. Different homotopy loop equivalence class corresponds to different winding numbers. For instance, the shape of the topological space studied by us is a torus and we draw three different loops on it (see Fig.\ref{10a}, \ref{10b}, and\ref{10c}). The winding numbers of these three loops are different. For loop A, it can be shrunk into a point, so the winding number of it is (0,0). But loop B and loop C can not be shrunk becaus, they circle holes of the torus. Particularly, holes interdicting loop B and loop C are vertical and horizonal two holes, thus their winding numbers are different, $W_B=(1,0)$ and $W_C=(0,1)$. Obviously, we can define direction of the loop to complete our range of winding number.	

	In our model, different winding number means different classes of physical quantities. Periodic boundary means the top of the lattice is connected to the bottom and the left of the lattice is connected to the right that yields the topological space of our model is equivalent to a torus. The loops through the lattice from left (right) to the right (left) (from top (bottom) to the bottom (top)) correspond to their own winding numbers. Topological sections are characterized by a pair of conserved winding numbers $(W_x,W_y)$. On $L\times L$ square lattice, allowed winding numbers are $-L/2 \le W_x,W_y \le L/2$. Two different equivalent definitions of topological winding number has been defined before \cite{2020Improved, 1996Columnar}. The first definition follows RK’s original definition \cite{1988Superconductivity}. But more widely used, we just introduce the following definition in detail.
	\begin{equation}
		W_x=N_y(A)-N_y(B)
	\end{equation} 	
	\begin{equation}
		W_y=N_x(A)-N_x(B)
	\end{equation} 	
	Where $N_x(A)$, $N_x(B)$, $N_y(A)$ and $N_y(B)$ are the numbers of dimers cut by the dashed line on A or B links shown in Fig.\ref{10d}. Vertical dimers are cut by the red dashed line in $x$ direction. No matter which row (column) along, we will derive same winding number $W_y(W_x)$. Every state of classical dimer corresponds to a pair of winding numbers. The whole ensemble of the microstates will be divided into different topological sections and every section has a characteristic physical quantities in average. 
\section{Detailed Balance}\label{app2}
\begin{figure}[h]
		\centering	
		\subfigure[]{ \label{13a}  
		\includegraphics[scale=0.4]{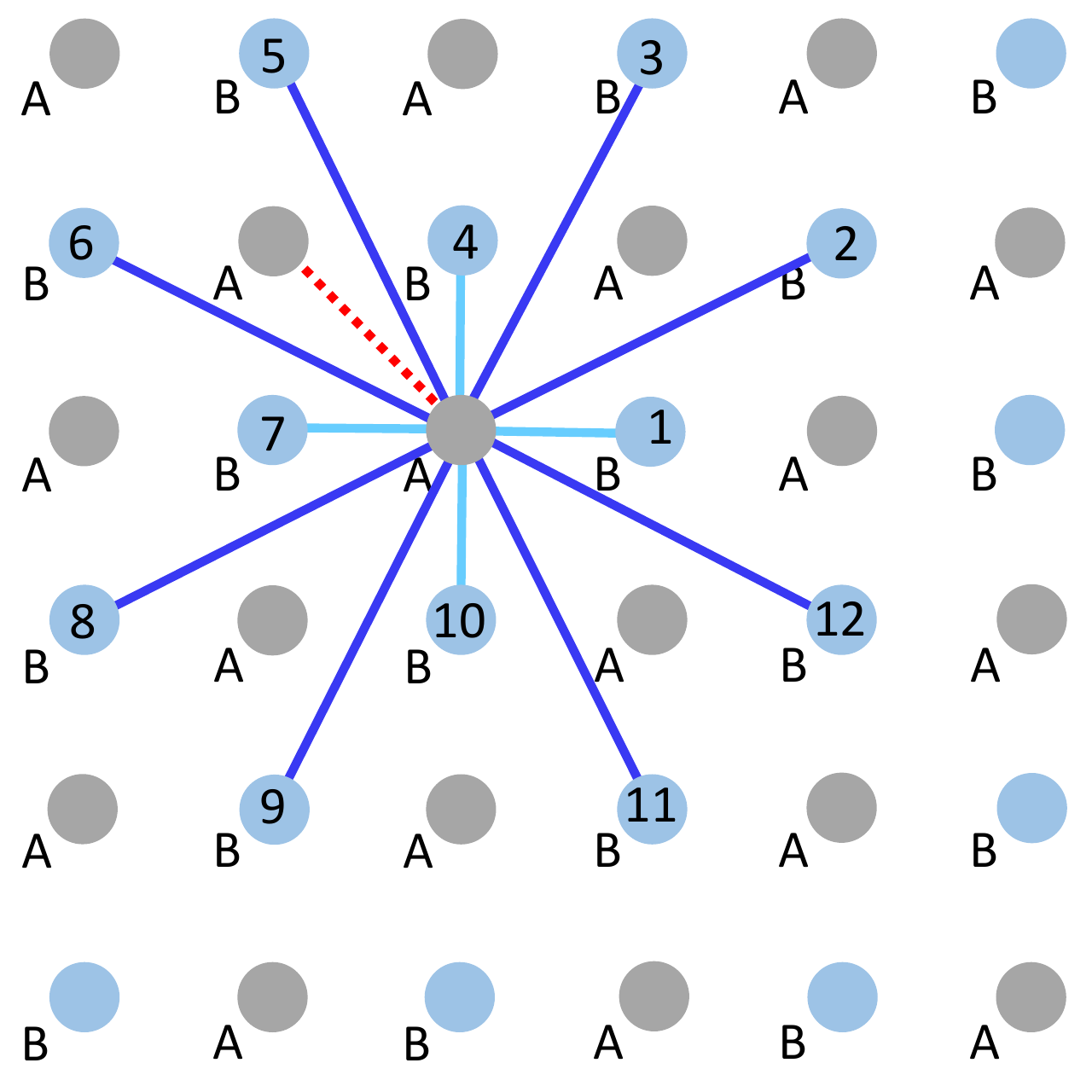}}
		\caption{twelve vertexes could bond and twelve possible bonds. For example, one term of bond is the directed loop update enter from 1 and exit from 2, and the other is enter from 2 and exit from1.}	
	\end{figure}

	Appendix B will show that the detailed balance of the directed loop algorithm used in the softed dimer model. Detailed balance in the classical dimer model has been proven in the early article \cite{2003The}, but if we adjust the weight of dimer bonds and snAB bonds, we must recompute the detailed balance equation. The key graph is shown in the above Fig.\ref{13a}, and we need to consider two special enter-exit directions. One is the loop enters from the point bonded as dimer, steps through the dimer and exits from the vertex bonded as snAB bond soon (see in Fig.\ref{13a}, for instance, enters from the vertex 1 through the dimer and exits from the vertex 2). Otherwise, the loop enters from the point bonded as snAB bond and through it, exits from the vertex bonded as dimer soon(see in Fig.\ref{13a}, for instance, enters from the vertex 2 through the dimer and exits from the vertex 1).

	The following is the computation process of the detailed balance equation. Suppose $w_1$ and $w_2$ are the weights of dimer bond and the snAB bond. Weight $a_{jk}$ is defined for the process in which $a$ vertex in state $j$ is entered at site $j$ and exited at $k$ shown in the Fig.\ref{13a}.
	The actual probabilities $P_{jk}=a_{jk}/w_j$  which means $\sum_k a_{jk}=w_j$. That is, in softed dimer model, we distinguish two classifications as $s$ (snAB) and $d$ (dimer). There exits four different weights: $a_{sd}$, $a_{ss}$, $a_{ds}$ and $a_{dd}$. For detailed balance, we have $a_{ds}=a_{sd}$, thus 
	\begin{equation}
		w_1=3a_{dd}+8a_{ds}
	\end{equation}
	\begin{equation}	
		w_2=7a_{ss}+4a_{sd}
	\end{equation}
	\begin{equation}
		a_{ds}=a_{sd}
	\end{equation}
	There are infinite number of positive-definite solutions and for convenience, we can set
	\begin{equation}
		a_{ds}=a_{sd}=1
	\end{equation}	
	Finally, we achieve the detailed balance in the softed dimer model.	
	
\bibliographystyle{unsrt}	
\bibliography{statisticalarticle}
\end{document}